\documentclass[twocolumn,english]{revtex4-1}
\usepackage[T1]{fontenc}
\usepackage[latin9]{inputenc}
\usepackage{color}
\usepackage{array}
\usepackage{amsmath}
\usepackage{graphicx}
\makeatletter

\providecommand{\tabularnewline}{\\}

\usepackage{times}
\usepackage{epstopdf}
\usepackage{epsfig}

\makeatother

\usepackage{babel}
\begin{document}

\title{Nonclassicality in non-degenerate hyper-Raman processes}

\author{{Kishore Thapliyal$^{\dagger}$, Anirban Pathak$^{\dagger}$,
Biswajit Sen$^{\ddagger}$, and Jan }Pe{\v{r}}ina{$^{\star,\mathsection}$ }}

\affiliation{$^{\dagger}$Jaypee Institute of Information Technology, A-10, Sector-62,
Noida, UP-201307, India~\\
$^{\ddagger}$Department of Physics, Vidyasagar Teachers' Training
College, Midnapore-721101, India~\\
$^{\star}$RCPTM, Joint Laboratory of Optics of Palacky University
and Institute of Physics of Academy of Science of the Czech Republic,
Faculty of Science, Palacky University, 17. listopadu 12, 771 46 Olomouc,
Czech Republic~\\
$^{\mathsection}$Department of Optics, Palacky University, 17. listopadu
12, 771 46 Olomouc, Czech Republic}
\begin{abstract}
{A perturbative analytic operator solution of a completely
quantum mechanical Hamiltonian of multi-photon pump
non-degenerate hyper-Raman process is obtained. It is shown that the
obtained solution is  general in nature as the solutions of non-degenerate hyper-Raman and stimulated Raman processes can be obtained as special
cases of the present solution. The analytic solutions obtained here
are used to investigate the nonclassical properties of the different
modes in the stimulated, spontaneous and partially spontaneous multi-photon pump non-degenerate hyper-Raman processes. The nonclassical nature of these processes is witnessed
by means of single mode and intermodal quadrature squeezing, intermodal
entanglement of different orders, lower order and higher order photon
antibunching. Interestingly, manifesting the multiphoton nature of the pump modes, a bunch of
nonclassicality involving them are observed due to self-interaction of various pump modes.}
\end{abstract}
\maketitle

\section{Introduction \label{sec:Intro}}

Applications of nonclassical states can only manifest the true power of quantum mechanics. This is so because the working of any technology that does not use nonclassical state(s) can be understood/explained classically (i.e., without using quantum mechanics). Recently, many applications of nonclassical states manifesting the power of quantum mechanics have been reported. Specifically, squeezed vacuum state has been used in the  detection of gravitational wave \cite{abbott2016observation,abbott2016gw151226}
at the Laser Interferometer Gravitational-Wave Observatory (LIGO). Further,  with the recent progresses in the field of quantum
computation and communication, the importance and necessity of nonclassical states have
been established strongly established. For example, it has been established that the entangled  states are essential for
the implementation of a set of schemes for quantum cryptography  \cite{ekert1991quantum,thapliyal2015applications,thapliyal2017quantum},
quantum teleportation \cite{bennett1993teleporting}, dense-coding
\cite{bennett1992communication}; Bell nonlocal states are required for device independent quantum key distribution (DI-QKD) \cite{acin2006bell};  squeezed states are useful for continuous
variable quantum cryptography \cite{hillery2000quantum} and  antibunched states are  useful in
building single photon sources \cite{yuan2002electrically,pathak2010recent}. These interesting applications of nonclassical states have motivated many groups to investigate the possibilities of generating nonclassical states using frequent and important physical processes. One such important physical process is Raman process, which has several variants (e.g., spontaneous Raman scattering, stimulated Raman scattering, degenerate and non-degenerate hyper-Raman scattering, coherent anti-Stokes Raman scattering, coherent anti-Stokes hyper-Raman scattering) and clubbing them together we refer to them as Raman processes. Among these Raman processes non-degenerate hyper-Raman process is most general in nature
 as the Hamiltonians of  spontaneous and stimulated Raman scattering and non-degenerate hyper-Raman process (in the simplest  case which can be viewed as a 3 photon analogue of  stimulated Raman scattering \cite{sen2007squeezing}) can be obtained as limiting cases of it. Nonclassicality  in spontaneous and stimulated Raman scattering (see \cite{sen2011sub,sen2007quantum,sen2005squeezed,sen2008amplitude}
and references therein, Section 10.4 of \cite{perina1991quantum}
and \cite{miranowicz1994quantum} for reviews) and non-degenerate hyper-Raman processes \cite{perinova1979quantum2,szlachetka1980photon,perinova1984sub,sen2007squeezing} has already been studied in detail. However, non-degenerate hyper-Raman process has not yet been investigated rigorously because of its inherent mathematical complexity and the potential difficulties associated with the experimental realization of this process. This is what motivated us to investigate the possibilities of observing nonclassical effects in the non-degenerate hyper-Raman process. 

We were further motivated by the fact that in Ref. \cite{olivik1995non}, Oliv{\'\i}k and Pe{\v{r}}ina noted
that higher order non-linearity present in the hyper-Raman process may lead to more significant nonclassical effects compared to the standard Raman process (at least in the context of the statistical properties of radiation fields
and quadrature squeezing). In fact, hyper-Raman scattering represents a very interesting nonlinear optical
process as it allows self-interaction of the pump modes and thus  leads to the generation of different types of nonclassicality.
Specifically, the presence of antibunched, sub-Poissonian and squeezed light in the degenerate hyper-Raman processes has  already been reported in the past \cite{perinova1979quantum2,szlachetka1980photon,perinova1984sub,sen2007squeezing}. However, only antibunching in the photon and phonon modes of its  non-degenerate counterpart have been reported until now \cite{perinova1979quantum}.

The fact that the limiting cases of the non-degenerate hyper-Raman process have found application in various spheres of modern science has also motivated us to perform the present study. To be precise, quantum repeater \cite{grangier2005quantum,duan2001long} has been built using the spontaneous Raman process; stimulated Raman scattering has been used to design  devices for laser cooling of solids \cite{rand2013raman}, highly sensitive
label-free biomedical imaging \cite{freudiger2008label}, imaging
of a degenerate Bose-Einstein gas \cite{sadler2007coherence} and to design a quantum random number generator (QRNG) \cite{bustard2011quantum} which is a true random number generator having no classical analogue. The multi-photon processes, such as hyper-Raman processes, may reveal many-body correlation functions and thus useful information regarding the nonlinear medium (see \cite{kielich1993multi} for a review). A set of  possibilities  for experimentally observing these processes have been discussed since long \cite{ziegler1990hyper}. One such possibility was reported in Ref. \cite{french1975versatile}, where the output of a hyper-Raman spectrometer was illustrated and analyzed.  Theoretical  proposals for studying hyper-Raman spectroscopy are still of prime interest \cite{valley2010theoretical,butet2015surface}. Specifically, these multi-photon processes possess a particular experimental advantage as their signals are spectrally well separated from the input laser \cite{butet2015surface}. It is also shown in the past that due to specific selection rules involved in these processes they can reveal information not accessible by Raman and infrared spectroscopy \cite{butet2015surface}. Further, nanosensors based on the surface-enhanced hyper-Raman processes enable measurement of wide range of pH circumventing use of multiple probes \cite{kneipp2007one}.
Also, due to wide applications of the hyper-Raman processes and other nonlinear optical phenomena in quantum information processing tasks, its analogues with single atoms and virtual photons are also proposed \cite{kockum2017deterministic}. In addition, with the recent growth in the experimental facilities a set of experimental results using hyper-Raman scattering has been presented \cite{kneipp2007one,kozich2007non}. A brief review of numerous applications and the future scopes of hyper-Raman processes may be found in \cite{madzharova2017surface}.

Motivated by the above, to investigate the possibilities of observing nonclassical features in non-degenerate hyper-Raman process (illustrated in Figure \ref{fig:scheme}),  a completely quantum
mechanical description of the system is used here to construct a Hamiltonian of the system. To obtain a closed analytic expression for the time evolution of each mode involved here, we have used Sen-Mandal perturbative technique (\cite{sen2005squeezed,thapliyal2014higher,thapliyal2014nonclassical,thapliyal2016linear}
and references therein), which is known to be a superior method compared to the corresponding short-time technique \cite{perina1991quantum,szlachetka1979dynamics,szlachetka1980photon}.  Further, we have established the general nature of the obtained solution by obtaining the existing Sen-Mandal solutions of Raman and degenerate hyper-Raman processes \cite{sen2005squeezed,sen2007quantum,sen2008amplitude,sen2011sub,sen2007squeezing}, (as limiting cases of the solution obtained here) which were already reduced to corresponding short-time solutions in the past \cite{perina1991quantum,szlachetka1979dynamics,szlachetka1980photon}. Subsequently, the obtained time evolution of all the photon and phonon modes has allowed us to use a finite set of moments-based  criteria \cite{miranowicz2010testing} to establish the highly nonclassical behavior of the hyper-Raman processes. Specifically, the model (Hamiltonian) used here is capable of dealing with
the stimulated, spontaneous and partially spontaneous non-degenerate hyper-Raman process considering some or all the modes as stimulated. In all these cases, we have analyzed the possibilities of generating lower and higher order single mode nonclassicality. Specifically, in what follows, we would investigate the possibilities of observing single mode antibunched and squeezed states, and compound mode nonclassicality as intermodal squeezing, antibunching and entanglement. Further, feasibility of higher order entanglement in the hyper-Raman processes is examined. 

The remaining part of the paper is organized as follows. The model Hamiltonian
for the non-degenerate hyper-Raman process and its solution is reported
in Section \ref{sec:The-model-Hamiltonian}. A list of criteria to be used for the 
study of the nonclassical properties of the non-degenerate hyper-Raman process are given
in Section \ref{sec:Criteria-of-nonclassicalities}. In Section \ref{sec:Nonclassicality-observed},
we summarize our results illustrating the presence and evolution of various types of nonclassicality and discuss the obtained results in detail
before finally concluding the paper in Section \ref{sec:Conclusion}.

\section{The model Hamiltonian and its solution \label{sec:The-model-Hamiltonian}}

\begin{figure}[h]
\begin{centering}
\includegraphics[scale=0.8]{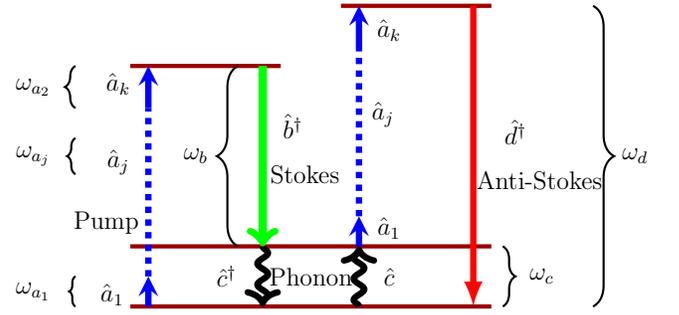}\caption{\label{fig:scheme}(Color online) The schematic energy diagram for
multi-photon non-degenerate hyper-Raman processes. Here, we have shown
$k$ pump $\left(a_{i}\right),$ Stokes $\left(b\right),$ vibrational
(phonon) $\left(c\right),$ and anti-Stokes $\left(d\right)$ modes.}
\par\end{centering}
\end{figure}

The most general Hamiltonian of the hyper-Raman processes is

\begin{equation}
\begin{array}{lcl}
H & = & \sum_{i=1}^{k}\omega_{i}a_{i}^{\dagger}a_{i}+\omega_{b}b^{\dagger}b+\omega_{c}c^{\dagger}c+\omega_{d}d^{\dagger}d\\
 & - & \left(g\prod_{i=1}^{k}a_{i}b^{\dagger}c^{\dagger}+\chi^{*}\prod_{i=1}^{k}a_{i}cd^{\dagger}+{\rm h.c.}\right),
\end{array}\label{eq:hamiltonian}
\end{equation}
where $a_{i}$ is the annihilation operator for $i$th laser (pump)
mode, $b,\,c,$ and $d$ are the annihilation operators corresponding
to Stokes, phonon (vibration) and anti-Stokes modes, respectively.
The Hamiltonian given in Eq. (\ref{eq:hamiltonian}) corresponds to
$k$-pump modes in the non-degenerate hyper-Raman process (shown in
Figure \ref{fig:scheme}). It is straightforward to obtain the Hamiltonian
corresponding to Raman or $k$-pump degenerate hyper-Raman process just by considering
$k=1$ or $\omega_{i}=\omega_{p}$, respectively.

Specifically, if we choose $a_{1}=a_{2}$ (i.e., $\omega_{1}=\omega_{2}$)
for $k=2$ then we would obtain the Hamiltonian of degenerate hyper-Raman
process which is already studied in a reasonably detailed manner in Ref. \cite{sen2007squeezing}.
Apart from this, 2-pump mode non-degenerate hyper-Raman process was
discussed in Ref. \cite{perina1984relations}. The present Hamiltonian
can be viewed as a generalization of this case to multi-mode pump
non-degenerate hyper-Raman process. However, nonclassical properties
of multi-mode pump non-degenerate Hamiltonian is not studied in that
detail. This is why we are interested in the operator solution of the
Hamiltonian of non-degenerate multi-photon pump hyper-Raman process.
To obtain the solution, first we write the Heisenberg's equations
of motion for different modes as 
\begin{equation}
\begin{array}{lcl}
\dot{a_{j}} & = & -i\omega_{j}a_{j}+\prod_{i=1:i\neq j}^{k}\left(ig^{*}a_{i}^{\dagger}bc+i\chi a_{i}^{\dagger}c^{\dagger}d\right),\\
\dot{b} & = & -i\omega_{b}b+ig\prod_{i=1}^{k}a_{i}c^{\dagger},\\
\dot{c} & = & -i\omega_{c}c+\prod_{i=1}^{k}\left(iga_{i}b^{\dagger}+i\chi a_{i}^{\dagger}d\right),\\
\dot{d} & = & -i\omega_{d}d+i\chi^{*}\prod_{i=1}^{k}a_{i}c,
\end{array}\label{eq:field operators}
\end{equation}
for which we derive the solution, using Sen-Mandal perturbative (\cite{thapliyal2014higher,thapliyal2014nonclassical,thapliyal2016linear,sen2005squeezed}
and references therein) approach,
as

\begin{widetext}

\begin{equation}
\begin{array}{lcl}
a_{j}(t) & = & \prod_{i=1:i\neq j}^{k}\left(f_{1_{j}}a_{j}(0)+f_{2_{j}}a_{i}^{\dagger}(0)b(0)c(0)+f_{3_{j}}a_{i}^{\dagger}(0)c^{\dagger}(0)d(0)\right.\\
 & + & f_{4_{j}}a_{j}(0)A_{l}b^{\dagger}(0)b(0)c^{\dagger}(0)c(0)+f_{5_{j}}a_{i}^{\dagger}(0)a_{i}(0)a_{j}(0)b(0)b^{\dagger}(0)\\
 & + & f_{6_{j}}a_{i}^{\dagger}(0)a_{i}(0)a_{j}(0)c^{\dagger}(0)c(0)+f_{7_{j}}a_{j}(0)A_{l}b(0)c^{2}(0)d^{\dagger}(0)\\
 & + & f_{8_{j}}a_{j}^{\dagger}(0)a_{i}^{\dagger2}(0)b(0)d(0)+f_{9_{j}}a_{j}(0)A_{l}b^{\dagger}(0)c^{\dagger2}(0)d(0)\\
 & + & f_{10_{j}}a_{j}(0)a_{i}^{\dagger}(0)a_{i}(0)d^{\dagger}(0)d(0)+f_{11_{j}}a_{j}(0)A_{l}c(0)c^{\dagger}(0)d^{\dagger}(0)d(0)\\
 & + & \left.f_{12_{j}}a_{i}^{\dagger}(0)a_{i}(0)a_{j}(0)c^{\dagger}(0)c(0)\right),\\
b(t) & = & \prod_{i=1}^{k}\left(g_{1}b(0)+g_{2}a_{i}(0)c^{\dagger}(0)+g_{3}a_{i}^{2}(0)d^{\dagger}(0)+g_{4}a_{i}^{\dagger}(0)a_{i}(0)b(0)\right.\\
 & + & \left.g_{5}A_{l}b(0)c^{\dagger}(0)c(0)+g_{6}A_{l}c^{\dagger2}(0)d(0)\right),\\
c(t) & = & \prod_{i=1}^{k}\left(h_{1}c(0)+h_{2}a_{i}(0)b^{\dagger}(0)+h_{3}a_{i}^{\dagger}(0)d(0)+h_{4}a_{i}^{\dagger}(0)a_{i}(0)c(0)\right.\\
 & + & h_{5}A_{l}b^{\dagger}(0)b(0)c(0)+h_{6}A_{l}b^{\dagger}(0)c^{\dagger}(0)d(0)+h_{7}A_{l}c(0)d^{\dagger}(0)d(0)\\
 & + & \left.h_{8}a_{i}^{\dagger}(0)a_{i}(0)c(0)\right),\\
d(t) & = & \prod_{i=1}^{k}\left(l_{1}d(0)+l_{2}a_{i}(0)c(0)+l_{3}a_{i}^{2}(0)b^{\dagger}(0)+l_{4}A_{l}b(0)c^{2}(0)\right.\\
 & + & \left.l_{5}A_{l}c^{\dagger}(0)c(0)d(0)+l_{6}a_{i}(0)a_{i}^{\dagger}(0)d(0)\right),
\end{array}\label{eq:solution}
\end{equation}

\end{widetext}

\noindent where $A_{l}=\prod_{i}\left(a_{i}(0)a_{i}^{\dagger}(0)-a_{i}^{\dagger}(0)a_{i}(0)\right)$ (which gives us $l=2^k-1$ terms in $k$ pump mode case). For example, for 2-pump non-degenerate hyper-Raman process, we obtain $l=3$ terms as follows, $A_{3}=\left(\left\{a_{1}^{\dagger}(0)a_{1}(0)+1\right\}\left\{a_{2}^{\dagger}(0)a_{2}(0)+1\right\}\right.-\left.a_{1}^{\dagger}(0)a_{1}(0)a_{2}^{\dagger}(0)a_{2}(0)\right)=\left(a_{1}^{\dagger}(0)a_{1}(0)+a_{2}^{\dagger}(0)a_{2}(0)+1\right).$ Further,
various terms in Eq. (\ref{eq:solution}) are given as Eqs. (\ref{eq:solutions of f})-(\ref{eq:eq:solutions of l})
in Appendix A. The details of obtaining the Sen-Mandal perturbative
solution are given in Appendix B. Here, it is also worth mentioning
that we have neglected all the terms higher than quadratic in coupling
constants $\chi$ and $g$ while obtaining the present solution. 

The most general nature of the Hamiltonian describing the hyper-Raman
processes used here has already been established. On top of that,
the obtained solution is also quite general in nature and it is imperative
to mention here that all the existing solutions of various Raman \cite{sen2005squeezed,sen2007quantum,sen2008amplitude,sen2011sub}
and degenerate-hyper-Raman \cite{sen2007squeezing} processes can be
obtained as the limiting cases of the present solution. It is also relevant
to mention here that the solution obtained in \cite{sen2005squeezed},
which is a limiting case of the present solution, has already been shown to be
reducible to the short-time solution reported till then \cite{perina1991quantum,szlachetka1979dynamics,szlachetka1980photon}.
It is also important to note here that in some of our recent works, it has been established that the Sen-Mandal perturbative solutions
are more general than the corresponding short-time solutions for the same
systems  (\cite{sen2005squeezed,sen2007quantum,sen2008amplitude,sen2011sub,sen2007squeezing,thapliyal2014higher,thapliyal2014nonclassical,thapliyal2016linear}
and references therein). To reduce our general solution to the solution for degenerate hyper-Raman
process reported in \cite{sen2007squeezing}  we need to consider $k=2$, with $a_{1}=a_{2}=a$, and $\omega_{1}=\omega_{2}=\omega_{a}$ (as the process is degenerate), 
and $\chi$ and $g$  to be  real (as $\chi$ and $g$  were considered as real in Ref. \cite{sen2007squeezing}). The coupling
constants $\chi$ and $g$ were treated as real in the case of Raman process, too \cite{sen2005squeezed}, but to be consistent with the convention used in Ref.  \cite{sen2005squeezed} and to reduce the solution reported here to the solution reported in \cite{sen2005squeezed}, we would require to replace $g$ by $-g$. Specifically, the
solution used in \cite{sen2005squeezed,sen2007quantum,sen2008amplitude,sen2011sub,sen2013intermodal,giri2016higher}
can be reproduced using $k=1$, $\omega_{1}=\omega_{a}$
in the present solution. The relation among various
time dependent functional coefficients in the evolution of the pump
mode in the present case and previous results \cite{sen2005squeezed,sen2007quantum,sen2008amplitude,sen2011sub,sen2007squeezing,sen2013intermodal,giri2016higher}
is summarized in Table \ref{tab:pump}. A similar correspondence among 
the functions for the remaining modes is mentioned explicitly in Table
\ref{tab:non-pump} (see Appendix B).

\begin{table}
\begin{centering}
\begin{tabular}{|>{\centering}p{2.5cm}|>{\centering}p{2.5cm}|>{\centering}p{2.5cm}|}
\hline 
Multi-photon pump hyper-Raman case & Degenerate hyper-Raman case \cite{sen2007squeezing} & Raman case \cite{sen2005squeezed}\tabularnewline
\hline 
$f_{1_{i}}$ & $f_{1}^{'}$ & $f_{1}^{''}$\tabularnewline
\hline 
$f_{2_{i}}$ & $f_{2}^{'}$ & $f_{2}^{''}$\tabularnewline
\hline 
$f_{3_{i}}$ & $f_{3}^{'}$ & $f_{3}^{''}$\tabularnewline
\hline 
$f_{4_{i}}$ & $f_{7}^{'}$ & \tabularnewline
\hline 
$f_{5_{i}}$ & $f_{9}^{'}$ & $f_{5}^{''}$\tabularnewline
\hline 
$f_{6_{i}}$ & $f_{8}^{'}$ & $f_{6}^{''}$\tabularnewline
\hline 
$f_{7_{i}}$ & $f_{4}^{'}$ & \tabularnewline
\hline 
$f_{8_{i}}$ & $f_{6}^{'}$ & $f_{4}^{''}$\tabularnewline
\hline 
$f_{9_{i}}$ & $f_{5}^{'}$ & \tabularnewline
\hline 
$f_{10_{i}}$ & $f_{11}^{'}$ & $f_{8}^{''}$\tabularnewline
\hline 
$f_{11_{i}}$ & $f_{10}^{'}$ & \tabularnewline
\hline 
$f_{12_{i}}$ & $f_{12}^{'}$ & $f_{7}^{''}$\tabularnewline
\hline 
\end{tabular}
\par\end{centering}
\caption{\label{tab:pump}The present solution is general in nature, and the
existing solutions for Raman and degenerate hyper-Raman processes
can be obtained as special cases of the present solution for the functions
in the evolution of pump modes. Here, we have used a (two) prime(s)
in the superscript of the functions $f_{i}$s to distinguish the present
solution from the degenerate hyper-Raman (Raman) process. }
\end{table}

\section{Criteria of nonclassicalities \label{sec:Criteria-of-nonclassicalities}}

Once we have the closed form analytic expressions for the evolution
of various field and phonon modes involved in the process (given in
Eq. (\ref{eq:solution})), we can test the nonclassical properties
of the process using various moments-based  criteria (such as listed
in \cite{miranowicz2010testing}), which are essentially the expectation
values of annihilation and creation operators of the modes under consideration.
Although an infinite set of moment-based  criteria would be essential to form a
necessary criterion of nonclassicality that would be equivalent to $P$-function \cite{richter2002nonclassicality}, here, we only use a small subset
of this infinite set, which is therefore only sufficient. However, this small set of noncassicality criteria is found to be good enough to establish
the highly nonclassical character of the non-degenarate hyper-Raman process. In this
section, we enlist the set of criteria that is used in the present
work to analyze the presence of lower and higher order nonclassicality.
With the advent of sophisticated experimental techniques, some exciting experimental results involving  a few of these moments-based  higher order nonclassicality
criteria have been reported in the recent past \cite{allevi2012measuring,allevi2012high,avenhaus2010accessing,hamar2014non}.
More recently, experimental detection of higher order nonclassicality up to ninth order has also been reported \cite{perina2017higher}. Further, in Section \ref{sec:Intro}, we have already mentioned several recent applications of nonclassical states and Raman processes. Because of the above mentioned facts, in what follows, we are interested in analyzing the nonclassical properties
of the process with specific attention to squeezed, antibunched and
entangled states.

\subsection{Lower and higher order squeezing }

In order to study the squeezing effects in the various modes, we define
the quadrature operators 
\begin{equation}
\begin{array}{lcl}
X_{a} & = & \frac{1}{2}\left(a(t)+a^{\dagger}(t)\right),\\
Y_{a} & = & -\frac{i}{2}\left(a(t)-a^{\dagger}(t)\right),
\end{array}\label{eq:quadrature}
\end{equation}
where $a\,(a^{^{\dagger}})$ is the annihilation (creation) operator
for a specific bosonic mode, and it satisfies $[a,a^{\dagger}]=1$. Squeezing in mode
$a$ is possible if the fluctuation in one of the quadrature operators
goes below the minimum uncertainty level, i.e., if 
\begin{equation}
\left(\Delta X_{a}\right)^{2}<\frac{1}{4}\,\mathrm{or}\,\left(\Delta Y_{a}\right)^{2}<\frac{1}{4}.\label{eq:condition for squeezing}
\end{equation}

Similarly, we may study intermodal squeezing in the compound mode
$ab$ using the following quadrature operator for the compound mode
introduced by Loudon and Knight \cite{loudon1987squeezed}: 
\begin{equation}
\begin{array}{lcl}
X_{ab} & = & \frac{1}{2\sqrt{2}}\left(a(t)+a^{\dagger}(t)+b(t)+b^{\dagger}(t)\right)\\
Y_{ab} & = & -\frac{i}{2\sqrt{2}}\left(a(t)-a^{\dagger}(t)+b(t)-b^{\dagger}(t)\right).
\end{array}\label{eq:two-mode quadrature}
\end{equation}

Usually, the higher order counterpart of squeezing is studied using two
different criteria, proposed by Hong and Mandel \cite{hong1985higher,hong1985generation}
and Hillery \cite{hillery1987amplitude}, independently. Hong-Mandel-type
squeezing takes into consideration the higher order moments of usual
quadrature defined in Eq. (\ref{eq:quadrature}), while Hillery's
squeezing criterion deals with amplitude powered quadratures. Here,
we have focused only on the latter type. For which, the amplitude
powered quadratures are defined as

\begin{equation}
Y_{1,a}=\frac{a^{n}+\left(a^{\dagger}\right)^{n}}{2}\label{eq:quadrature-power1}
\end{equation}
and 
\begin{equation}
Y_{2,a}=i\left(\frac{\left(a^{\dagger}\right)^{n}-a^{n}}{2}\right).\label{eq:quadrature-power2}
\end{equation}
As the quadratures fail to commute, we can obtain a criterion for amplitude
powered squeezing from the uncertainty principle as 
\begin{equation}
A_{i,a}=\left(\Delta Y_{i,a}\right)^{2}-\frac{1}{2}\left|\left\langle \left[Y_{1,a},Y_{2,a}\right]\right\rangle \right|<0\label{eq:HOS}
\end{equation}
for each quadrature $i\in\left\{ 1,2\right\}$, where $\left[A,B\right]=AB-BA$ is the commutator,  and $\left[Y_{1,a},Y_{2,a}\right]\neq0 \forall n$.

\subsection{Lower and higher order antibunching}

Higher order antibunching criterion was introduced by Lee \cite{lee1990higher}.
With time, several variants of this criterion, which are essentially
equivalent, have been proposed. One such criterion was proposed by Pathak
and Garcia \cite{pathak2006control} as
\begin{equation}
\begin{array}{lcl}
D_{a}(n-1)=\left\langle a^{\dagger n}a^{n}\right\rangle -\left\langle a^{\dagger}a\right\rangle ^{n} & < & 0.\end{array}\label{hoa}
\end{equation}
Importantly, for $n=2$, it reduces to lower order antibunching, while
for all $n\geq3$ we obtain the higher order counterpart. Therefore,
here we have calculated and reported $\left(n-1\right)$th order antibunching
and also inferred corresponding lower order results from it.

Similarly, in order to study the intermodal antibunching, one can use the solution reported here and the following criterion
\begin{equation}
D_{ab}=(\Delta N_{ab})^{2}=\left\langle a^{\dagger}b^{\dagger}ba\right\rangle -\left\langle a^{\dagger}a\right\rangle \left\langle b^{\dagger}b\right\rangle <0.\label{eq:In-ant}
\end{equation}

\subsection{Lower and higher order entanglement}

Along the same line, two higher order entanglement criteria were proposed
by Hillery and Zubairy \cite{hillery2006entanglement,hillery2006entanglementapplications}
as inseparability criteria. It may be noted that each of these criteria are sufficient
but not necessary and thus the entanglement (nonclassical nature) not detected by a particular criterion may be detected by the other one, and  in some situations both of them may fail to detect entanglement. A quantum
state can be verified to be entangled using Hillery-Zubairy's HZ-I
criterion
\begin{equation}
E_{ab}^{m,n}=\left\langle \left(a^{\dagger}\right)^{m}a^{m}\left(b^{\dagger}\right)^{n}b^{n}\right\rangle -\left\vert \left\langle a^{m}\left(b^{\dagger}\right)^{n}\right\rangle \right\vert ^{2}<0\label{hoe-criteria}
\end{equation}
or Hillery-Zubairy's HZ-II criterion
\begin{equation}
E_{ab}^{\prime m,n}=\left\langle \left(a^{\dagger}\right)^{m}a^{m}\right\rangle \left\langle \left(b^{\dagger}\right)^{n}b^{n}\right\rangle -\left\vert \left\langle a^{m}b^{n}\right\rangle \right\vert ^{2}<0.\label{hoe-criteria-2}
\end{equation}
An arbitrary quantum state is higher order entangled if it satisfies
HZ-I and/or HZ-II criteria for $m+n\geq3$. Importantly, the lower order
entanglement can also be verified from Eqs. (\ref{hoe-criteria})
and (\ref{hoe-criteria-2}), by considering $m=n=1$.

\section{Nonclassicality observed \label{sec:Nonclassicality-observed}}

All the nonclassicality criteria listed in Eqs. (\ref{eq:condition for squeezing})-(\ref{hoe-criteria-2})
contain average values of functions of time evolved annihilation
and creation operators given in Eq. (\ref{eq:solution}). To calculate
the average values, we have to consider an initial state of the system.
Without any loss of generality, the initial state is chosen to be
a product state of coherent states in each mode
\begin{equation}
|\psi(0)\rangle=|\alpha_{i}\rangle\otimes|\beta\rangle\otimes|\gamma\rangle\otimes|\delta\rangle,\label{eq:initial state}
\end{equation}
where $|\alpha_{i}\rangle=|\alpha_{1}\rangle\otimes|\alpha_{2}\rangle\otimes\cdots|\alpha_{i}\rangle\cdots\otimes|\alpha_{k}\rangle$
is the initial state of the pump modes in the product state of $k$ coherent states. 

Further, we have considered a detuning of $\left|\frac{\Delta\omega_{1}}{g}\right|=10$
and $\left|\frac{\Delta\omega_{2}}{g}\right|=19$ in Stokes and anti-Stokes
hyper-Raman processes, respectively. In the stimulated case, we have
considered non-zero photon number initially in each mode, i.e., $\left|\beta\right|=8,$
$\left|\gamma\right|=0.01,$ and $\left|\delta\right|=1$, while all
these values are initially zero in the spontaneous case. Additionally, we have considered
$\left|\alpha_{i}\right|=10$ for all the pump modes, unless stated
otherwise, in both the stimulated and spontaneous cases. 

\subsection{Lower and higher order squeezing}

Using Eqs. (\ref{eq:solution}) and (\ref{eq:initial state}) in criterion
of squeezing (\ref{eq:condition for squeezing}), we have obtained
the closed form analytic expressions for the witnesses of squeezing in all the modes
involved. Specifically, the witnesses for the single mode squeezing in an arbitrary pump
mode is calculated to be \begin{subequations} 
\begin{equation}
\begin{array}{lcl}
\left[\begin{array}{c}
\left(\Delta X_{a_{j}}\right)^{2}\\
\left(\Delta Y_{a_{j}}\right)^{2}
\end{array}\right] & = & \frac{1}{4}\left[1+2\left|f_{2}\right|^{2}\sigma_{l}\left|\beta\right|^{2}\left|\gamma\right|^{2}+2\left|f_{3}\right|^{2}\left|\delta\right|^{2}\right.\\
 & \times & \left(\left|\alpha_{i}\right|^{2}+\sigma_{l}\left(\left|\gamma\right|^{2}+1\right)\right)+2\left\{ f_{2}^{*}f_{3}\sigma_{l}\right.\\
 & \times & \left.\left.\beta^{*}\gamma^{*2}\delta\mp f_{1}^{2}g_{3}^{*}g_{1}\alpha_{i}^{*2}\beta\delta+{\rm c.c.}\right\} \right],
\end{array}\label{eq:Sq-a}
\end{equation}
while the witnesses of squeezing in Stokes and vibration (phonon) modes are obtained as 
\begin{equation}
\begin{array}{lcl}
\left[\begin{array}{c}
\left(\Delta X_{b}\right)^{2}\\
\left(\Delta Y_{b}\right)^{2}
\end{array}\right] & = & \frac{1}{4}\left[1+2\left|g_{2}\right|^{2}\left|\alpha_{i}\right|^{2}\right]\end{array}\label{eq:Sq-b}
\end{equation}
and 

\begin{equation}
\begin{array}{lcl}
\left[\begin{array}{c}
\left(\Delta X_{c}\right)^{2}\\
\left(\Delta Y_{c}\right)^{2}
\end{array}\right] & = & \frac{1}{4}\left[1+2\left|h_{2}\right|^{2}\left|\alpha_{i}\right|^{2}+2\left|h_{3}\right|^{2}\sigma_{l}\left|\delta\right|^{2}\right.\\
 & \pm & \left.\left\{ 2h_{1}^{2}g_{1}^{*}g_{6}\sigma_{l}\beta^{*}\delta+{\rm c.c.}\right\} \right],
\end{array}\label{eq:sq-c}
\end{equation}
respectively. Using the obtained Sen-Mandal perturbative solution,
squeezing in anti-Stokes mode was not observed, i.e.,
\begin{equation}
\begin{array}{lcl}
\left[\begin{array}{c}
\left(\Delta X_{d}\right)^{2}\\
\left(\Delta Y_{d}\right)^{2}
\end{array}\right] & = & \frac{1}{4}.\end{array}\label{eq:Sq-d}
\end{equation}
\end{subequations} Here, and in what follows we have used $\sigma_{l}=\langle A_{l}\rangle$.
In Eq. (\ref{eq:Sq-b}), we can observe a positive quantity is added
to $\frac{1}{4}$, therefore, none of these quadratures can show variance
less than $\frac{1}{4}$, and consequently, squeezing cannot be observed in these quadratures. However, unlike Stokes and anti-Stokes modes, the
compact expressions for squeezing in all the remaining modes are complex
enough to infer directly from them. To analyze the dependence of squeezing
in all these modes on various parameters we performed a rigorous numerical
analysis of the obtained expressions. To do so, we have used
$\underset{i}{\sum}\frac{\omega_{i}}{g}=1000.0001\times10^{5},$ $\frac{\omega_{b}}{g}=999.999\times10^{5},$
$\frac{\omega_{c}}{g}=0.001\times10^{5},$ and $\frac{\omega_{d}}{g}=1000.00091\times10^{5}.$
In 2-pump mode non-degenerate hyper-Raman processes, $\frac{\omega_{1}}{g}=600.0001\times10^{5}$
and $\frac{\omega_{2}}{g}=400\times10^{5}$; while in 3-pump mode
non-degenerate hyper-Raman processes, $\frac{\omega_{1}}{g}=100.0001\times10^{5},$
$\frac{\omega_{2}}{g}=700\times10^{5}$ and $\frac{\omega_{3}}{g}=200\times10^{5}.$  Further, for the sake of simplicity, in the following discussion we have subtracted $\frac{1}{4}$ from both sides of all the expressions of squeezing. This helps us to plot the variation of squeezing parameter in a manner consistent with the remaining illustrations where the negative regions of the plots depict nonclassicality.

The study revealed that squeezing in the stimulated case of non-degenerate
$k$-pump hyper-Raman process is observed only in the pump mode, which
is shown to vary with various parameters. Thus, in turn, these parameters
can be used to control the amount of squeezing. Specifically, witnesses of squeezing are found to be independent of the phases of different pump modes.
However, the amount of squeezing is found to depend on the frequency of the pump
modes. This fact can be established from Figure \ref{fig:SqA} (a)-(c),
where we can observe different amount of squeezing for each mode with
different frequency, which becomes the same in the degenerate case. Further,
different natures of squeezing for degenerate and non-degenerate cases
have been observed (cf. Figure \ref{fig:SqA} (a) and (d)). This point
is also established in context of the spontaneous case in Figure \ref{fig:Spont}
(d) discussed later. Additionally, the amount of squeezing in a particular
pump mode can also be controlled by the intensity of one of the other
pump modes (cf. Figure \ref{fig:SqA} (a) and (b)).  Note that we have shown the variation in quadrature squeezing for relatively smaller time domain to establish its dependence on various independent parameters. Only due to this reason, the amount of squeezing appears to be very small (in the order of $10^{-12}$ in Figure \ref{fig:SqA} (a)). However, we observed relatively higher amount of squeezing for a larger rescaled time as shown in inset in Figure \ref{fig:SqA} (a) in case of $X_{a_{1}}$ quadrature. Similar highly oscillating nature is also observed in all other cases of single mode and intermodal squeezing, too, but being repetitive, such illustrations are not included in the subsequent plots.

Similarly, intermodal squeezing in the compound two-mode cases can be
studied using Loudon and Knight's criterion given in Eq. (\ref{eq:two-mode quadrature})
with Eqs. (\ref{eq:solution}) and (\ref{eq:initial state}). We are
reporting here the analytic expressions of two-mode squeezing for
compound pump-pump mode as \begin{subequations}

\begin{widetext}
\begin{equation}
\begin{array}{lcl}
\left[\begin{array}{c}
\left(\Delta X_{a_{j}a_{r}}\right)^{2}\\
\left(\Delta Y_{a_{j}a_{r}}\right)^{2}
\end{array}\right] & = & \frac{1}{4}\left[2\left(\left[\begin{array}{c}
\left(\Delta X_{a_{j}}\right)^{2}\\
\left(\Delta Y_{a_{j}}\right)^{2}
\end{array}\right]+\left[\begin{array}{c}
\left(\Delta X_{a_{r}}\right)^{2}\\
\left(\Delta Y_{a_{r}}\right)^{2}
\end{array}\right]\right)+\left\{ f_{2}r_{2}^{*}\left|\beta\right|^{2}\left|\gamma\right|^{2}\alpha_{j}\alpha_{r}^{*}\sigma_{l}+f_{3}r_{2}^{*}\alpha_{j}\alpha_{r}^{*}\sigma_{l}\beta^{*}\gamma^{*2}\delta\right.\right.\\
 & + & f_{2}r_{3}^{*}\alpha_{j}\alpha_{r}^{*}\sigma_{l}\beta\gamma^{2}\delta^{*}\pm\left\{ f_{1}r_{2}\alpha_{i}^{*}\beta\gamma+f_{1}r_{3}\alpha_{i}^{*}\beta^{*}\delta+f_{1}r_{4}\left|\beta\right|^{2}\left|\gamma\right|^{2}\alpha_{j}\alpha_{r}\sigma_{l}+f_{1}r_{5}\left|\alpha_{i}\right|^{2}\left(\left|\beta\right|^{2}+1\right)\alpha_{j}\alpha_{r}\right.\\
 & + & \left(f_{1}r_{6}+f_{1}r_{12}\right)\left|\alpha_{i}\right|^{2}\left|\gamma\right|^{2}\alpha_{j}\alpha_{r}+f_{1}r_{10}\left(\left|\alpha_{i}\right|^{2}+\sigma_{l}\left(\left|\gamma\right|^{2}+1\right)\right)\left|\delta\right|^{2}\alpha_{j}\alpha_{r}+f_{1}r_{7}\alpha_{j}\alpha_{r}\sigma_{l}\beta\gamma^{2}\delta^{*}\\
 & + & \left.\left.\left.\left(2f_{1}r_{8}+f_{2}r_{3}\right)\alpha_{j}^{*}\alpha_{r}^{*}\alpha_{i}^{*2}\beta\delta+f_{1}r_{9}\alpha_{j}\alpha_{r}\sigma_{l}\beta^{*}\gamma^{*2}\delta\right\} +f_{3}r_{3}^{*}\left(\left|\alpha_{i}\right|^{2}+\sigma_{l}\left(\left|\gamma\right|^{2}+1\right)\right)\left|\delta\right|^{2}\alpha_{j}\alpha_{r}^{*}+{\rm c.c.}\right\} \right],
\end{array}\label{eq:Int-sq-aij}
\end{equation}

\begin{figure}
\centering{}\includegraphics[scale=0.8]{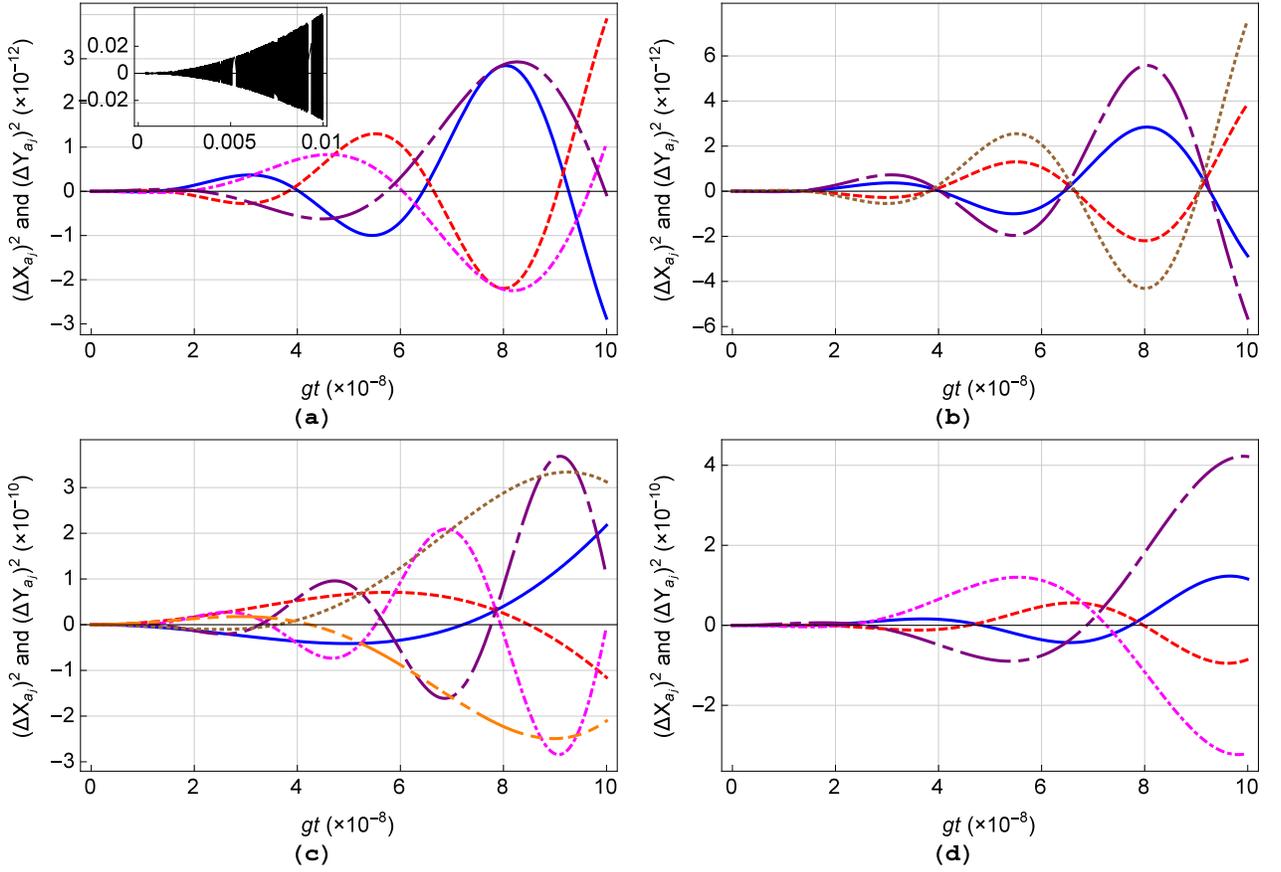}\caption{\label{fig:SqA}(Color online) Squeezing in $j$th pump mode. Squeezing
in both modes of the 2-pump modes non-degenerate stimulated hyper-Raman
processes with (a) $\left|\alpha_{1}\right|=\left|\alpha_{2}\right|=10$,
and (b) $\left|\alpha_{1}\right|=10,$ $\left|\alpha_{2}\right|=12$.
(c) Squeezing in all three modes of the 3-pump modes non-degenerate
hyper-Raman processes with $\left|\alpha_{1}\right|=\left|\alpha_{2}\right|=\left|\alpha_{3}\right|=10$.
(d) Squeezing in degenerate hyper-Raman processes for the cases with
2 and 3-pump modes is compared. Here, we have amplified the variation
in the case of 2-pump modes 30 times to show it with the squeezing
in 3-pump modes. In all the cases, the solid-blue, dot-dashed-magenta,
and dotted-black lines represent the quadrature $\left(\Delta X_{a_{j}}\right)^{2}$;
and dashed-red, purple-large-dot-dashed, and orange-double-dotted-dashed
lines correspond to the quadrature $\left(\Delta Y_{a_{j}}\right)^{2}$.}
\end{figure}

\noindent which is applicable to any arbitrary pump modes $a_{j}$ and $a_{r}$.
Compound pump-Stokes, pump-vibration, and pump-anti-Stokes modes squeezing
are obtained as 
\begin{equation}
\begin{array}{lcl}
\left[\begin{array}{c}
\left(\Delta X_{a_{j}b}\right)^{2}\\
\left(\Delta Y_{a_{j}b}\right)^{2}
\end{array}\right] & = & \frac{1}{4}\left[2\left(\left[\begin{array}{c}
\left(\Delta X_{a_{j}}\right)^{2}\\
\left(\Delta Y_{a_{j}}\right)^{2}
\end{array}\right]+\left[\begin{array}{c}
\left(\Delta X_{b}\right)^{2}\\
\left(\Delta Y_{b}\right)^{2}
\end{array}\right]\right)+\left\{ f_{3}g_{2}^{*}\alpha_{j}^{*}\alpha_{i}^{*2}\delta\mp f_{4}g_{1}\left|\alpha_{i}\right|^{2}\alpha_{j}\beta\mp f_{1}g_{4}\left|\gamma\right|^{2}\sigma_{l}\alpha_{j}\beta\right.\right.\\
 & \pm & \left.\left.f_{1}g_{6}\sigma_{l}\alpha_{j}\gamma^{*2}\delta+{\rm c.c.}\right\} \right],
\end{array}\label{eq:Int-sq-ab}
\end{equation}
\begin{equation}
\begin{array}{lcl}
\left[\begin{array}{c}
\left(\Delta X_{a_{j}c}\right)^{2}\\
\left(\Delta Y_{a_{j}c}\right)^{2}
\end{array}\right] & = & \frac{1}{4}\left[2\left(\left[\begin{array}{c}
\left(\Delta X_{a_{j}}\right)^{2}\\
\left(\Delta Y_{a_{j}}\right)^{2}
\end{array}\right]+\left[\begin{array}{c}
\left(\Delta X_{c}\right)^{2}\\
\left(\Delta Y_{c}\right)^{2}
\end{array}\right]\right)+\left\{ f_{2}h_{3}^{*}\sigma_{l}\alpha_{j}\beta\gamma\delta^{*}+f_{3}h_{3}^{*}\left|\delta\right|^{2}\sigma_{l}\alpha_{j}\gamma^{*}\pm\left(f_{1}h_{7}-f_{4}h_{1}\right)\left|\alpha_{i}\right|^{2}\alpha_{j}\gamma\right.\right.\\
 & \pm & \left.\left.f_{1}h_{3}\alpha_{i}^{*}\delta\pm f_{1}h_{4}\left|\beta\right|^{2}\sigma_{l}\alpha_{j}\gamma\pm f_{1}h_{6}\sigma_{l}\alpha_{j}\beta^{*}\gamma^{*}\delta\pm f_{1}h_{7}\left|\delta\right|^{2}\sigma_{l}\alpha_{j}\gamma+{\rm c.c.}\right\} \right],
\end{array}\label{eq:Int-sq-ac}
\end{equation}
and 
\begin{equation}
\begin{array}{lcl}
\left[\begin{array}{c}
\left(\Delta X_{a_{j}d}\right)^{2}\\
\left(\Delta Y_{a_{j}d}\right)^{2}
\end{array}\right] & = & \frac{1}{4}\left[2\left(\frac{1}{2}+\left[\begin{array}{c}
\left(\Delta X_{a_{j}}\right)^{2}\\
\left(\Delta Y_{a_{j}}\right)^{2}
\end{array}\right]\right)+\left\{ f_{1}l_{4}\sigma_{l}\alpha_{j}\beta\gamma^{2}\pm f_{1}l_{5}\alpha_{j}\delta\left(\left|\alpha_{i}\right|^{2}+\sigma_{l}\left(\left|\gamma\right|^{2}+1\right)\right)+{\rm c.c.}\right\} \right],\end{array}\label{eq:Int-sq-ad}
\end{equation}
respectively. We have also considered two-mode squeezing among Stokes-vibration
mode 
\begin{equation}
\begin{array}{lcl}
\left[\begin{array}{c}
\left(\Delta X_{bc}\right)^{2}\\
\left(\Delta Y_{bc}\right)^{2}
\end{array}\right] & = & \frac{1}{4}\left[2\left(\left[\begin{array}{c}
\left(\Delta X_{b}\right)^{2}\\
\left(\Delta Y_{b}\right)^{2}
\end{array}\right]+\left[\begin{array}{c}
\left(\Delta X_{c}\right)^{2}\\
\left(\Delta Y_{c}\right)^{2}
\end{array}\right]\right)+\left\{ g_{2}h_{2}^{*}\sigma_{l}\beta\gamma^{*}\pm\left(g_{1}h_{2}\alpha_{i}+g_{1}h_{5}\sigma_{l}\beta\gamma+2g_{6}h_{1}\sigma_{l}\gamma^{*}\delta\right)+{\rm c.c.}\right\} \right],\end{array}\label{eq:Int-sq-bc}
\end{equation}

\begin{figure}
\centering{}\includegraphics[scale=0.8]{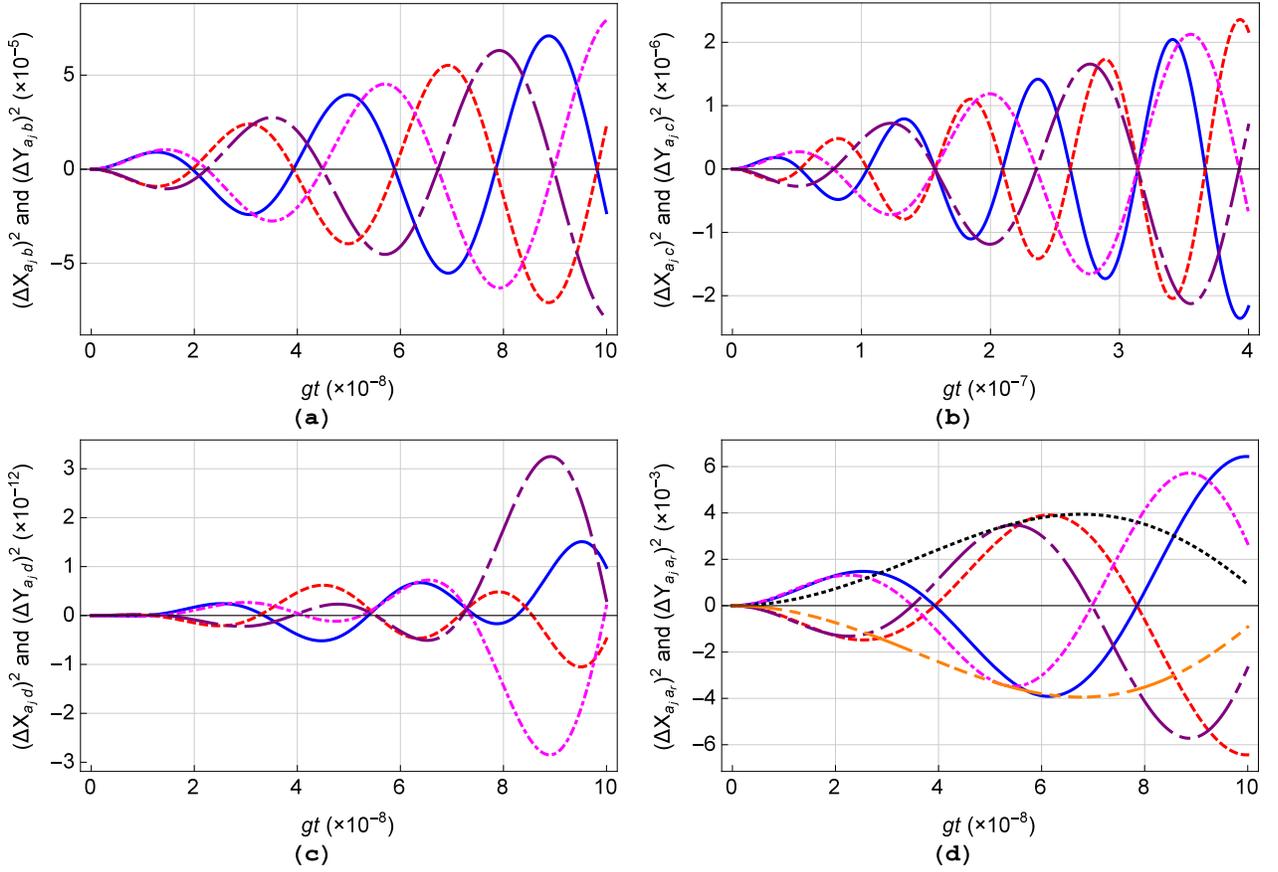}\caption{\label{fig:IntSq}(Color online) Intermodal squeezing is observed
in 2-pump modes non-degenerate stimulated hyper-Raman processes in
(a) $j$th pump-Stokes mode, (b) $j$th pump-vibration mode, and (c)
$j$th pump-anti-Stokes mode. In all three cases, the solid-blue (dot-dashed-magenta)
and dashed-red (purple-large-dot-dashed) lines correspond to the quadratures
$\left(\Delta X_{a_{j}K}\right)^{2}$ and $\left(\Delta Y_{a_{j}K}\right)^{2}$
for compound $a_{1}K$ ($a_{2}K$) mode, respectively. While in (d),
intermodal squeezing in compound $a_{j}a_{k}$ mode for 3-pump modes
non-degenerate stimulated hyper-Raman processes is shown for all three
cases, i.e., $a_{1}a_{2}$ mode shown in solid-blue and dashed-red
lines, $a_{2}a_{3}$ mode shown in dot-dashed-magenta and purple-large-dot-dashed
lines, and $a_{1}a_{3}$ mode as dotted-black and orange-double-dotted-dashed
lines for the quadratures $\left(\Delta X_{a_{j}a_{r}}\right)^{2}$
and $\left(\Delta Y_{a_{j}a_{r}}\right)^{2}$, respectively. In all
the cases, $\left|\alpha_{i}\right|=10\,\forall i\in\left\{ 1,2,3\right\} $. }
\end{figure}

\noindent Stokes-anti-Stokes mode 
\begin{equation}
\begin{array}{lcl}
\left[\begin{array}{c}
\left(\Delta X_{bd}\right)^{2}\\
\left(\Delta Y_{bd}\right)^{2}
\end{array}\right] & = & \frac{1}{4}\left[1+\frac{1}{2}\left[\begin{array}{c}
\left(\Delta X_{b}\right)^{2}\\
\left(\Delta Y_{b}\right)^{2}
\end{array}\right]\pm\left\{ g_{1}l_{3}\alpha_{i}^{2}+{\rm c.c.}\right\} \right],\end{array}\label{eq:Int-sq-bd}
\end{equation}
and vibration-anti-Stokes mode 
\begin{equation}
\begin{array}{lcl}
\left[\begin{array}{c}
\left(\Delta X_{cd}\right)^{2}\\
\left(\Delta Y_{cd}\right)^{2}
\end{array}\right] & = & \frac{1}{4}\left[1+\frac{1}{2}\left[\begin{array}{c}
\left(\Delta X_{c}\right)^{2}\\
\left(\Delta Y_{c}\right)^{2}
\end{array}\right]\pm\left\{ h_{1}l_{5}\sigma_{l}\gamma\delta+{\rm c.c.}\right\} \right].\end{array}\label{eq:Int-sq-cd}
\end{equation}

\end{widetext}\end{subequations}

In all the expressions obtained
for two-mode squeezing (i.e., Eqs. (\ref{eq:Int-sq-aij})-(\ref{eq:Int-sq-cd})),
 the single mode squeezing witnesses (i.e., variance $\left(\Delta X_{i}\right)^{2}$, and $\left(\Delta Y_{i}\right)^{2}$, with $i\in{a,b,c}$) that appear in the  right hand sides are to be substituted by the corresponding expressions
reported for single mode squeezing in Eqs. (\ref{eq:Sq-a})-(\ref{eq:Sq-d}).

Finally, we analyzed the expressions for the compound mode squeezing, and
variation is shown in Figure \ref{fig:IntSq}. Interestingly, intermodal
squeezing is observed in all the compound modes involving pump mode.
In the analogy of quadrature squeezing illustrated in Figure \ref{fig:SqA}, 
the observed nonclassicality is shown to depend on the frequency of
the pump mode. The same fact has been established here using non-degenerate
2-pump and 3-pump hyper-Raman processes, where the amount of intermodal
squeezing are found to be different for various pump modes.

Hillary's amplitude powered squeezing for all the modes involved is
calculated using Eqs. (\ref{eq:solution}) and (\ref{eq:initial state})
in criterion of squeezing (\ref{eq:HOS}). Specifically, the analytic
expression for an arbitrary pump mode is obtained as follows \begin{subequations}
\begin{equation}
\begin{array}{lcl}
\left[\begin{array}{c}
A_{1,a_{j}}\\
A_{2,a_{j}}
\end{array}\right] & = & \frac{1}{2}k^{2}\left|\alpha_{j}\right|^{2\left(k-1\right)}\left[\left|f_{2}\right|^{2}\sigma_{l}\left|\beta\right|^{2}\left|\gamma\right|^{2}\right.\\
 & + & \left|f_{3}\right|^{2}\left|\delta\right|^{2}\left(\left|\alpha_{i}\right|^{2}+\sigma_{l}\left(\left|\gamma\right|^{2}+1\right)\right)\\
 & + & \left.\left\{ f_{2}^{*}f_{3}\sigma_{l}\beta^{*}\gamma^{*2}\delta\mp f_{1}^{2}g_{3}^{*}g_{1}\alpha_{i}^{*2}\beta\delta+{\rm c.c.}\right\} \right].
\end{array}\label{eq:asq-aj}
\end{equation}

\begin{figure}
\centering{}\includegraphics[scale=0.6]{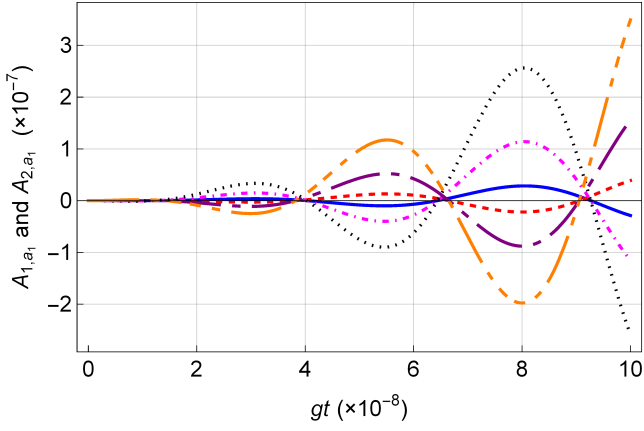}\caption{\label{fig:AmSq}(Color online) Amplitude powered squeezing is observed
in $a_{1}$ pump mode in 2-pump modes non-degenerate hyper-Raman processes
with $\left|\alpha_{1}\right|=\left|\alpha_{2}\right|=10$. The solid-blue
and dashed-red lines, dot-dashed-magenta and purple-large-dot-dashed
lines, and dotted-black and orange-double-dotted-dashed lines correspond
to the amplitude powered quadratures $A_{1,a_{1}}$ and $A_{1,a_{2}}$
for $k=1,2,$ and 3, respectively. To accommodate all the variations
in the same plot we have amplified the values for $k=1$ and 2 with $10^{4}$
and $10^{2}$, respectively. }
\end{figure}

Similar study for Stokes, vibration, and anti-Stokes modes resulted
in 
\begin{equation}
\begin{array}{lcl}
\left[\begin{array}{c}
A_{1,b}\\
A_{2,b}
\end{array}\right] & = & \frac{1}{2}\left[k^{2}\left|g_{2}\right|^{2}\left|\alpha_{i}\right|^{2}\left|\beta\right|^{2\left(k-1\right)}\right],\end{array}\label{eq:as-q-b}
\end{equation}
\begin{equation}
\begin{array}{lcl}
\left[\begin{array}{c}
A_{1,c}\\
A_{2,c}
\end{array}\right] & = & \frac{1}{2}k^{2}\left|\gamma\right|^{2\left(k-1\right)}\left[\left(\left|h_{2}\right|^{2}\left|\alpha_{i}\right|^{2}+\left|h_{3}\right|^{2}\sigma_{l}\left|\delta\right|^{2}\right)\right.\\
 & \pm & \left.\left\{ h_{1}^{2k}g_{1}^{*}g_{6}\sigma_{l}\beta^{*}\delta+{\rm c.c.}\right\} \right],
\end{array}\label{eq:as-q-c}
\end{equation}
 and 
\begin{equation}
\begin{array}{lcl}
\left[\begin{array}{c}
A_{1,d}\\
A_{2,d}
\end{array}\right] & = & 0,\end{array}\label{eq:as-d}
\end{equation}
respectively. \end{subequations}\textbf{ }From the obtained expressions,
the presence of amplitude powered squeezing in the pump mode has been
observed. Similar to the quadrature squeezing (shown in Figure \ref{fig:SqA}),
the nonclassicality is found to be absent in the remaining modes. Further, with
increase in higher orders of squeezing, depth of the witness of amplitude powered squeezing
is also observed to increase, which is in accordance with some of
our recent observations (\cite{thapliyal2014higher,thapliyal2014nonclassical,giri2017nonclassicality}
and references therein).

\subsection{Lower and higher order antibunching}

Higher order antibunching criterion given in Eq. (\ref{hoa}), used
with Eqs. (\ref{eq:solution}) and (\ref{eq:initial state}) leads
to the closed analytic expressions for the pump, Stokes, vibration
and anti-Stokes modes as \begin{subequations}
\begin{equation}
\begin{array}{lcl}
D_{a_{j}}(n-1) & = & n\left(n-1\right)\left[\left|\alpha_{j}\right|^{2\left(n-1\right)}\left(\left|f_{2}\right|^{2}\left|\beta\right|^{2}\left|\gamma\right|^{2}\sigma_{l}\right.\right.\\
 & + & \left.\left|f_{3}\right|^{2}\left|\delta\right|^{2}\left\{ \left|\alpha_{i}\right|^{2}+\sigma_{l}\left(\left|\gamma\right|^{2}+1\right)\right\} \right)\\
 & + & \left|\alpha_{j}\right|^{2\left(n-2\right)}\left\{ f_{2}^{*}f_{3}\sigma_{l}\beta^{*}\gamma^{*2}\delta\right.\\
 & - & \left.\left.g_{1}^{*}g_{3}\alpha_{j}^{2}\alpha_{i}^{2}\beta^{*}\delta^{*}+{\rm c.c.}\right\} \right],
\end{array}\label{eq:dan}
\end{equation}
\begin{equation}
\begin{array}{lcl}
D_{b}(n-1) & = & n\left(n-1\right)\left|g_{2}\right|^{2}\left|\alpha_{i}\right|^{2}\left|\beta\right|^{2\left(n-1\right)},\end{array}\label{eq:dbn}
\end{equation}
\begin{equation}
\begin{array}{lcl}
D_{c}(n-1) & = & n\left(n-1\right)\left[\left(\left|h_{2}\right|^{2}\left|\alpha_{i}\right|^{2}+\left|h_{3}\right|^{2}\sigma_{l}\left|\delta\right|^{2}\right)\right.\\
 & \times & \left|\gamma\right|^{2\left(n-1\right)}+\left\{ g_{6}^{*}g_{1}\left|\gamma\right|^{2\left(n-2\right)}\sigma_{l}\beta\gamma^{2}\delta^{*}\right.\\
 & + & \left.\left.{\rm c.c.}\right\} \right],
\end{array}\label{eq:dcn}
\end{equation}
and
\begin{equation}
D_{d}(n)=0,\label{eq:ddn}
\end{equation}
respectively. \end{subequations}Using these expressions we have analyzed
the possibilities  of observing both lower  and higher order antibunching
in all the modes except anti-Stokes mode (as $D_{d}(n)$ is always
zero). The presence of lower and higher order antibunching in
the pump mode has been observed and shown in Figure \ref{fig:Ant}
(a). Antibunching is not observed in the other modes, i.e., in the modes other than the pump modes. It is also important to note here that the second-order
correlation computed here for obtaining the signature of antibunching depends on the number of photons
in certain mode while it is independent of its frequency.

\begin{subequations}

Similarly, intermodal antibunching defined in Eq. (\ref{eq:In-ant})
can be calculated for all the possible two-mode cases, i.e., pump-pump
mode 
\begin{widetext}

\begin{figure}[t]
\centering{}\includegraphics[scale=0.8]{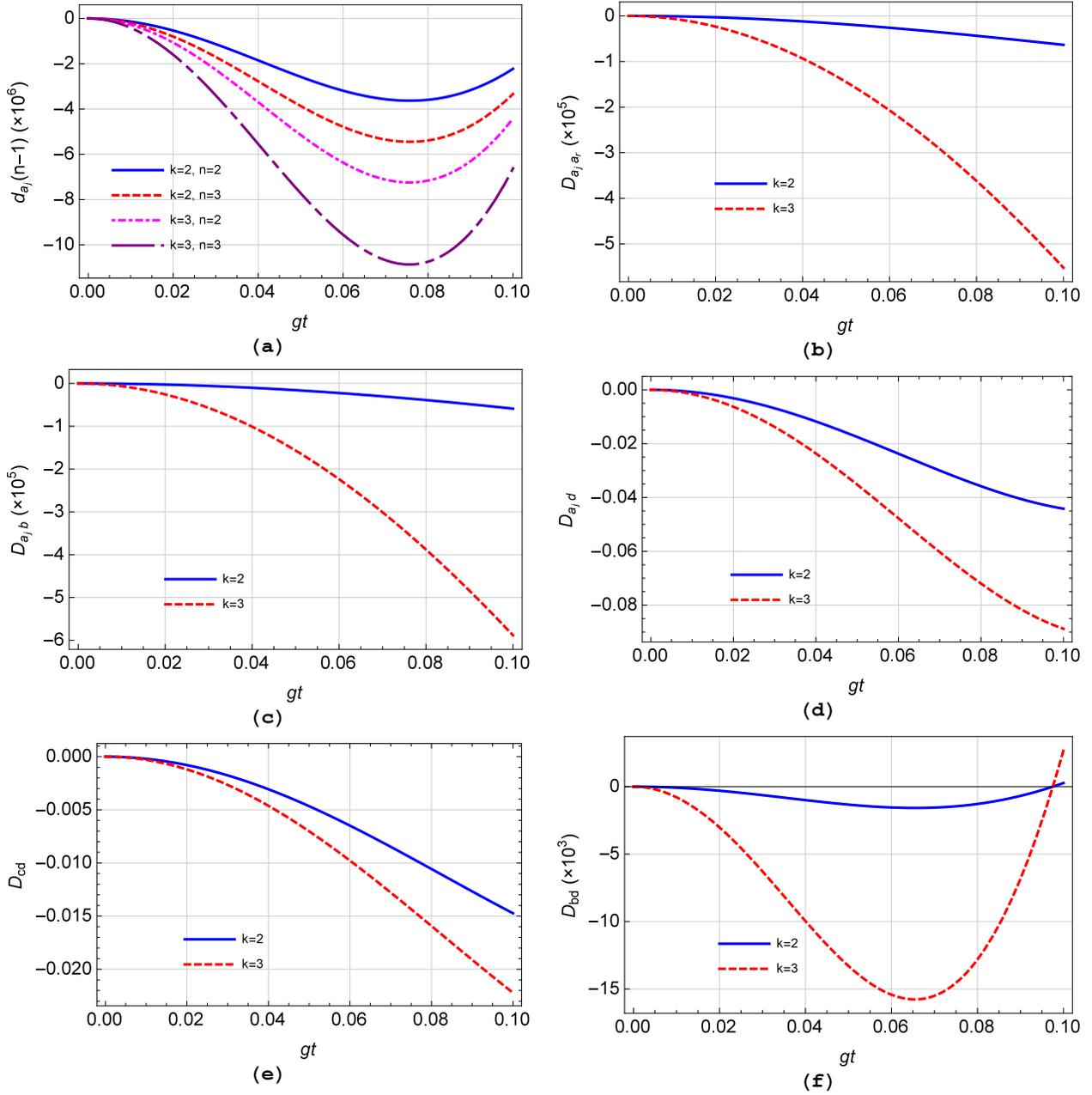}\caption{\label{fig:Ant}(Color online) Higher order antibunching and intermodal
antibunching in $k$-pump modes non-degenerate hyper-Raman processes
with (a) The values for lower order antibunching ($n=2$) in 2-pump
(3-pump) modes non-degenerate hyper-Raman processes is multiplied
by $10^{4}$ ($2\times10^{2}$) and higher order antibunching in 2-pump
modes non-degenerate hyper-Raman processes is amplified 50 times. Intermodal
antibunching in (b) pump-pump, (c) pump-Stokes, (d) pump-anti-Stokes,
(e) vibration-anti-Stokes, and (f) Stokes-anti-Stokes modes for 2-pump
and 3-pump modes non-degenerate hyper-Raman processes. The variation
in case of 2-pump modes is amplified 10 times in (b),  (c), and (f), while
100 times in (d) and (e). In all the cases, we have used $\left|\alpha_{i}\right|=10$.}
\end{figure}

\begin{equation}
\begin{array}{lcl}
D_{a_{j}a_{r}} & = & \left|\alpha_{j}\right|^{2}\left|\alpha_{r}\right|^{2}\left|\alpha_{i}\right|^{2}\left\{ -\left|f_{2}\right|^{2}\left(\left|\beta\right|^{2}+\left|\gamma\right|^{2}+1\right)\right.+\left.\left|f_{3}\right|^{2}\left(3\left|\delta\right|^{2}-\left|\gamma\right|^{2}\right)\right\} +3\left|\alpha_{j}\right|^{2}\left|\alpha_{r}\right|^{2}\sigma_{l}\left\{ \left|f_{2}\right|^{2}\right.\\
 & \times & \left|\beta\right|^{2}\left|\gamma\right|^{2}+\left|f_{3}\right|^{2}\left|\delta\right|^{2}\left(\left|\gamma\right|^{2}+1\right)+ \left.\left(f_{2}^{*}f_{3}\beta^{*}\gamma^{*2}\delta+{\rm c.c.}\right)\right\} +2\left(\left|\alpha_{i}\right|^{2}+\sigma_{l}\right)\\
 & + & \left(\left|\alpha_{j}\right|^{2}+\left|\alpha_{r}\right|^{2}+1\right)\left\{ \left|f_{2}\right|^{2}\left|\beta\right|^{2}\left|\gamma\right|^{2}\right.+ \left.\left|f_{3}\right|^{2}\left|\delta\right|^{2}\left(\left|\gamma\right|^{2}+1\right)+\left(f_{2}^{*}f_{3}\beta^{*}\gamma^{*2}\delta+{\rm c.c.}\right)\right\} \\
 & + & 2\left|\alpha_{j}\right|^{2}\left(\left|\alpha_{i}\right|^{2}+\sigma_{l}\right)\left\{ \left|f_{2}\right|^{2}\left|\beta\right|^{2}\left|\gamma\right|^{2}+\left|f_{3}\right|^{2}\left|\delta\right|^{2}\right. \left.\left(\left|\gamma\right|^{2}+1\right)+\left(f_{2}^{*}f_{3}\beta^{*}\gamma^{*2}\delta+{\rm c.c.}\right)\right\} +\left\{ f_{1}^{*}f_{2}\right.\\
 & \times & \alpha_{j}^{*}\alpha_{r}^{*}\alpha_{i}^{*}\beta\gamma+\left(2f_{1}^{*}f_{8}+f_{1}^{*2}f_{2}f_{3}\right) \left.\alpha_{j}^{*2}\alpha_{r}^{*2}\alpha_{i}^{*2}\beta\delta+f_{1}^{*}f_{3}\alpha_{j}^{*}\alpha_{r}^{*}\alpha_{i}^{*}\gamma^{*}\delta+{\rm c.c.}\right\} ,
\end{array}\label{eq:dai-aj}
\end{equation}

\begin{figure}[t]
\centering{}\includegraphics[scale=0.8]{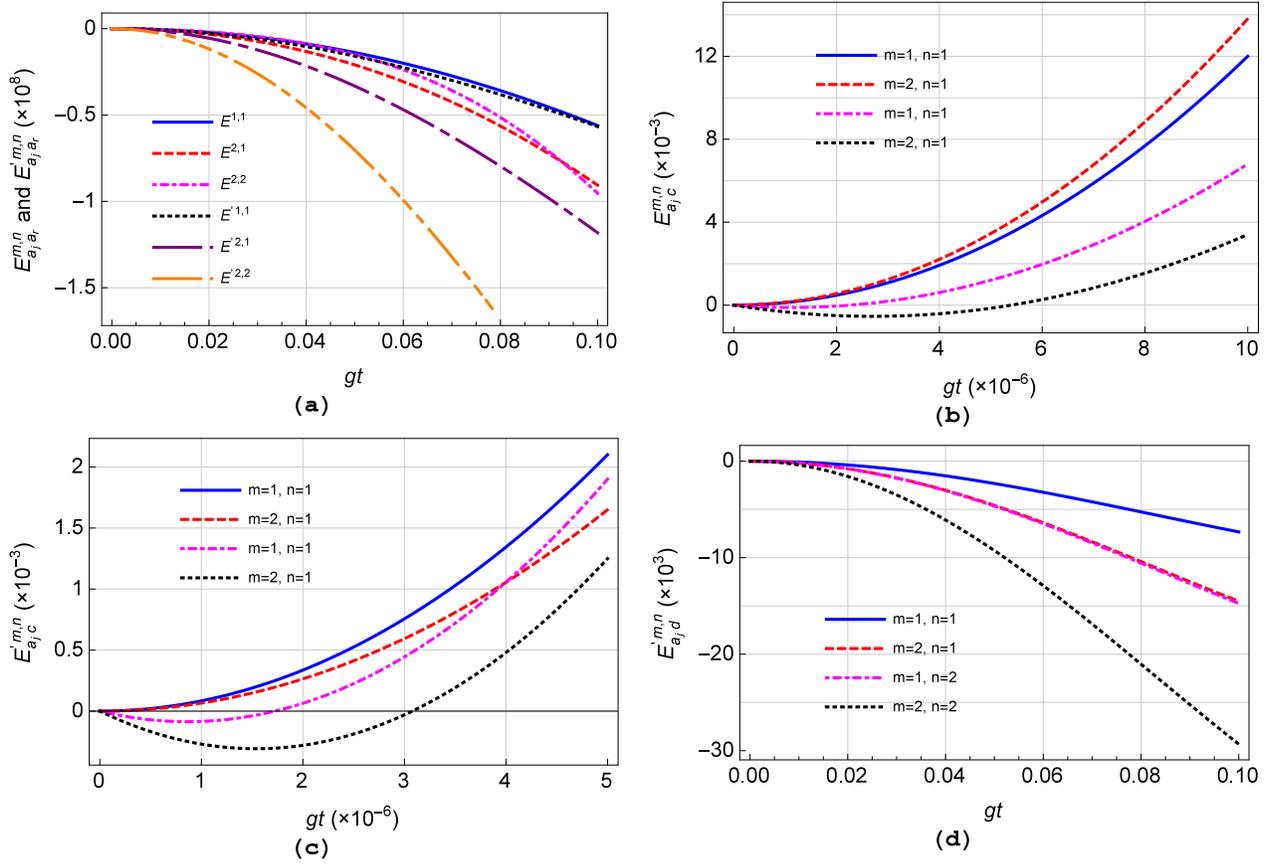}\caption{\label{fig:Ent-A}(Color online) The presence of lower and higher
order entanglement in two pump modes is shown in (a) using HZ1 and
HZ2 criteria. In (b) and (c), both lower and higher order entanglement
between pump and vibration modes is established using HZ1 and HZ2
criteria, respectively. The lower order entanglement in (a) is amplified
$10^{4}$ times, while higher order entanglement with $m=2,\,n=1$
(which is same as for $m=1,\,n=2$) is amplified $10^{2}$ times. The
dot-dashed-magenta and dotted-black lines in (b) and (c) correspond
to the phase angle $\phi_{1}=\frac{\pi}{2}$ and $-\frac{\pi}{2}$,
respectively, in $\alpha_{1}=\left|\alpha_{1}\right|\exp\left(i\phi_{1}\right)$.
In (b) and (c), the lower order entanglement is amplified 100 times, whereas
in (d), the lower order entanglement and higher order entanglement
with $m=1,\,n=2$ are amplified by 100 times, respectively. In all
the cases, $\left|\alpha_{i}\right|=10$ has been used. }
\end{figure}

\end{widetext}

\noindent pump-Stokes mode 
\begin{equation}
\begin{array}{lcl}
D_{a_{j}b} & = & -\left|f_{2}\right|^{2}\left|\alpha_{j}\right|^{2}\left|\beta\right|^{2}\left(\left|\alpha_{i}\right|^{2}+\sigma_{l}\left|\gamma\right|^{2}\right)\\
 & + & \left\{ \left(f_{1}^{*}f_{9}+g_{1}^{*}g_{2}f_{1}^{*}f_{3}\right)\left|\alpha_{j}\right|^{2}\sigma_{l}\beta^{*}\gamma^{*2}\delta+{\rm c.c.}\right\} ,
\end{array}\label{eq:dab}
\end{equation}
pump-vibration mode
\begin{equation}
\begin{array}{lcl}
D_{a_{j}c} & = & -\left|\alpha_{j}\right|^{2}\left|\gamma\right|^{2}\left\{ \left|f_{2}\right|^{2}\left(\left|\alpha_{i}\right|^{2}+\sigma_{l}\left|\beta\right|^{2}\right)-\left|f_{3}\right|^{2}\right.\\
 & \times & \left.\left(\left|\alpha_{i}\right|^{2}+\sigma_{l}\left|\delta\right|^{2}\right)\right\} +\left|f_{3}\right|^{2}\left|\delta\right|^{2}\left(\left|\alpha_{i}\right|^{2}+\sigma_{l}\right)\\
 & \times & \left(2\left|\gamma\right|^{2}+1\right)+\left\{ f_{1}^{*}f_{3}\alpha_{j}^{*}\alpha_{i}^{*}\gamma^{*}\delta\right.\\
 & + & h_{2}^{*}h_{1}f_{1}^{*}f_{3}\alpha_{j}^{*2}\alpha_{i}^{*2}\beta\delta+f_{2}^{*}f_{3}\left(\left|\alpha_{i}\right|^{2}+\sigma_{l}\right)\beta^{*}\gamma^{*2}\delta\\
 & + & \left.\left(2f_{1}^{*}f_{9}+h_{1}^{*}h_{2}f_{1}^{*}f_{3}\right)\left|\alpha_{j}\right|^{2}\sigma_{l}\beta^{*}\gamma^{*2}\delta+{\rm c.c.}\right\} ,
\end{array}\label{eq:dac}
\end{equation}
pump-anti-Stokes mode
\begin{equation}
\begin{array}{lcl}
D_{a_{j}d} & = & \left|f_{3}\right|^{2}\left|\alpha_{j}\right|^{2}\left|\delta\right|^{2}\left\{ \left|\alpha_{i}\right|^{2}+\sigma_{l}\left(\left|\gamma\right|^{2}+\left(\left|\delta\right|^{2}-1\right)\right)\right\} \\
 & + & \left\{ \left(f_{1}^{*}f_{7}+l_{1}^{*}l_{2}f_{1}^{*}f_{2}\right)\left|\alpha_{j}\right|^{2}\sigma_{l}\beta\gamma^{2}\delta^{*}+{\rm c.c.}\right\} ,
\end{array}\label{eq:dad}
\end{equation}
Stokes-vibration mode



\begin{equation}
\begin{array}{lcl}
D_{bc} & = & \left|g_{2}\right|^{2}\left|\alpha_{i}\right|^{2}\left(2\left|\gamma\right|^{2}+1\right)+\left\{ g_{1}^{*}g_{2}\alpha_{i}\beta^{*}\gamma^{*}\right.\\
 & + & g_{1}^{*}g_{5}\left|\beta\right|^{2}\left|\gamma\right|^{2}\sigma_{l}+g_{1}^{*}g_{6}\sigma_{l}\beta^{*}\gamma^{*2}\delta\\
 & + & \left.h_{2}^{*}h_{1}g_{1}^{*}g_{2}\left|\alpha_{i}\right|^{2}\left|\beta\right|^{2}+h_{3}^{*}h_{1}g_{1}^{*}g_{2}\alpha_{i}^{2}\beta^{*}\delta^{*}+{\rm c.c.}\right\} ,
\end{array}\label{eq:dbc}
\end{equation}
Stokes-anti-Stokes mode
\begin{equation}
\begin{array}{lcl}
D_{bd} & = & \left\{ l_{1}^{*}l_{3}\alpha_{i}^{2}\beta^{*}\delta^{*}+{\rm c.c.}\right\} ,\end{array}\label{eq:dbd}
\end{equation}
and vibration-anti-Stokes mode 
\begin{equation}
\begin{array}{lcc}
D_{cd} & = & -\left|l_{2}\right|^{2}\left|\gamma\right|^{2}\left|\delta\right|^{2}\sigma_{l}.\end{array}\label{eq:dcd}
\end{equation}

\end{subequations}

From the obtained expressions, the presence of intermodal antibunching
in various compound modes is shown in Figure \ref{fig:Ant}. We could
detect antibunching in all possible compound modes except pump-vibration
and Stokes-vibration modes. It is interesting to observe that the depth
of nonclassicality witness increases with the increase in number of pump modes in hyper-Raman
process for the same values of the coupling constants. On top of that, two arbitrary
pump modes are also found to possess intermodal antibunching as shown
in Figure \ref{fig:Ant} (b).

\subsection{Lower and higher order entanglement}

Inseparability of various modes can be analyzed using HZ-I and HZ-II
criteria of entanglement given in Eqs. (\ref{hoe-criteria}) and (\ref{hoe-criteria-2}).
For the two arbitrary pump modes the compact expression is obtained as follows
\begin{subequations}

\begin{widetext}

\begin{equation}
\begin{array}{lcl}
\left(\begin{array}{c}
E_{a_{j},a_{r}}^{m,n}\\
E_{a_{j},a_{r}}^{\prime m,n}
\end{array}\right) & = & \left|f_{2}\right|^{2}\left|\alpha_{j}\right|^{2\left(m-1\right)}\left|\alpha_{r}\right|^{2\left(n-1\right)}\left[\left|\alpha_{i}\right|^{2}\left\{ -mn\left|\alpha_{j}\right|^{2}\left|\alpha_{r}\right|^{2}\left(1+\left|\beta\right|^{2}+\left|\gamma\right|^{2}\right)\pm\left|\beta\right|^{2}\left|\gamma\right|^{2}\right.\right.\\
 & \times & \left.\left(m^{2}n^{2}+m^{2}\left(2n\pm1\right)\left|\alpha_{r}\right|^{2}+n^{2}\left(2m\pm1\right)\left|\alpha_{j}\right|^{2}\right)\right\} +\left\{ m^{2}\left(1+\left|\alpha_{r}\right|^{2}\right)\left|\alpha_{r}\right|^{2}+n^{2}\left(1+\left|\alpha_{j}\right|^{2}\right)\left|\alpha_{j}\right|^{2}\right.\\
 & \pm & \left.\left.mn\left|\alpha_{j}\right|^{2}\left|\alpha_{r}\right|^{2}\right\} \sigma_{l}\left|\beta\right|^{2}\left|\gamma\right|^{2}\right]+\left|f_{3}\right|^{2}\left|\alpha_{j}\right|^{2\left(m-1\right)}\left|\alpha_{r}\right|^{2\left(n-1\right)}\left[\left|\alpha_{i}\right|^{2}\left\{ \left|\delta\right|^{2}\left(m^{2}\left(1+\left|\alpha_{r}\right|^{2}\right)\left|\alpha_{r}\right|^{2}\right.\right.\right.\\
 & + & \left.n^{2}\left(1+\left|\alpha_{j}\right|^{2}\right)\left|\alpha_{j}\right|^{2}+\left(m^{2}\left(1\pm2n\right)\left|\alpha_{r}\right|^{2}+n^{2}\left(1\pm2m\right)\left|\alpha_{j}\right|^{2}\pm m^{2}n^{2}\right)\left|\gamma\right|^{2}\right)\pm mn\left|\alpha_{j}\right|^{2}\left|\alpha_{r}\right|^{2}\\
 & \times & \left.\left(\left|\delta\right|^{2}-\left|\gamma\right|^{2}\right)\right\} +\sigma_{l}\left(\left|\gamma\right|^{2}+1\right)\left|\delta\right|^{2}\left\{ m^{2}\left|\alpha_{r}\right|^{2}\left(1+\left|\alpha_{r}\right|^{2}\right)+n^{2}\left(1+\left|\alpha_{j}\right|^{2}\right)\left|\alpha_{j}\right|^{2}\pm mn\left|\alpha_{j}\right|^{2}\right.\\
 & \times & \left.\left.\left|\alpha_{r}\right|^{2}\right\} \right]+F_{a\pm}\pm mn\left|\alpha_{j}\right|^{2\left(m-2\right)}\left|\alpha_{r}\right|^{2\left(n-2\right)}\left[f_{2}^{*}f_{1}\alpha_{j}\alpha_{r}\alpha_{i}\beta^{*}\gamma^{*}+f_{3}^{*}f_{1}\alpha_{j}\alpha_{r}\alpha_{i}\gamma\delta^{*}+\left(f_{2}^{*2}f_{1}^{2}\right.\right.\\
 & \times & \left.\alpha_{j}^{2}\alpha_{r}^{2}\alpha_{i}^{2}\beta^{*2}\gamma^{*2}+f_{3}^{*2}f_{1}^{2}\alpha_{j}^{2}\alpha_{r}^{2}\alpha_{i}^{2}\gamma^{2}\delta^{*2}\right)\left(\left(m-1\right)\left|\alpha_{r}\right|^{2}+\left(n-1\right)\left|\alpha_{j}\right|^{2}+\frac{\left(m-1\right)\left(n-1\right)}{2}\right)\\
 & + & \alpha_{j}^{2}\alpha_{r}^{2}\alpha_{i}^{2}\beta^{*}\delta^{*}\left\{ f_{8}^{*}f_{1}\left(\left(2\left|\alpha_{j}\right|^{2}+m-1\right)\left|\alpha_{r}\right|^{2}+\left(n-1\right)\left(\left|\alpha_{j}\right|^{2}+\frac{m-1}{2}\right)\right)+f_{1}^{2}f_{2}^{*}f_{3}^{*}\left(\left|\gamma\right|^{2}\left(n-1\right)\right.\right.\\
 & \times & \left.\left.\left(2\left|\alpha_{j}\right|^{2}+m-1\right)+\left(n-1\right)\left(\left|\alpha_{j}\right|^{2}+\frac{\left(m-1\right)}{2}\right)+2\left(m-1\right)\left|\alpha_{r}\right|^{2}\left(2\left|\gamma\right|^{2}+1\right)+\left|\alpha_{r}\right|^{2}\left|\gamma\right|^{2}\right)\right\} \\
 & + & f_{2}^{*}f_{3}\left|\alpha_{j}\right|^{2}\left|\alpha_{r}\right|^{2}\beta^{*}\gamma^{*2}\delta\left\{ \left|\alpha_{i}\right|^{2}\left(n^{2}\left(1\pm2m\right)\left|\alpha_{j}\right|^{2}+m^{2}\left(1\pm2n\right)\left|\alpha_{r}\right|^{2}\pm m^{2}n^{2}\right)\right.\\
 & + & \left.\left.\left.\sigma_{l}\left(n^{2}\left(1+\left|\alpha_{j}\right|^{2}\right)\left|\alpha_{j}\right|^{2}\right.+m^{2}\left(1+\left|\alpha_{r}\right|^{2}\right)\left|\alpha_{r}\right|^{2}\pm mn\left|\alpha_{j}\right|^{2}\left|\alpha_{r}\right|^{2}\right)\right\} +{\rm c.c.}\right],
\end{array}\label{eq:aiaj}
\end{equation}
\end{widetext}

\noindent where 
\[
\begin{array}{lcl}
F_{a+} & = & mn\left|\alpha_{j}\right|^{2\left(m-1\right)}\left|\alpha_{r}\right|^{2\left(n-1\right)}\left(2n\left|\alpha_{j}\right|^{2}+2m\left|\alpha_{r}\right|^{2}\right.\\
 & + & \left.mn\right)\left\{ \left|f_{2}\right|^{2}\sigma_{l}+\left|f_{3}\right|^{2}\left|\delta\right|^{2}\left(\left|\alpha_{i}\right|^{2}+\sigma_{l}\left(\left|\gamma\right|^{2}+1\right)\right)\right.\\
 & + & \left.\left(f_{2}^{*}f_{3}\sigma_{l}\beta^{*}\gamma^{*2}\delta+{\rm c.c.}\right)\right\} 
\end{array}
\]
 and $F_{a-}=0$. While the analytic expression of HZ-I and HZ-II
for pump-Stokes mode is obtained as 
\begin{equation}
\begin{array}{lcl}
\left(\begin{array}{c}
E_{a_{j},b}^{m,n}\\
E_{a_{j},b}^{\prime m,n}
\end{array}\right) & = & \left|f_{2}\right|^{2}\left|\alpha_{j}\right|^{2\left(m-1\right)}\left|\beta\right|^{2\left(n-1\right)}\left(n\left|\alpha_{j}\right|^{2}\mp m\left|\beta\right|^{2}\right)\\
 & \times & \left(n\left|\alpha_{j}\right|^{2}\left|\alpha_{i}\right|^{2}-m\sigma_{l}\left|\beta\right|^{2}\left|\gamma\right|^{2}\right)+m^{2}\left|f_{3}\right|^{2}\\
 & \times & \left|\alpha_{j}\right|^{2\left(m-1\right)}\left|\beta\right|^{2n}\left|\delta\right|^{2}\left(\left|\alpha_{i}\right|^{2}+\sigma_{l}\left(\left|\gamma\right|^{2}+1\right)\right)\\
 & + & m\left|\alpha_{j}\right|^{2\left(m-1\right)}\left|\beta\right|^{2\left(n-1\right)}\sigma_{l}\\
 & \times & \left\{ \left(mf_{2}^{*}f_{3}\left|\beta\right|^{2}\pm ng_{1}^{*}g_{6}\left|\alpha_{j}\right|^{2}\right)\beta^{*}\gamma^{*2}\delta+{\rm c.c.}\right\} .
\end{array}\label{ab}
\end{equation}
A similar study for pump-vibration and pump-anti-Stokes modes are
obtained as

\begin{equation}
\begin{array}{lcl}
\left(\begin{array}{c}
E_{a_{j},c}^{m,n}\\
E_{a_{j},c}^{\prime m,n}
\end{array}\right) & = & \left|f_{2}\right|^{2}\left(n\left|\alpha_{j}\right|^{2}\left|\alpha_{i}\right|^{2}-m\sigma_{l}\left|\beta\right|^{2}\left|\gamma\right|^{2}\right)\\
 & \times & \left(n\left|\alpha_{j}\right|^{2}\mp m\left|\gamma\right|^{2}\right)\left|\alpha_{j}\right|^{2\left(m-1\right)}\left|\gamma\right|^{2\left(n-1\right)}\\
 & + & \left|f_{3}\right|^{2}\left|\alpha_{j}\right|^{2\left(m-1\right)}\left|\gamma\right|^{2\left(n-1\right)}\left[\left|\alpha_{i}\right|^{2}\right.\\
 & \times & \left\{ n^{2}\left(1\pm2m\right)\left|\alpha_{j}\right|^{2}\left|\delta\right|^{2}+m^{2}\left(1\pm2n\right)\left|\gamma\right|^{2}\right.\\
 & \times & \left.\left|\delta\right|^{2}\mp mn\left|\alpha_{j}\right|^{2}\left|\gamma\right|^{2}\pm m^{2}n^{2}\left|\delta\right|^{2}\right\} \\
 & + & \left\{ n^{2}\left(1+\left|\alpha_{j}\right|^{2}\right)\left|\alpha_{j}\right|^{2}+m^{2}\left(1+\left|\gamma\right|^{2}\right)\left|\gamma\right|^{2}\right.\\
 & \pm & \left.\left.mn\left|\alpha_{j}\right|^{2}\left|\gamma\right|^{2}\right\} \sigma_{l}\left|\delta\right|^{2}\right]+F_{c\pm}\\
 & \pm & mn\left|\alpha_{j}\right|^{2\left(m-2\right)}\left|\gamma\right|^{2\left(n-2\right)}\left[f_{3}^{*}f_{1}\left|\alpha_{j}\right|^{2}\left|\gamma\right|^{2}\right.\\
 & \times & \alpha_{j}\alpha_{i}\gamma\delta^{*}+f_{2}^{*}f_{3}\left|\alpha_{j}\right|^{2}\left|\alpha_{i}\right|^{2}\beta^{*}\gamma^{*2}\delta\\
 & \times & \left\{ m\left|\gamma\right|^{2}-\left(n-1\right)\left|\alpha_{j}\right|^{2}\right\} +h_{3}^{*}h_{2}\left|\gamma\right|^{2}\alpha_{j}^{2}\alpha_{i}^{2}\\
 & \times & \beta^{*}\delta^{*}\left\{ n\left|\alpha_{j}\right|^{2}-\left(m-1\right)\left|\gamma\right|^{2}\right\} +\left|\alpha_{j}\right|^{2}\sigma_{l}\\
 & \times & \beta^{*}\gamma^{*2}\delta\left\{ \frac{m}{n}f_{2}^{*}f_{3}\left|\gamma\right|^{4}\pm h_{1}^{*}h_{6}\left|\alpha_{j}\right|^{2}\left|\gamma\right|^{2}\right.\\
 & \pm & \left.\left.\left(n-1\right)\left|\alpha_{j}\right|^{2}\left(f_{1}^{*}f_{9}-f_{2}^{*}f_{3}\right)\right\} +{\rm c.c.}\right],
\end{array}\label{ac}
\end{equation}

\begin{widetext}

\begin{figure}
\centering{}\includegraphics[scale=0.8]{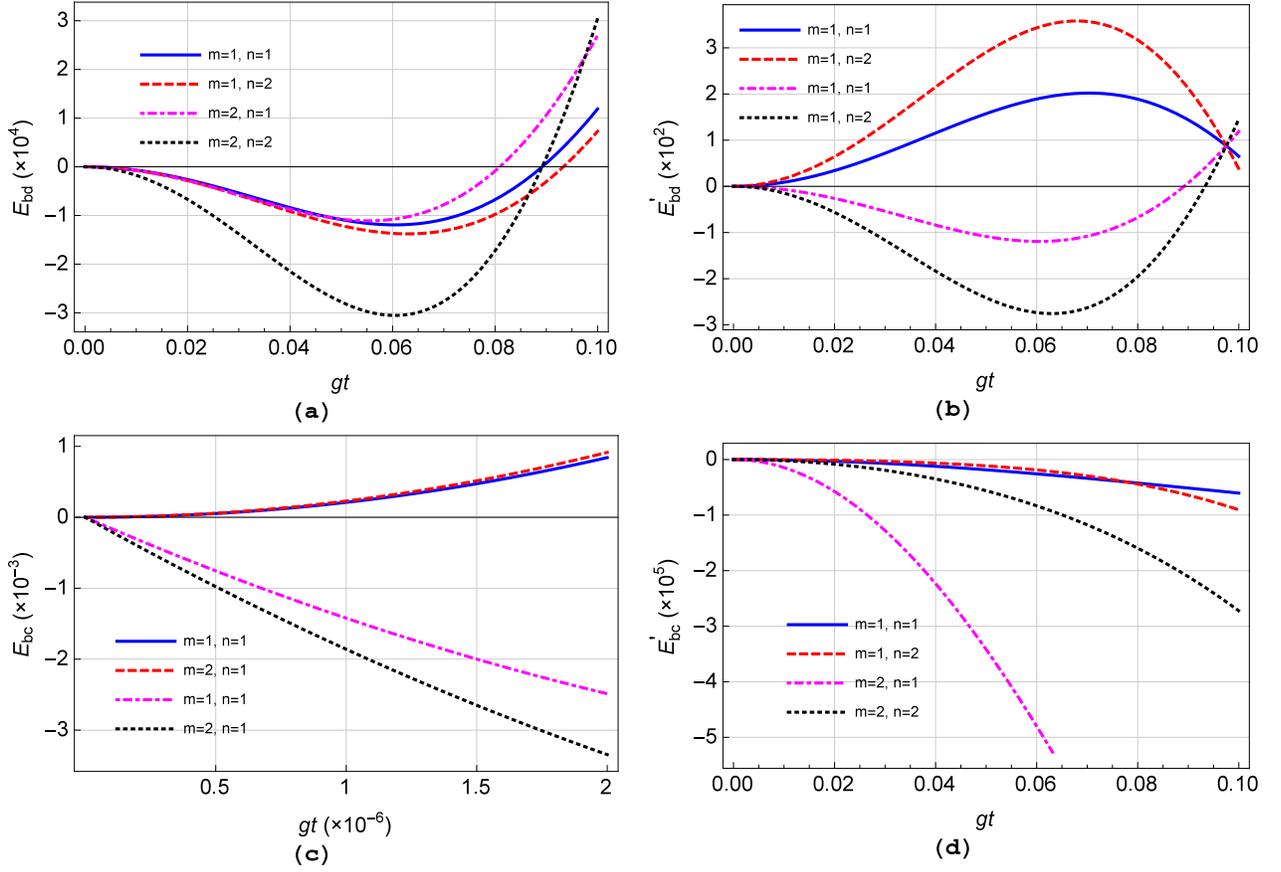}\caption{\label{fig:Ent}(Color online) The presence of lower and higher order
entanglement in Stokes-anti-Stokes ((a)-(b)) and Stokes-vibration
((c)-(d)) modes is established using HZ1 ((a) and (c)) and HZ2 ((b)
and (d)) criteria. The lower order entanglement in (a) is amplified
100 times, while higher order entanglement with $m=1,\,n=2$ is amplified
50 times. The dot-dashed-magenta and dotted-black lines in (b) and
(c) correspond to the phase angle $\phi_{1}=\frac{\pi}{2}$ and $-\frac{\pi}{2}$,
respectively, in $\alpha_{1}=\left|\alpha_{1}\right|\exp\left(i\phi_{1}\right)$.
In (c), the lower order entanglement is shown after multiplying with 100, whereas in
(d), the lower order entanglement is amplified 10 times and $m=1,\,n=2$
and $m=2,\,n=2$ are $10^{5}$ and $10^{3}$ times amplified, respectively.
In all the cases, $\left|\alpha_{i}\right|=10$ has been used. }
\end{figure}

\begin{figure}
\centering{}\includegraphics{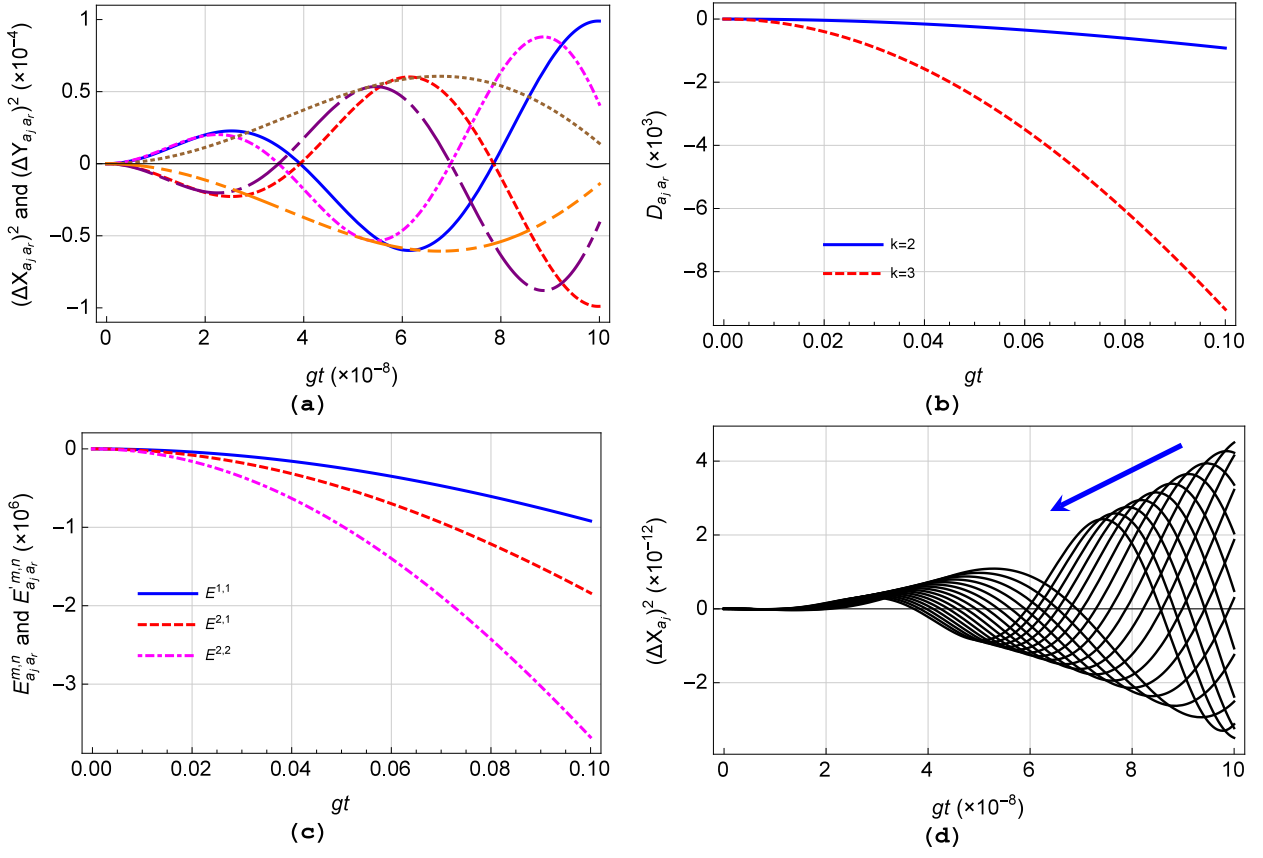}\caption{\label{fig:Spont}(Color online) The presence of lower and higher
order nonclassicality in pump modes in spontaneous case. Intermodal
(a) squeezing, (b) antibunching, and (c) entanglement for compound
mode $a_{j}a_{r}$ is shown for the same values of the corresponding
plots in Figures \ref{fig:SqA}-\ref{fig:Ent-A} for stimulated case.
It is important to note that the values for HZ1 and HZ2 are obtained
to be the same. In (d), squeezing in quadrature $X_{a_{j}}$
for 2-pump modes stimulated case with the value of frequency $\left(\frac{\omega_{i}}{g}\right)$
for $a_{j}$ mode varied between $35\times10^{6}$ and $65\times10^{6}$
in the steps of $2\times10^{6}$. The arrow indicates the variation in quadrature squeezing due to increase in the frequency.}
\end{figure}

\end{widetext}

\noindent where 
\[
\begin{array}{lcl}
F_{c+} & = & mn\sigma_{l}\left|\alpha_{j}\right|^{2\left(m-1\right)}\left|\gamma\right|^{2\left(n-1\right)}\left\{ \left(n\left|\alpha_{j}\right|^{2}+m\left|\gamma\right|^{2}\right)\right.\\
 & \times & \left.2\left|f_{3}\right|^{2}\left|\delta\right|^{2}+\left(mf_{2}^{*}f_{3}\beta^{*}\gamma^{*2}\delta+{\rm c.c.}\right)\right\} 
\end{array}
\]
 and $F_{c-}=0$; and 
\begin{equation}
\begin{array}{lcl}
\left(\begin{array}{c}
E_{a_{j},d}^{m,n}\\
E_{a_{j},d}^{\prime m,n}
\end{array}\right) & = & m\left|f_{3}\right|^{2}\left|\alpha_{j}\right|^{2\left(m-1\right)}\left(\left|\alpha_{i}\right|^{2}+\sigma_{l}\left(\left|\gamma\right|^{2}+1\right)\right)\\
 & \times & \left|\delta\right|^{2n}\left(m\left|\delta\right|^{2}\mp n\left|\alpha_{j}\right|^{2}\right)+m^{2}\left|f_{2}\right|^{2}\left|\alpha_{j}\right|^{2\left(m-1\right)}\\
 & \times & \left|\delta\right|^{2n}\left|\beta\right|^{2}\left|\gamma\right|^{2}\sigma_{l}+m\left|\alpha_{j}\right|^{2\left(m-1\right)}\left|\delta\right|^{2\left(n-1\right)}\sigma_{l}\\
 & \times & \left\{ \left(mf_{2}^{*}f_{3}\left|\delta\right|^{2}+nl_{4}^{*}l_{1}\left|\alpha_{j}\right|^{2}\right)\beta^{*}\gamma^{*2}\delta+{\rm c.c.}\right\} ,
\end{array}\label{ad}
\end{equation}
respectively.

The analysis of the obtained analytic expressions of entanglement
of an arbitrary pump mode with all the remaining modes revealed some
interesting results. Specifically, all the pump modes are found to
be entangled with vibration and anti-Stokes modes as shown in Figure
\ref{fig:Ent-A} (b), (c) and (d). Importantly, the bipartite entanglement
between a pump and vibration modes could only be ensured for initial
evolution of the system. However, as the criteria used here are only
sufficient not necessary, the separability of these two modes can
not be deduced. One significant result, which would be absent in Raman
or degenerate hyper-Raman process due to the existence of single pump
mode, is entanglement between two pump modes (cf. Figure \ref{fig:Ent-A}
(a)). The present results establish that two pump modes are always
entangled in the non-degenarate hyper-Raman process. This interesting result would be in continuation of a set
of systems able to produce always entangled pump modes \cite{thapliyal2014higher,thapliyal2014nonclassical}
and bosonic modes \cite{giri2017nonclassicality}.

A similar study for all the modes, except pump mode, resulted in following
compact analytic expressions for Stokes-vibration, vibration-anti-Stokes,
and Stokes-anti-Stokes modes 

\begin{widetext}

\begin{equation}
\begin{array}{lcl}
\left(\begin{array}{c}
E_{b,c}^{m,n}\\
E_{b,c}^{\prime m,n}
\end{array}\right) & = & \left|g_{2}\right|^{2}\left|\beta\right|^{2\left(m-1\right)}\left|\gamma\right|^{2\left(n-1\right)}\left\{ m^{2}\left(1\pm2n\right)\left|\alpha_{i}\right|^{2}\left|\gamma\right|^{2}+n^{2}\left(1\pm2m\right)\left|\alpha_{i}\right|^{2}\left|\beta\right|^{2}\right.\\
 & \pm & \left.m^{2}n^{2}\left|\alpha_{i}\right|^{2}\mp mn\sigma_{l}\left|\beta\right|^{2}\left|\gamma\right|^{2}\right\} +n^{2}\left|h_{3}\right|^{2}\sigma_{l}\left|\beta\right|^{2m}\left|\gamma\right|^{2\left(n-1\right)}\left|\delta\right|^{2}\\
 & \pm & mn\left|\beta\right|^{2\left(m-2\right)}\left|\gamma\right|^{2\left(n-2\right)}\left[g_{2}^{*}g_{1}\left|\beta\right|^{2}\left|\gamma\right|^{2}\alpha_{i}^{*}\beta\gamma+nh_{3}^{*}h_{2}\left|\beta\right|^{2}\left|\gamma\right|^{2}\alpha_{i}^{2}\beta^{*}\delta^{*}\right.\\
 & + & g_{6}^{*}g_{1}\sigma_{l}\left|\beta\right|^{2}\beta\gamma^{2}\delta^{*}\left(2\left|\gamma\right|^{2}+n-1\right)+\left(n-1\right)h_{1}^{*2}h_{2}h_{3}\left|\alpha_{i}\right|^{2}\left|\beta\right|^{2}\beta^{*}\gamma^{*2}\delta\\
 & + & \left.g_{2}^{*2}g_{1}^{2}\alpha_{i}^{*2}\beta^{2}\gamma^{2}\left\{ \frac{1}{2}\left(m-1\right)\left(n-1\right)+\left(n-1\right)\left|\beta\right|^{2}+\left(m-1\right)\left|\gamma\right|^{2}\right\} +{\rm c.c.}\right],
\end{array}\label{bc}
\end{equation}

\end{widetext}

\begin{equation}
\begin{array}{lcl}
\left(\begin{array}{c}
E_{c,d}^{m,n}\\
E_{c,d}^{\prime m,n}
\end{array}\right) & = & m^{2}\left|h_{2}\right|^{2}\left|\alpha_{i}\right|^{2}\left|\gamma\right|^{2\left(m-1\right)}\left|\delta\right|^{2n}\\
 & + & \left|l_{2}\right|^{2}\left|\gamma\right|^{2\left(m-1\right)}\left|\delta\right|^{2n}\sigma_{l}\left[m^{2}\left|\delta\right|^{2}\mp mn\left|\gamma\right|^{2}\right],
\end{array}\label{cd}
\end{equation}
and
\begin{equation}
\begin{array}{lcl}
\left(\begin{array}{c}
E_{b,d}^{m,n}\\
E_{b,d}^{\prime m,n}
\end{array}\right) & = & m^{2}\left|g_{2}\right|^{2}\left|\alpha_{i}\right|^{2}\left|\beta\right|^{2\left(m-1\right)}\left|\delta\right|^{2n}\\
 & \pm & mn\left|\beta\right|^{2\left(m-1\right)}\left|\delta\right|^{2\left(n-1\right)}\left[l_{1}^{*}l_{3}\alpha_{i}^{2}\beta^{*}\delta^{*}+{\rm c.c.}\right],
\end{array}\label{bd}
\end{equation}
respectively. \end{subequations}

Entanglement between the modes except pump mode is also a topic
of prime interest in some of the recent studies on Raman or degenerate
hyper-Raman process \cite{sen2013intermodal,giri2016higher}. The
present results clearly reestablish that the non-separability criteria
are only sufficient as one of the criteria (either HZ1 or HZ2) detects
entanglement while the other one fails to detect entanglement in the same  regimes of various
parameters (cf. Figure \ref{fig:Ent} (a)-(b) or (c)-(d)). The present
results show Stokes-anti-Stokes and Stokes-vibration modes are both
lower and higher order entangled.

Finally, before we conclude the paper it is customary to check the
possibility of nonclassical behavior that can be observed even under
the spontaneous condition. The present results show that intermodal
squeezing, antibunching and entanglement between different pump modes
can be observed in the spontaneous case, too (cf. Figure \ref{fig:Spont}). Here, in Figure \ref{fig:Spont}
(d), we also establish the effect of change in frequency of input
pump beams in spontaneous case, but it should be noted that a similar
nature can be observed in stimulated case as well.

It is also worth noting here that in partial spontaneous case, when
one (or two) of the modes except the pump modes has non-zero photons initially,
all the nonclassicality observed in the spontaneous case will also
survive. On top of that, certain other nonclassical behaviors may appear.
Specifically, for non-zero photons in the Stokes mode, intermodal
squeezing and antibunching in the pump-Stokes compound mode can also
be observed.

\section{Conclusion \label{sec:Conclusion}}

Here, we have obtained a completely quantum mechanical solution of
the most general case of hyper-Raman process, i.e., with $k$ non-degenerate
pump modes. Our endeavor to obtain the Sen-Mandal perturbative solution
for this most general Hamiltonian, describing the multi-mode non-degenerate
hyper-Raman process, resulted in a solution quite general in its nature.
This general nature of the present Hamiltonian and corresponding solution
insinuated to deduce all the existing Sen-Mandal and short time solutions
for Raman and 2-pump mode degenerate hyper-Raman processes. This reduction
establishes the wide applicability of the present results for all the
Raman and hyper-Raman processes. 

Further, the present study also revealed various interesting results.
Specifically, the most significant property of the present system is more
than one non-degenerate pump modes. Therefore, the 
nonclassical features reported in an arbitrary single pump and compound two-pump modes
specify our most significant contribution. The present study revealed that
an arbitrary single pump mode shows both lower and higher order
squeezing and antibunching; while the compound pump-pump
mode possesses all the nonclassical properties studied here, i.e.,
intermodal squeezing, antibunching, and entanglement. 

The pump mode also shows compound mode nonclassicalities with Stokes,
vibration, anti-Stokes modes as well. Specifically, compound pump-Stokes
mode shows both intermodal squeezing and antibunching; compound pump-vibration
mode exhibits intermodal squeezing and lower and higher order entanglement;
intermodal squeezing antibunching, and lower and higher order entanglement
are present in compound pump-anti-Stokes mode. Higher order entanglement in terms of multimode entanglement (as studied in Refs. \cite{thapliyal2014higher,thapliyal2014nonclassical}) is not studied here as it is already shown that due to self-interaction two arbitrary pump modes are always entangled. Therefore, it is expected that all the pump modes would form a $k$-partite entangled state. In addition, all
the nonclassical properties observed in Raman or degenerate hyper-Raman
processes (\cite{sen2005squeezed,sen2007quantum,sen2007squeezing,sen2008amplitude,perina1991quantum}
and references therein) are also found to be present in the multi-mode non-degenerate
hyper-Raman process. Precisely, the presence of intermodal antibunching
and both lower and higher order entanglement in compound vibration-anti-Stokes
and Stokes-anti-Stokes modes have been established. 

Interestingly, most of the nonclassical properties of the hyper-Raman
process under consideration survive even in the spontaneous case.
To be specific, intermodal squeezing, antibunching, and lower and
higher order entanglement between two arbitrary pump modes are observed in the spontaneous case. Further,
squeezing and
intermodal squeezing involving the pump mode are found to depend on
the frequency and number of photons in the pump mode under consideration.
It is also observed to vary with the number of non-degenerate pump
modes. Additionally, intermodal antibunching and entanglement are
phase dependent properties and can be controlled by the phases of
the pump modes. The nonclassical behavior of hyper-Raman processes can also be established with the help of quasidistribution functions \cite{thapliyal2015quasiprobability}, which will be performed in the near future.

We conclude this paper with a hope that the growing experimental facilities and techniques would lead to  experimental realization of the single mode and
intermodal nonclassical properties observed in the pump and other modes in the present work.

\textbf{Acknowledgment:} KT acknowledges support from the Council
of Scientific and Industrial Research, Government of India. AP thanks
Department of Science and Technology (DST), India for the support
provided through the project number EMR/2015/000393. JP thanks the support from LO1305 of the Ministry of Education,
Youth and Sports of the Czech Republic.

\bibliographystyle{apsrev4-1}
\bibliography{hyRa}

\begin{thebibliography}{62}%
\makeatletter
\providecommand \@ifxundefined [1]{%
 \@ifx{#1\undefined}
}%
\providecommand \@ifnum [1]{%
 \ifnum #1\expandafter \@firstoftwo
 \else \expandafter \@secondoftwo
 \fi
}%
\providecommand \@ifx [1]{%
 \ifx #1\expandafter \@firstoftwo
 \else \expandafter \@secondoftwo
 \fi
}%
\providecommand \natexlab [1]{#1}%
\providecommand \enquote  [1]{``#1''}%
\providecommand \bibnamefont  [1]{#1}%
\providecommand \bibfnamefont [1]{#1}%
\providecommand \citenamefont [1]{#1}%
\providecommand \href@noop [0]{\@secondoftwo}%
\providecommand \href [0]{\begingroup \@sanitize@url \@href}%
\providecommand \@href[1]{\@@startlink{#1}\@@href}%
\providecommand \@@href[1]{\endgroup#1\@@endlink}%
\providecommand \@sanitize@url [0]{\catcode `\\12\catcode `\$12\catcode
  `\&12\catcode `\#12\catcode `\^12\catcode `\_12\catcode `\%12\relax}%
\providecommand \@@startlink[1]{}%
\providecommand \@@endlink[0]{}%
\providecommand \url  [0]{\begingroup\@sanitize@url \@url }%
\providecommand \@url [1]{\endgroup\@href {#1}{\urlprefix }}%
\providecommand \urlprefix  [0]{URL }%
\providecommand \Eprint [0]{\href }%
\providecommand \doibase [0]{http://dx.doi.org/}%
\providecommand \selectlanguage [0]{\@gobble}%
\providecommand \bibinfo  [0]{\@secondoftwo}%
\providecommand \bibfield  [0]{\@secondoftwo}%
\providecommand \translation [1]{[#1]}%
\providecommand \BibitemOpen [0]{}%
\providecommand \bibitemStop [0]{}%
\providecommand \bibitemNoStop [0]{.\EOS\space}%
\providecommand \EOS [0]{\spacefactor3000\relax}%
\providecommand \BibitemShut  [1]{\csname bibitem#1\endcsname}%
\let\auto@bib@innerbib\@empty
\bibitem [{\citenamefont {Abbott}\ \emph
  {et~al.}(2016{\natexlab{a}})\citenamefont {Abbott}, \citenamefont {Abbott},
  \citenamefont {Abbott}, \citenamefont {Abernathy}, \citenamefont {Acernese},
  \citenamefont {Ackley}, \citenamefont {Adams}, \citenamefont {Adams},
  \citenamefont {Addesso}, \citenamefont {Adhikari} \emph
  {et~al.}}]{abbott2016observation}%
  \BibitemOpen
  \bibfield  {author} {\bibinfo {author} {\bibfnamefont {B.~P.}\ \bibnamefont
  {Abbott}}, \bibinfo {author} {\bibfnamefont {R.}~\bibnamefont {Abbott}},
  \bibinfo {author} {\bibfnamefont {T.~D.}\ \bibnamefont {Abbott}}, \bibinfo
  {author} {\bibfnamefont {M.~R.}\ \bibnamefont {Abernathy}}, \bibinfo {author}
  {\bibfnamefont {F.}~\bibnamefont {Acernese}}, \bibinfo {author}
  {\bibfnamefont {K.}~\bibnamefont {Ackley}}, \bibinfo {author} {\bibfnamefont
  {C.}~\bibnamefont {Adams}}, \bibinfo {author} {\bibfnamefont
  {T.}~\bibnamefont {Adams}}, \bibinfo {author} {\bibfnamefont
  {P.}~\bibnamefont {Addesso}}, \bibinfo {author} {\bibfnamefont {R.~X.}\
  \bibnamefont {Adhikari}},  \emph {et~al.},\ }\href@noop {} {\bibfield
  {journal} {\bibinfo  {journal} {Phys. Rev. Lett.}\ }\textbf {\bibinfo
  {volume} {116}},\ \bibinfo {pages} {061102} (\bibinfo {year}
  {2016}{\natexlab{a}})}\BibitemShut {NoStop}%
\bibitem [{\citenamefont {Abbott}\ \emph
  {et~al.}(2016{\natexlab{b}})\citenamefont {Abbott}, \citenamefont {Abbott},
  \citenamefont {Abbott}, \citenamefont {Abernathy}, \citenamefont {Acernese},
  \citenamefont {Ackley}, \citenamefont {Adams}, \citenamefont {Adams},
  \citenamefont {Addesso}, \citenamefont {Adhikari} \emph
  {et~al.}}]{abbott2016gw151226}%
  \BibitemOpen
  \bibfield  {author} {\bibinfo {author} {\bibfnamefont {B.~P.}\ \bibnamefont
  {Abbott}}, \bibinfo {author} {\bibfnamefont {R.}~\bibnamefont {Abbott}},
  \bibinfo {author} {\bibfnamefont {T.~D.}\ \bibnamefont {Abbott}}, \bibinfo
  {author} {\bibfnamefont {M.~R.}\ \bibnamefont {Abernathy}}, \bibinfo {author}
  {\bibfnamefont {F.}~\bibnamefont {Acernese}}, \bibinfo {author}
  {\bibfnamefont {K.}~\bibnamefont {Ackley}}, \bibinfo {author} {\bibfnamefont
  {C.}~\bibnamefont {Adams}}, \bibinfo {author} {\bibfnamefont
  {T.}~\bibnamefont {Adams}}, \bibinfo {author} {\bibfnamefont
  {P.}~\bibnamefont {Addesso}}, \bibinfo {author} {\bibfnamefont {R.~X.}\
  \bibnamefont {Adhikari}},  \emph {et~al.},\ }\href@noop {} {\bibfield
  {journal} {\bibinfo  {journal} {Phys. Rev. Lett.}\ }\textbf {\bibinfo
  {volume} {116}},\ \bibinfo {pages} {241103} (\bibinfo {year}
  {2016}{\natexlab{b}})}\BibitemShut {NoStop}%
\bibitem [{\citenamefont {Ekert}(1991)}]{ekert1991quantum}%
  \BibitemOpen
  \bibfield  {author} {\bibinfo {author} {\bibfnamefont {A.~K.}\ \bibnamefont
  {Ekert}},\ }\href@noop {} {\bibfield  {journal} {\bibinfo  {journal} {Phys.
  Rev. Lett.}\ }\textbf {\bibinfo {volume} {67}},\ \bibinfo {pages} {661}
  (\bibinfo {year} {1991})}\BibitemShut {NoStop}%
\bibitem [{\citenamefont {Thapliyal}\ and\ \citenamefont
  {Pathak}(2015)}]{thapliyal2015applications}%
  \BibitemOpen
  \bibfield  {author} {\bibinfo {author} {\bibfnamefont {K.}~\bibnamefont
  {Thapliyal}}\ and\ \bibinfo {author} {\bibfnamefont {A.}~\bibnamefont
  {Pathak}},\ }\href@noop {} {\bibfield  {journal} {\bibinfo  {journal}
  {Quantum Inf. Process.}\ }\textbf {\bibinfo {volume} {14}},\ \bibinfo {pages}
  {2599} (\bibinfo {year} {2015})}\BibitemShut {NoStop}%
\bibitem [{\citenamefont {Thapliyal}\ \emph {et~al.}(2017)\citenamefont
  {Thapliyal}, \citenamefont {Pathak},\ and\ \citenamefont
  {Banerjee}}]{thapliyal2017quantum}%
  \BibitemOpen
  \bibfield  {author} {\bibinfo {author} {\bibfnamefont {K.}~\bibnamefont
  {Thapliyal}}, \bibinfo {author} {\bibfnamefont {A.}~\bibnamefont {Pathak}}, \
  and\ \bibinfo {author} {\bibfnamefont {S.}~\bibnamefont {Banerjee}},\
  }\href@noop {} {\bibfield  {journal} {\bibinfo  {journal} {Quantum Inf.
  Process.}\ }\textbf {\bibinfo {volume} {16}},\ \bibinfo {pages} {115}
  (\bibinfo {year} {2017})}\BibitemShut {NoStop}%
\bibitem [{\citenamefont {Bennett}\ \emph {et~al.}(1993)\citenamefont
  {Bennett}, \citenamefont {Brassard}, \citenamefont {Cr{\'e}peau},
  \citenamefont {Jozsa}, \citenamefont {Peres},\ and\ \citenamefont
  {Wootters}}]{bennett1993teleporting}%
  \BibitemOpen
  \bibfield  {author} {\bibinfo {author} {\bibfnamefont {C.~H.}\ \bibnamefont
  {Bennett}}, \bibinfo {author} {\bibfnamefont {G.}~\bibnamefont {Brassard}},
  \bibinfo {author} {\bibfnamefont {C.}~\bibnamefont {Cr{\'e}peau}}, \bibinfo
  {author} {\bibfnamefont {R.}~\bibnamefont {Jozsa}}, \bibinfo {author}
  {\bibfnamefont {A.}~\bibnamefont {Peres}}, \ and\ \bibinfo {author}
  {\bibfnamefont {W.~K.}\ \bibnamefont {Wootters}},\ }\href@noop {} {\bibfield
  {journal} {\bibinfo  {journal} {Phys. Rev. Lett.}\ }\textbf {\bibinfo
  {volume} {70}},\ \bibinfo {pages} {1895} (\bibinfo {year}
  {1993})}\BibitemShut {NoStop}%
\bibitem [{\citenamefont {Bennett}\ and\ \citenamefont
  {Wiesner}(1992)}]{bennett1992communication}%
  \BibitemOpen
  \bibfield  {author} {\bibinfo {author} {\bibfnamefont {C.~H.}\ \bibnamefont
  {Bennett}}\ and\ \bibinfo {author} {\bibfnamefont {S.~J.}\ \bibnamefont
  {Wiesner}},\ }\href@noop {} {\bibfield  {journal} {\bibinfo  {journal} {Phys.
  Rev. Lett.}\ }\textbf {\bibinfo {volume} {69}},\ \bibinfo {pages} {2881}
  (\bibinfo {year} {1992})}\BibitemShut {NoStop}%
\bibitem [{\citenamefont {Acin}\ \emph {et~al.}(2006)\citenamefont {Acin},
  \citenamefont {Gisin},\ and\ \citenamefont {Masanes}}]{acin2006bell}%
  \BibitemOpen
  \bibfield  {author} {\bibinfo {author} {\bibfnamefont {A.}~\bibnamefont
  {Acin}}, \bibinfo {author} {\bibfnamefont {N.}~\bibnamefont {Gisin}}, \ and\
  \bibinfo {author} {\bibfnamefont {L.}~\bibnamefont {Masanes}},\ }\href@noop
  {} {\bibfield  {journal} {\bibinfo  {journal} {Phys. Rev. Lett.}\ }\textbf
  {\bibinfo {volume} {97}},\ \bibinfo {pages} {120405} (\bibinfo {year}
  {2006})}\BibitemShut {NoStop}%
\bibitem [{\citenamefont {Hillery}(2000)}]{hillery2000quantum}%
  \BibitemOpen
  \bibfield  {author} {\bibinfo {author} {\bibfnamefont {M.}~\bibnamefont
  {Hillery}},\ }\href@noop {} {\bibfield  {journal} {\bibinfo  {journal} {Phys.
  Rev. A}\ }\textbf {\bibinfo {volume} {61}},\ \bibinfo {pages} {022309}
  (\bibinfo {year} {2000})}\BibitemShut {NoStop}%
\bibitem [{\citenamefont {Yuan}\ \emph {et~al.}(2002)\citenamefont {Yuan},
  \citenamefont {Kardynal}, \citenamefont {Stevenson}, \citenamefont {Shields},
  \citenamefont {Lobo}, \citenamefont {Cooper}, \citenamefont {Beattie},
  \citenamefont {Ritchie},\ and\ \citenamefont
  {Pepper}}]{yuan2002electrically}%
  \BibitemOpen
  \bibfield  {author} {\bibinfo {author} {\bibfnamefont {Z.}~\bibnamefont
  {Yuan}}, \bibinfo {author} {\bibfnamefont {B.~E.}\ \bibnamefont {Kardynal}},
  \bibinfo {author} {\bibfnamefont {R.~M.}\ \bibnamefont {Stevenson}}, \bibinfo
  {author} {\bibfnamefont {A.~J.}\ \bibnamefont {Shields}}, \bibinfo {author}
  {\bibfnamefont {C.~J.}\ \bibnamefont {Lobo}}, \bibinfo {author}
  {\bibfnamefont {K.}~\bibnamefont {Cooper}}, \bibinfo {author} {\bibfnamefont
  {N.~S.}\ \bibnamefont {Beattie}}, \bibinfo {author} {\bibfnamefont {D.~A.}\
  \bibnamefont {Ritchie}}, \ and\ \bibinfo {author} {\bibfnamefont
  {M.}~\bibnamefont {Pepper}},\ }\href@noop {} {\bibfield  {journal} {\bibinfo
  {journal} {Science}\ }\textbf {\bibinfo {volume} {295}},\ \bibinfo {pages}
  {102} (\bibinfo {year} {2002})}\BibitemShut {NoStop}%
\bibitem [{\citenamefont {Pathak}\ and\ \citenamefont
  {Verma}(2010)}]{pathak2010recent}%
  \BibitemOpen
  \bibfield  {author} {\bibinfo {author} {\bibfnamefont {A.}~\bibnamefont
  {Pathak}}\ and\ \bibinfo {author} {\bibfnamefont {A.}~\bibnamefont {Verma}},\
  }\href@noop {} {\bibfield  {journal} {\bibinfo  {journal} {Ind. J. Phys.}\
  }\textbf {\bibinfo {volume} {84}},\ \bibinfo {pages} {1005} (\bibinfo {year}
  {2010})}\BibitemShut {NoStop}%
\bibitem [{\citenamefont {Sen}\ and\ \citenamefont
  {Mandal}(2007)}]{sen2007squeezing}%
  \BibitemOpen
  \bibfield  {author} {\bibinfo {author} {\bibfnamefont {B.}~\bibnamefont
  {Sen}}\ and\ \bibinfo {author} {\bibfnamefont {S.}~\bibnamefont {Mandal}},\
  }\href@noop {} {\bibfield  {journal} {\bibinfo  {journal} {J. Phys. B}\
  }\textbf {\bibinfo {volume} {40}},\ \bibinfo {pages} {2901} (\bibinfo {year}
  {2007})}\BibitemShut {NoStop}%
\bibitem [{\citenamefont {Sen}\ \emph {et~al.}(2011)\citenamefont {Sen},
  \citenamefont {Pe{\v{r}}inov{\'a}}, \citenamefont {Luk{\v{s}}}, \citenamefont
  {Pe{\v{r}}ina},\ and\ \citenamefont {K{\v{r}}epelka}}]{sen2011sub}%
  \BibitemOpen
  \bibfield  {author} {\bibinfo {author} {\bibfnamefont {B.}~\bibnamefont
  {Sen}}, \bibinfo {author} {\bibfnamefont {V.}~\bibnamefont
  {Pe{\v{r}}inov{\'a}}}, \bibinfo {author} {\bibfnamefont {A.}~\bibnamefont
  {Luk{\v{s}}}}, \bibinfo {author} {\bibfnamefont {J.}~\bibnamefont
  {Pe{\v{r}}ina}}, \ and\ \bibinfo {author} {\bibfnamefont {J.}~\bibnamefont
  {K{\v{r}}epelka}},\ }\href@noop {} {\bibfield  {journal} {\bibinfo  {journal}
  {J. Phys. B}\ }\textbf {\bibinfo {volume} {44}},\ \bibinfo {pages} {105503}
  (\bibinfo {year} {2011})}\BibitemShut {NoStop}%
\bibitem [{\citenamefont {Sen}\ \emph {et~al.}(2007)\citenamefont {Sen},
  \citenamefont {Mandal},\ and\ \citenamefont {Pe{\v{r}}ina}}]{sen2007quantum}%
  \BibitemOpen
  \bibfield  {author} {\bibinfo {author} {\bibfnamefont {B.}~\bibnamefont
  {Sen}}, \bibinfo {author} {\bibfnamefont {S.}~\bibnamefont {Mandal}}, \ and\
  \bibinfo {author} {\bibfnamefont {J.}~\bibnamefont {Pe{\v{r}}ina}},\
  }\href@noop {} {\bibfield  {journal} {\bibinfo  {journal} {J. Phys. B}\
  }\textbf {\bibinfo {volume} {40}},\ \bibinfo {pages} {1417} (\bibinfo {year}
  {2007})}\BibitemShut {NoStop}%
\bibitem [{\citenamefont {Sen}\ and\ \citenamefont
  {Mandal}(2005)}]{sen2005squeezed}%
  \BibitemOpen
  \bibfield  {author} {\bibinfo {author} {\bibfnamefont {B.}~\bibnamefont
  {Sen}}\ and\ \bibinfo {author} {\bibfnamefont {S.}~\bibnamefont {Mandal}},\
  }\href@noop {} {\bibfield  {journal} {\bibinfo  {journal} {J. Mod. Opt.}\
  }\textbf {\bibinfo {volume} {52}},\ \bibinfo {pages} {1789} (\bibinfo {year}
  {2005})}\BibitemShut {NoStop}%
\bibitem [{\citenamefont {Sen}\ and\ \citenamefont
  {Mandal}(2008)}]{sen2008amplitude}%
  \BibitemOpen
  \bibfield  {author} {\bibinfo {author} {\bibfnamefont {B.}~\bibnamefont
  {Sen}}\ and\ \bibinfo {author} {\bibfnamefont {S.}~\bibnamefont {Mandal}},\
  }\href@noop {} {\bibfield  {journal} {\bibinfo  {journal} {J. Mod. Opt.}\
  }\textbf {\bibinfo {volume} {55}},\ \bibinfo {pages} {1697} (\bibinfo {year}
  {2008})}\BibitemShut {NoStop}%
\bibitem [{\citenamefont {Pe{\v{r}}ina}(1991)}]{perina1991quantum}%
  \BibitemOpen
  \bibfield  {author} {\bibinfo {author} {\bibfnamefont {J.}~\bibnamefont
  {Pe{\v{r}}ina}},\ }\href@noop {} {\emph {\bibinfo {title} {Quantum Statistics
  of Linear and Nonlinear Optical Phenomena}}}\ (\bibinfo  {publisher} {Kluwer
  Academic, Dordrecht-Boston},\ \bibinfo {year} {1991})\BibitemShut {NoStop}%
\bibitem [{\citenamefont {Miranowicz}\ and\ \citenamefont
  {Kielich}(1994)}]{miranowicz1994quantum}%
  \BibitemOpen
  \bibfield  {author} {\bibinfo {author} {\bibfnamefont {A.}~\bibnamefont
  {Miranowicz}}\ and\ \bibinfo {author} {\bibfnamefont {S.}~\bibnamefont
  {Kielich}},\ }\enquote {\bibinfo {title} {Quantum-statistical theory of
  {R}aman scattering processes},}\ in\ \href@noop {} {\emph {\bibinfo
  {booktitle} {Modern Nonlinear Optics}}},\ Vol.~\bibinfo {volume} {3}\
  (\bibinfo  {publisher} {John Wiley \& Sons, New York},\ \bibinfo {year}
  {1994})\ pp.\ \bibinfo {pages} {531--626}\BibitemShut {NoStop}%
\bibitem [{\citenamefont {Pe{\v{r}}inov{\'a}}\ \emph
  {et~al.}(1979{\natexlab{a}})\citenamefont {Pe{\v{r}}inov{\'a}}, \citenamefont
  {Pe{\v{r}}ina}, \citenamefont {Szlachetka},\ and\ \citenamefont
  {Kielich}}]{perinova1979quantum2}%
  \BibitemOpen
  \bibfield  {author} {\bibinfo {author} {\bibfnamefont {V.}~\bibnamefont
  {Pe{\v{r}}inov{\'a}}}, \bibinfo {author} {\bibfnamefont {J.}~\bibnamefont
  {Pe{\v{r}}ina}}, \bibinfo {author} {\bibfnamefont {P.}~\bibnamefont
  {Szlachetka}}, \ and\ \bibinfo {author} {\bibfnamefont {S.}~\bibnamefont
  {Kielich}},\ }\href@noop {} {\bibfield  {journal} {\bibinfo  {journal} {Acta
  Phys. Polonica A}\ }\textbf {\bibinfo {volume} {56}},\ \bibinfo {pages} {275}
  (\bibinfo {year} {1979}{\natexlab{a}})}\BibitemShut {NoStop}%
\bibitem [{\citenamefont {Szlachetka}\ \emph {et~al.}(1980)\citenamefont
  {Szlachetka}, \citenamefont {Kielich}, \citenamefont {Pe{\v{r}}ina},\ and\
  \citenamefont {Pe{\v{r}}inov{\'a}}}]{szlachetka1980photon}%
  \BibitemOpen
  \bibfield  {author} {\bibinfo {author} {\bibfnamefont {P.}~\bibnamefont
  {Szlachetka}}, \bibinfo {author} {\bibfnamefont {S.}~\bibnamefont {Kielich}},
  \bibinfo {author} {\bibfnamefont {J.}~\bibnamefont {Pe{\v{r}}ina}}, \ and\
  \bibinfo {author} {\bibfnamefont {V.}~\bibnamefont {Pe{\v{r}}inov{\'a}}},\
  }\href@noop {} {\bibfield  {journal} {\bibinfo  {journal} {J. Mod. Opt.}\
  }\textbf {\bibinfo {volume} {27}},\ \bibinfo {pages} {1609} (\bibinfo {year}
  {1980})}\BibitemShut {NoStop}%
\bibitem [{\citenamefont {Pe{\v{r}}inov{\'a}}\ and\ \citenamefont
  {Tiebel}(1984)}]{perinova1984sub}%
  \BibitemOpen
  \bibfield  {author} {\bibinfo {author} {\bibfnamefont {V.}~\bibnamefont
  {Pe{\v{r}}inov{\'a}}}\ and\ \bibinfo {author} {\bibfnamefont
  {R.}~\bibnamefont {Tiebel}},\ }\href@noop {} {\bibfield  {journal} {\bibinfo
  {journal} {Opt. Comm.}\ }\textbf {\bibinfo {volume} {50}},\ \bibinfo {pages}
  {401} (\bibinfo {year} {1984})}\BibitemShut {NoStop}%
\bibitem [{\citenamefont {Oliv{\'\i}k}\ and\ \citenamefont
  {Pe{\v{r}}ina}(1995)}]{olivik1995non}%
  \BibitemOpen
  \bibfield  {author} {\bibinfo {author} {\bibfnamefont {M.}~\bibnamefont
  {Oliv{\'\i}k}}\ and\ \bibinfo {author} {\bibfnamefont {J.}~\bibnamefont
  {Pe{\v{r}}ina}},\ }\href@noop {} {\bibfield  {journal} {\bibinfo  {journal}
  {J. Mod. Opt.}\ }\textbf {\bibinfo {volume} {42}},\ \bibinfo {pages} {197}
  (\bibinfo {year} {1995})}\BibitemShut {NoStop}%
\bibitem [{\citenamefont {Pe{\v{r}}inov{\'a}}\ \emph
  {et~al.}(1979{\natexlab{b}})\citenamefont {Pe{\v{r}}inov{\'a}}, \citenamefont
  {Pe{\v{r}}ina}, \citenamefont {Szlachetka},\ and\ \citenamefont
  {Kielich}}]{perinova1979quantum}%
  \BibitemOpen
  \bibfield  {author} {\bibinfo {author} {\bibfnamefont {V.}~\bibnamefont
  {Pe{\v{r}}inov{\'a}}}, \bibinfo {author} {\bibfnamefont {J.}~\bibnamefont
  {Pe{\v{r}}ina}}, \bibinfo {author} {\bibfnamefont {P.}~\bibnamefont
  {Szlachetka}}, \ and\ \bibinfo {author} {\bibfnamefont {S.}~\bibnamefont
  {Kielich}},\ }\href@noop {} {\bibfield  {journal} {\bibinfo  {journal} {Acta
  Phys. Polonica A}\ }\textbf {\bibinfo {volume} {56}},\ \bibinfo {pages} {267}
  (\bibinfo {year} {1979}{\natexlab{b}})}\BibitemShut {NoStop}%
\bibitem [{\citenamefont {Grangier}(2005)}]{grangier2005quantum}%
  \BibitemOpen
  \bibfield  {author} {\bibinfo {author} {\bibfnamefont {P.}~\bibnamefont
  {Grangier}},\ }\href@noop {} {\bibfield  {journal} {\bibinfo  {journal}
  {Nature}\ }\textbf {\bibinfo {volume} {438}},\ \bibinfo {pages} {749}
  (\bibinfo {year} {2005})}\BibitemShut {NoStop}%
\bibitem [{\citenamefont {Duan}\ \emph {et~al.}(2001)\citenamefont {Duan},
  \citenamefont {Lukin}, \citenamefont {Cirac},\ and\ \citenamefont
  {Zoller}}]{duan2001long}%
  \BibitemOpen
  \bibfield  {author} {\bibinfo {author} {\bibfnamefont {L.-M.}\ \bibnamefont
  {Duan}}, \bibinfo {author} {\bibfnamefont {M.~D.}\ \bibnamefont {Lukin}},
  \bibinfo {author} {\bibfnamefont {J.~I.}\ \bibnamefont {Cirac}}, \ and\
  \bibinfo {author} {\bibfnamefont {P.}~\bibnamefont {Zoller}},\ }\href@noop {}
  {\bibfield  {journal} {\bibinfo  {journal} {Nature}\ }\textbf {\bibinfo
  {volume} {414}},\ \bibinfo {pages} {413} (\bibinfo {year}
  {2001})}\BibitemShut {NoStop}%
\bibitem [{\citenamefont {Rand}(2013)}]{rand2013raman}%
  \BibitemOpen
  \bibfield  {author} {\bibinfo {author} {\bibfnamefont {S.~C.}\ \bibnamefont
  {Rand}},\ }\href@noop {} {\bibfield  {journal} {\bibinfo  {journal} {J.
  Lumin.}\ }\textbf {\bibinfo {volume} {133}},\ \bibinfo {pages} {10} (\bibinfo
  {year} {2013})}\BibitemShut {NoStop}%
\bibitem [{\citenamefont {Freudiger}\ \emph {et~al.}(2008)\citenamefont
  {Freudiger}, \citenamefont {Min}, \citenamefont {Saar}, \citenamefont {Lu},
  \citenamefont {Holtom}, \citenamefont {He}, \citenamefont {Tsai},
  \citenamefont {Kang},\ and\ \citenamefont {Xie}}]{freudiger2008label}%
  \BibitemOpen
  \bibfield  {author} {\bibinfo {author} {\bibfnamefont {C.~W.}\ \bibnamefont
  {Freudiger}}, \bibinfo {author} {\bibfnamefont {W.}~\bibnamefont {Min}},
  \bibinfo {author} {\bibfnamefont {B.~G.}\ \bibnamefont {Saar}}, \bibinfo
  {author} {\bibfnamefont {S.}~\bibnamefont {Lu}}, \bibinfo {author}
  {\bibfnamefont {G.~R.}\ \bibnamefont {Holtom}}, \bibinfo {author}
  {\bibfnamefont {C.}~\bibnamefont {He}}, \bibinfo {author} {\bibfnamefont
  {J.~C.}\ \bibnamefont {Tsai}}, \bibinfo {author} {\bibfnamefont {J.~X.}\
  \bibnamefont {Kang}}, \ and\ \bibinfo {author} {\bibfnamefont {X.~S.}\
  \bibnamefont {Xie}},\ }\href@noop {} {\bibfield  {journal} {\bibinfo
  {journal} {Science}\ }\textbf {\bibinfo {volume} {322}},\ \bibinfo {pages}
  {1857} (\bibinfo {year} {2008})}\BibitemShut {NoStop}%
\bibitem [{\citenamefont {Sadler}\ \emph {et~al.}(2007)\citenamefont {Sadler},
  \citenamefont {Higbie}, \citenamefont {Leslie}, \citenamefont
  {Vengalattore},\ and\ \citenamefont {Stamper-Kurn}}]{sadler2007coherence}%
  \BibitemOpen
  \bibfield  {author} {\bibinfo {author} {\bibfnamefont {L.~E.}\ \bibnamefont
  {Sadler}}, \bibinfo {author} {\bibfnamefont {J.~M.}\ \bibnamefont {Higbie}},
  \bibinfo {author} {\bibfnamefont {S.~R.}\ \bibnamefont {Leslie}}, \bibinfo
  {author} {\bibfnamefont {M.}~\bibnamefont {Vengalattore}}, \ and\ \bibinfo
  {author} {\bibfnamefont {D.~M.}\ \bibnamefont {Stamper-Kurn}},\ }\href@noop
  {} {\bibfield  {journal} {\bibinfo  {journal} {Phys. Rev. Lett.}\ }\textbf
  {\bibinfo {volume} {98}},\ \bibinfo {pages} {110401} (\bibinfo {year}
  {2007})}\BibitemShut {NoStop}%
\bibitem [{\citenamefont {Bustard}\ \emph {et~al.}(2011)\citenamefont
  {Bustard}, \citenamefont {Moffatt}, \citenamefont {Lausten}, \citenamefont
  {Wu}, \citenamefont {Walmsley},\ and\ \citenamefont
  {Sussman}}]{bustard2011quantum}%
  \BibitemOpen
  \bibfield  {author} {\bibinfo {author} {\bibfnamefont {P.~J.}\ \bibnamefont
  {Bustard}}, \bibinfo {author} {\bibfnamefont {D.}~\bibnamefont {Moffatt}},
  \bibinfo {author} {\bibfnamefont {R.}~\bibnamefont {Lausten}}, \bibinfo
  {author} {\bibfnamefont {G.}~\bibnamefont {Wu}}, \bibinfo {author}
  {\bibfnamefont {I.~A.}\ \bibnamefont {Walmsley}}, \ and\ \bibinfo {author}
  {\bibfnamefont {B.~J.}\ \bibnamefont {Sussman}},\ }\href@noop {} {\bibfield
  {journal} {\bibinfo  {journal} {Opt. Exp.}\ }\textbf {\bibinfo {volume}
  {19}},\ \bibinfo {pages} {25173} (\bibinfo {year} {2011})}\BibitemShut
  {NoStop}%
\bibitem [{\citenamefont {Kielich}(1993)}]{kielich1993multi}%
  \BibitemOpen
  \bibfield  {author} {\bibinfo {author} {\bibfnamefont {S.}~\bibnamefont
  {Kielich}},\ }\enquote {\bibinfo {title} {Multi-photon scattering molecular
  spectroscopy},}\ in\ \href@noop {} {\emph {\bibinfo {booktitle} {Progress in
  Optics}}},\ Vol.~\bibinfo {volume} {20},\ \bibinfo {editor} {edited by\
  \bibinfo {editor} {\bibfnamefont {E.}~\bibnamefont {Wolf}}}\ (\bibinfo
  {publisher} {Elsevier, Amsterdam},\ \bibinfo {year} {1993})\ pp.\ \bibinfo
  {pages} {155--155}\BibitemShut {NoStop}%
\bibitem [{\citenamefont {Ziegler}(1990)}]{ziegler1990hyper}%
  \BibitemOpen
  \bibfield  {author} {\bibinfo {author} {\bibfnamefont {L.~D.}\ \bibnamefont
  {Ziegler}},\ }\href@noop {} {\bibfield  {journal} {\bibinfo  {journal} {J.
  Raman Spec.}\ }\textbf {\bibinfo {volume} {21}},\ \bibinfo {pages} {769}
  (\bibinfo {year} {1990})}\BibitemShut {NoStop}%
\bibitem [{\citenamefont {French}\ and\ \citenamefont
  {Long}(1975)}]{french1975versatile}%
  \BibitemOpen
  \bibfield  {author} {\bibinfo {author} {\bibfnamefont {M.~J.}\ \bibnamefont
  {French}}\ and\ \bibinfo {author} {\bibfnamefont {D.~A.}\ \bibnamefont
  {Long}},\ }\href@noop {} {\bibfield  {journal} {\bibinfo  {journal} {J. Raman
  Spec.}\ }\textbf {\bibinfo {volume} {3}},\ \bibinfo {pages} {391} (\bibinfo
  {year} {1975})}\BibitemShut {NoStop}%
\bibitem [{\citenamefont {Valley}\ \emph {et~al.}(2010)\citenamefont {Valley},
  \citenamefont {Jensen}, \citenamefont {Autschbach},\ and\ \citenamefont
  {Schatz}}]{valley2010theoretical}%
  \BibitemOpen
  \bibfield  {author} {\bibinfo {author} {\bibfnamefont {N.}~\bibnamefont
  {Valley}}, \bibinfo {author} {\bibfnamefont {L.}~\bibnamefont {Jensen}},
  \bibinfo {author} {\bibfnamefont {J.}~\bibnamefont {Autschbach}}, \ and\
  \bibinfo {author} {\bibfnamefont {G.~C.}\ \bibnamefont {Schatz}},\
  }\href@noop {} {\bibfield  {journal} {\bibinfo  {journal} {J. Chem. Phys.}\
  }\textbf {\bibinfo {volume} {133}},\ \bibinfo {pages} {054103} (\bibinfo
  {year} {2010})}\BibitemShut {NoStop}%
\bibitem [{\citenamefont {Butet}\ and\ \citenamefont
  {Martin}(2015)}]{butet2015surface}%
  \BibitemOpen
  \bibfield  {author} {\bibinfo {author} {\bibfnamefont {J.}~\bibnamefont
  {Butet}}\ and\ \bibinfo {author} {\bibfnamefont {O.~J.~F.}\ \bibnamefont
  {Martin}},\ }\href@noop {} {\bibfield  {journal} {\bibinfo  {journal} {J.
  Phys. Chem. C}\ }\textbf {\bibinfo {volume} {119}},\ \bibinfo {pages} {15547}
  (\bibinfo {year} {2015})}\BibitemShut {NoStop}%
\bibitem [{\citenamefont {Kneipp}\ \emph {et~al.}(2007)\citenamefont {Kneipp},
  \citenamefont {Kneipp}, \citenamefont {Wittig},\ and\ \citenamefont
  {Kneipp}}]{kneipp2007one}%
  \BibitemOpen
  \bibfield  {author} {\bibinfo {author} {\bibfnamefont {J.}~\bibnamefont
  {Kneipp}}, \bibinfo {author} {\bibfnamefont {H.}~\bibnamefont {Kneipp}},
  \bibinfo {author} {\bibfnamefont {B.}~\bibnamefont {Wittig}}, \ and\ \bibinfo
  {author} {\bibfnamefont {K.}~\bibnamefont {Kneipp}},\ }\href@noop {}
  {\bibfield  {journal} {\bibinfo  {journal} {Nano Lett.}\ }\textbf {\bibinfo
  {volume} {7}},\ \bibinfo {pages} {2819} (\bibinfo {year} {2007})}\BibitemShut
  {NoStop}%
\bibitem [{\citenamefont {Kockum}\ \emph {et~al.}(2017)\citenamefont {Kockum},
  \citenamefont {Miranowicz}, \citenamefont {Macr\`{\i}}, \citenamefont
  {Savasta},\ and\ \citenamefont {Nori}}]{kockum2017deterministic}%
  \BibitemOpen
  \bibfield  {author} {\bibinfo {author} {\bibfnamefont {A.~F.}\ \bibnamefont
  {Kockum}}, \bibinfo {author} {\bibfnamefont {A.}~\bibnamefont {Miranowicz}},
  \bibinfo {author} {\bibfnamefont {V.}~\bibnamefont {Macr\`{\i}}}, \bibinfo
  {author} {\bibfnamefont {S.}~\bibnamefont {Savasta}}, \ and\ \bibinfo
  {author} {\bibfnamefont {F.}~\bibnamefont {Nori}},\ }\href@noop {} {\bibfield
   {journal} {\bibinfo  {journal} {Phys. Rev. A}\ }\textbf {\bibinfo {volume}
  {95}},\ \bibinfo {pages} {063849} (\bibinfo {year} {2017})}\BibitemShut
  {NoStop}%
\bibitem [{\citenamefont {Kozich}\ and\ \citenamefont
  {Werncke}(2007)}]{kozich2007non}%
  \BibitemOpen
  \bibfield  {author} {\bibinfo {author} {\bibfnamefont {V.}~\bibnamefont
  {Kozich}}\ and\ \bibinfo {author} {\bibfnamefont {W.}~\bibnamefont
  {Werncke}},\ }\href@noop {} {\bibfield  {journal} {\bibinfo  {journal} {J.
  Raman Spec.}\ }\textbf {\bibinfo {volume} {38}},\ \bibinfo {pages} {1180}
  (\bibinfo {year} {2007})}\BibitemShut {NoStop}%
\bibitem [{\citenamefont {Madzharova}\ \emph {et~al.}(2017)\citenamefont
  {Madzharova}, \citenamefont {Heiner},\ and\ \citenamefont
  {Kneipp}}]{madzharova2017surface}%
  \BibitemOpen
  \bibfield  {author} {\bibinfo {author} {\bibfnamefont {F.}~\bibnamefont
  {Madzharova}}, \bibinfo {author} {\bibfnamefont {Z.}~\bibnamefont {Heiner}},
  \ and\ \bibinfo {author} {\bibfnamefont {J.}~\bibnamefont {Kneipp}},\
  }\href@noop {} {\bibfield  {journal} {\bibinfo  {journal} {Chem. Soc. Rev.}\
  } (\bibinfo {year} {2017})}\BibitemShut {NoStop}%
\bibitem [{\citenamefont {Thapliyal}\ \emph
  {et~al.}(2014{\natexlab{a}})\citenamefont {Thapliyal}, \citenamefont
  {Pathak}, \citenamefont {Sen},\ and\ \citenamefont
  {Pe{\v{r}}ina}}]{thapliyal2014higher}%
  \BibitemOpen
  \bibfield  {author} {\bibinfo {author} {\bibfnamefont {K.}~\bibnamefont
  {Thapliyal}}, \bibinfo {author} {\bibfnamefont {A.}~\bibnamefont {Pathak}},
  \bibinfo {author} {\bibfnamefont {B.}~\bibnamefont {Sen}}, \ and\ \bibinfo
  {author} {\bibfnamefont {J.}~\bibnamefont {Pe{\v{r}}ina}},\ }\href@noop {}
  {\bibfield  {journal} {\bibinfo  {journal} {Phys. Rev. A}\ }\textbf {\bibinfo
  {volume} {90}},\ \bibinfo {pages} {013808} (\bibinfo {year}
  {2014}{\natexlab{a}})}\BibitemShut {NoStop}%
\bibitem [{\citenamefont {Thapliyal}\ \emph
  {et~al.}(2014{\natexlab{b}})\citenamefont {Thapliyal}, \citenamefont
  {Pathak}, \citenamefont {Sen},\ and\ \citenamefont
  {Pe{\v{r}}ina}}]{thapliyal2014nonclassical}%
  \BibitemOpen
  \bibfield  {author} {\bibinfo {author} {\bibfnamefont {K.}~\bibnamefont
  {Thapliyal}}, \bibinfo {author} {\bibfnamefont {A.}~\bibnamefont {Pathak}},
  \bibinfo {author} {\bibfnamefont {B.}~\bibnamefont {Sen}}, \ and\ \bibinfo
  {author} {\bibfnamefont {J.}~\bibnamefont {Pe{\v{r}}ina}},\ }\href@noop {}
  {\bibfield  {journal} {\bibinfo  {journal} {Phys. Lett. A}\ }\textbf
  {\bibinfo {volume} {378}},\ \bibinfo {pages} {3431} (\bibinfo {year}
  {2014}{\natexlab{b}})}\BibitemShut {NoStop}%
\bibitem [{\citenamefont {Thapliyal}\ \emph {et~al.}(2016)\citenamefont
  {Thapliyal}, \citenamefont {Pathak},\ and\ \citenamefont
  {Pe{\v{r}}ina}}]{thapliyal2016linear}%
  \BibitemOpen
  \bibfield  {author} {\bibinfo {author} {\bibfnamefont {K.}~\bibnamefont
  {Thapliyal}}, \bibinfo {author} {\bibfnamefont {A.}~\bibnamefont {Pathak}}, \
  and\ \bibinfo {author} {\bibfnamefont {J.}~\bibnamefont {Pe{\v{r}}ina}},\
  }\href@noop {} {\bibfield  {journal} {\bibinfo  {journal} {Phys. Rev. A}\
  }\textbf {\bibinfo {volume} {93}},\ \bibinfo {pages} {022107} (\bibinfo
  {year} {2016})}\BibitemShut {NoStop}%
\bibitem [{\citenamefont {Szlachetka}\ \emph {et~al.}(1979)\citenamefont
  {Szlachetka}, \citenamefont {Kielich}, \citenamefont {Pe{\v{r}}ina},\ and\
  \citenamefont {Pe{\v{r}}inov{\'a}}}]{szlachetka1979dynamics}%
  \BibitemOpen
  \bibfield  {author} {\bibinfo {author} {\bibfnamefont {P.}~\bibnamefont
  {Szlachetka}}, \bibinfo {author} {\bibfnamefont {S.}~\bibnamefont {Kielich}},
  \bibinfo {author} {\bibfnamefont {J.}~\bibnamefont {Pe{\v{r}}ina}}, \ and\
  \bibinfo {author} {\bibfnamefont {V.}~\bibnamefont {Pe{\v{r}}inov{\'a}}},\
  }\href@noop {} {\bibfield  {journal} {\bibinfo  {journal} {J. Phys. A}\
  }\textbf {\bibinfo {volume} {12}},\ \bibinfo {pages} {1921} (\bibinfo {year}
  {1979})}\BibitemShut {NoStop}%
\bibitem [{\citenamefont {Miranowicz}\ \emph {et~al.}(2010)\citenamefont
  {Miranowicz}, \citenamefont {Bartkowiak}, \citenamefont {Wang}, \citenamefont
  {Liu},\ and\ \citenamefont {Nori}}]{miranowicz2010testing}%
  \BibitemOpen
  \bibfield  {author} {\bibinfo {author} {\bibfnamefont {A.}~\bibnamefont
  {Miranowicz}}, \bibinfo {author} {\bibfnamefont {M.}~\bibnamefont
  {Bartkowiak}}, \bibinfo {author} {\bibfnamefont {X.}~\bibnamefont {Wang}},
  \bibinfo {author} {\bibfnamefont {Y.-x.}\ \bibnamefont {Liu}}, \ and\
  \bibinfo {author} {\bibfnamefont {F.}~\bibnamefont {Nori}},\ }\href@noop {}
  {\bibfield  {journal} {\bibinfo  {journal} {Phys. Rev. A}\ }\textbf {\bibinfo
  {volume} {82}},\ \bibinfo {pages} {013824} (\bibinfo {year}
  {2010})}\BibitemShut {NoStop}%
\bibitem [{\citenamefont {Pe{\v{r}}ina}\ \emph {et~al.}(1984)\citenamefont
  {Pe{\v{r}}ina}, \citenamefont {Pe{\v{r}}inov{\'a}},\ and\ \citenamefont
  {Ko{\v{d}}ousek}}]{perina1984relations}%
  \BibitemOpen
  \bibfield  {author} {\bibinfo {author} {\bibfnamefont {J.}~\bibnamefont
  {Pe{\v{r}}ina}}, \bibinfo {author} {\bibfnamefont {V.}~\bibnamefont
  {Pe{\v{r}}inov{\'a}}}, \ and\ \bibinfo {author} {\bibfnamefont
  {J.}~\bibnamefont {Ko{\v{d}}ousek}},\ }\href@noop {} {\bibfield  {journal}
  {\bibinfo  {journal} {Opt. Comm.}\ }\textbf {\bibinfo {volume} {49}},\
  \bibinfo {pages} {210} (\bibinfo {year} {1984})}\BibitemShut {NoStop}%
\bibitem [{\citenamefont {Sen}\ \emph {et~al.}(2013)\citenamefont {Sen},
  \citenamefont {Giri}, \citenamefont {Mandal}, \citenamefont {Ooi},\ and\
  \citenamefont {Pathak}}]{sen2013intermodal}%
  \BibitemOpen
  \bibfield  {author} {\bibinfo {author} {\bibfnamefont {B.}~\bibnamefont
  {Sen}}, \bibinfo {author} {\bibfnamefont {S.~K.}\ \bibnamefont {Giri}},
  \bibinfo {author} {\bibfnamefont {S.}~\bibnamefont {Mandal}}, \bibinfo
  {author} {\bibfnamefont {C.~H.~R.}\ \bibnamefont {Ooi}}, \ and\ \bibinfo
  {author} {\bibfnamefont {A.}~\bibnamefont {Pathak}},\ }\href@noop {}
  {\bibfield  {journal} {\bibinfo  {journal} {Phys. Rev. A}\ }\textbf {\bibinfo
  {volume} {87}},\ \bibinfo {pages} {022325} (\bibinfo {year}
  {2013})}\BibitemShut {NoStop}%
\bibitem [{\citenamefont {Giri}\ \emph {et~al.}(2016)\citenamefont {Giri},
  \citenamefont {Sen}, \citenamefont {Pathak},\ and\ \citenamefont
  {Jana}}]{giri2016higher}%
  \BibitemOpen
  \bibfield  {author} {\bibinfo {author} {\bibfnamefont {S.~K.}\ \bibnamefont
  {Giri}}, \bibinfo {author} {\bibfnamefont {B.}~\bibnamefont {Sen}}, \bibinfo
  {author} {\bibfnamefont {A.}~\bibnamefont {Pathak}}, \ and\ \bibinfo {author}
  {\bibfnamefont {P.~C.}\ \bibnamefont {Jana}},\ }\href@noop {} {\bibfield
  {journal} {\bibinfo  {journal} {Phys. Rev. A}\ }\textbf {\bibinfo {volume}
  {93}},\ \bibinfo {pages} {012340} (\bibinfo {year} {2016})}\BibitemShut
  {NoStop}%
\bibitem [{\citenamefont {Richter}\ and\ \citenamefont
  {Vogel}(2002)}]{richter2002nonclassicality}%
  \BibitemOpen
  \bibfield  {author} {\bibinfo {author} {\bibfnamefont {T.}~\bibnamefont
  {Richter}}\ and\ \bibinfo {author} {\bibfnamefont {W.}~\bibnamefont
  {Vogel}},\ }\href@noop {} {\bibfield  {journal} {\bibinfo  {journal} {Phys.
  Rev. Lett.}\ }\textbf {\bibinfo {volume} {89}},\ \bibinfo {pages} {283601}
  (\bibinfo {year} {2002})}\BibitemShut {NoStop}%
\bibitem [{\citenamefont {Allevi}\ \emph
  {et~al.}(2012{\natexlab{a}})\citenamefont {Allevi}, \citenamefont
  {Olivares},\ and\ \citenamefont {Bondani}}]{allevi2012measuring}%
  \BibitemOpen
  \bibfield  {author} {\bibinfo {author} {\bibfnamefont {A.}~\bibnamefont
  {Allevi}}, \bibinfo {author} {\bibfnamefont {S.}~\bibnamefont {Olivares}}, \
  and\ \bibinfo {author} {\bibfnamefont {M.}~\bibnamefont {Bondani}},\
  }\href@noop {} {\bibfield  {journal} {\bibinfo  {journal} {Phys. Rev. A}\
  }\textbf {\bibinfo {volume} {85}},\ \bibinfo {pages} {063835} (\bibinfo
  {year} {2012}{\natexlab{a}})}\BibitemShut {NoStop}%
\bibitem [{\citenamefont {Allevi}\ \emph
  {et~al.}(2012{\natexlab{b}})\citenamefont {Allevi}, \citenamefont
  {Olivares},\ and\ \citenamefont {Bondani}}]{allevi2012high}%
  \BibitemOpen
  \bibfield  {author} {\bibinfo {author} {\bibfnamefont {A.}~\bibnamefont
  {Allevi}}, \bibinfo {author} {\bibfnamefont {S.}~\bibnamefont {Olivares}}, \
  and\ \bibinfo {author} {\bibfnamefont {M.}~\bibnamefont {Bondani}},\
  }\href@noop {} {\bibfield  {journal} {\bibinfo  {journal} {Int. J. Quantum
  Inf.}\ }\textbf {\bibinfo {volume} {10}},\ \bibinfo {pages} {1241003}
  (\bibinfo {year} {2012}{\natexlab{b}})}\BibitemShut {NoStop}%
\bibitem [{\citenamefont {Avenhaus}\ \emph {et~al.}(2010)\citenamefont
  {Avenhaus}, \citenamefont {Laiho}, \citenamefont {Chekhova},\ and\
  \citenamefont {Silberhorn}}]{avenhaus2010accessing}%
  \BibitemOpen
  \bibfield  {author} {\bibinfo {author} {\bibfnamefont {M.}~\bibnamefont
  {Avenhaus}}, \bibinfo {author} {\bibfnamefont {K.}~\bibnamefont {Laiho}},
  \bibinfo {author} {\bibfnamefont {M.~V.}\ \bibnamefont {Chekhova}}, \ and\
  \bibinfo {author} {\bibfnamefont {C.}~\bibnamefont {Silberhorn}},\
  }\href@noop {} {\bibfield  {journal} {\bibinfo  {journal} {Phys. Rev. Lett.}\
  }\textbf {\bibinfo {volume} {104}},\ \bibinfo {pages} {063602} (\bibinfo
  {year} {2010})}\BibitemShut {NoStop}%
\bibitem [{\citenamefont {Hamar}\ \emph {et~al.}(2014)\citenamefont {Hamar},
  \citenamefont {Mich{\'a}lek},\ and\ \citenamefont {Pathak}}]{hamar2014non}%
  \BibitemOpen
  \bibfield  {author} {\bibinfo {author} {\bibfnamefont {M.}~\bibnamefont
  {Hamar}}, \bibinfo {author} {\bibfnamefont {V.}~\bibnamefont {Mich{\'a}lek}},
  \ and\ \bibinfo {author} {\bibfnamefont {A.}~\bibnamefont {Pathak}},\
  }\href@noop {} {\bibfield  {journal} {\bibinfo  {journal} {Meas. Sci. Rev.}\
  }\textbf {\bibinfo {volume} {14}},\ \bibinfo {pages} {227} (\bibinfo {year}
  {2014})}\BibitemShut {NoStop}%
\bibitem [{\citenamefont {Pe{\v{r}}ina~Jr.}\ \emph {et~al.}(2017)\citenamefont
  {Pe{\v{r}}ina~Jr.}, \citenamefont {Mich{\'a}lek},\ and\ \citenamefont
  {Haderka}}]{perina2017higher}%
  \BibitemOpen
  \bibfield  {author} {\bibinfo {author} {\bibfnamefont {J.}~\bibnamefont
  {Pe{\v{r}}ina~Jr.}}, \bibinfo {author} {\bibfnamefont {V.}~\bibnamefont
  {Mich{\'a}lek}}, \ and\ \bibinfo {author} {\bibfnamefont {O.}~\bibnamefont
  {Haderka}},\ }\href@noop {} {\bibfield  {journal} {\bibinfo  {journal}
  {Physical Review A}\ }\textbf {\bibinfo {volume} {96}},\ \bibinfo {pages}
  {033852} (\bibinfo {year} {2017})}\BibitemShut {NoStop}%
\bibitem [{\citenamefont {Loudon}\ and\ \citenamefont
  {Knight}(1987)}]{loudon1987squeezed}%
  \BibitemOpen
  \bibfield  {author} {\bibinfo {author} {\bibfnamefont {R.}~\bibnamefont
  {Loudon}}\ and\ \bibinfo {author} {\bibfnamefont {P.~L.}\ \bibnamefont
  {Knight}},\ }\href@noop {} {\bibfield  {journal} {\bibinfo  {journal} {J.
  Mod. Opt.}\ }\textbf {\bibinfo {volume} {34}},\ \bibinfo {pages} {709}
  (\bibinfo {year} {1987})}\BibitemShut {NoStop}%
\bibitem [{\citenamefont {Hong}\ and\ \citenamefont
  {Mandel}(1985{\natexlab{a}})}]{hong1985higher}%
  \BibitemOpen
  \bibfield  {author} {\bibinfo {author} {\bibfnamefont {C.~K.}\ \bibnamefont
  {Hong}}\ and\ \bibinfo {author} {\bibfnamefont {L.}~\bibnamefont {Mandel}},\
  }\href@noop {} {\bibfield  {journal} {\bibinfo  {journal} {Phys. Rev. Lett.}\
  }\textbf {\bibinfo {volume} {54}},\ \bibinfo {pages} {323} (\bibinfo {year}
  {1985}{\natexlab{a}})}\BibitemShut {NoStop}%
\bibitem [{\citenamefont {Hong}\ and\ \citenamefont
  {Mandel}(1985{\natexlab{b}})}]{hong1985generation}%
  \BibitemOpen
  \bibfield  {author} {\bibinfo {author} {\bibfnamefont {C.~K.}\ \bibnamefont
  {Hong}}\ and\ \bibinfo {author} {\bibfnamefont {L.}~\bibnamefont {Mandel}},\
  }\href@noop {} {\bibfield  {journal} {\bibinfo  {journal} {Phys. Rev. A}\
  }\textbf {\bibinfo {volume} {32}},\ \bibinfo {pages} {974} (\bibinfo {year}
  {1985}{\natexlab{b}})}\BibitemShut {NoStop}%
\bibitem [{\citenamefont {Hillery}(1987)}]{hillery1987amplitude}%
  \BibitemOpen
  \bibfield  {author} {\bibinfo {author} {\bibfnamefont {M.}~\bibnamefont
  {Hillery}},\ }\href@noop {} {\bibfield  {journal} {\bibinfo  {journal} {Phys.
  Rev. A}\ }\textbf {\bibinfo {volume} {36}},\ \bibinfo {pages} {3796}
  (\bibinfo {year} {1987})}\BibitemShut {NoStop}%
\bibitem [{\citenamefont {Lee}(1990)}]{lee1990higher}%
  \BibitemOpen
  \bibfield  {author} {\bibinfo {author} {\bibfnamefont {C.~T.}\ \bibnamefont
  {Lee}},\ }\href@noop {} {\bibfield  {journal} {\bibinfo  {journal} {Phys.
  Rev. A}\ }\textbf {\bibinfo {volume} {41}},\ \bibinfo {pages} {1721}
  (\bibinfo {year} {1990})}\BibitemShut {NoStop}%
\bibitem [{\citenamefont {Pathak}\ and\ \citenamefont
  {Garcia}(2006)}]{pathak2006control}%
  \BibitemOpen
  \bibfield  {author} {\bibinfo {author} {\bibfnamefont {A.}~\bibnamefont
  {Pathak}}\ and\ \bibinfo {author} {\bibfnamefont {M.~E.}\ \bibnamefont
  {Garcia}},\ }\href@noop {} {\bibfield  {journal} {\bibinfo  {journal} {Appl.
  Phys. B}\ }\textbf {\bibinfo {volume} {84}},\ \bibinfo {pages} {479}
  (\bibinfo {year} {2006})}\BibitemShut {NoStop}%
\bibitem [{\citenamefont {Hillery}\ and\ \citenamefont
  {Zubairy}(2006{\natexlab{a}})}]{hillery2006entanglement}%
  \BibitemOpen
  \bibfield  {author} {\bibinfo {author} {\bibfnamefont {M.}~\bibnamefont
  {Hillery}}\ and\ \bibinfo {author} {\bibfnamefont {M.~S.}\ \bibnamefont
  {Zubairy}},\ }\href@noop {} {\bibfield  {journal} {\bibinfo  {journal} {Phys.
  Rev. Lett.}\ }\textbf {\bibinfo {volume} {96}},\ \bibinfo {pages} {050503}
  (\bibinfo {year} {2006}{\natexlab{a}})}\BibitemShut {NoStop}%
\bibitem [{\citenamefont {Hillery}\ and\ \citenamefont
  {Zubairy}(2006{\natexlab{b}})}]{hillery2006entanglementapplications}%
  \BibitemOpen
  \bibfield  {author} {\bibinfo {author} {\bibfnamefont {M.}~\bibnamefont
  {Hillery}}\ and\ \bibinfo {author} {\bibfnamefont {M.~S.}\ \bibnamefont
  {Zubairy}},\ }\href@noop {} {\bibfield  {journal} {\bibinfo  {journal} {Phys.
  Rev. A}\ }\textbf {\bibinfo {volume} {74}},\ \bibinfo {pages} {032333}
  (\bibinfo {year} {2006}{\natexlab{b}})}\BibitemShut {NoStop}%
\bibitem [{\citenamefont {Giri}\ \emph {et~al.}(2017)\citenamefont {Giri},
  \citenamefont {Thapliyal}, \citenamefont {Sen},\ and\ \citenamefont
  {Pathak}}]{giri2017nonclassicality}%
  \BibitemOpen
  \bibfield  {author} {\bibinfo {author} {\bibfnamefont {S.~K.}\ \bibnamefont
  {Giri}}, \bibinfo {author} {\bibfnamefont {K.}~\bibnamefont {Thapliyal}},
  \bibinfo {author} {\bibfnamefont {B.}~\bibnamefont {Sen}}, \ and\ \bibinfo
  {author} {\bibfnamefont {A.}~\bibnamefont {Pathak}},\ }\href@noop {}
  {\bibfield  {journal} {\bibinfo  {journal} {Physica A}\ }\textbf {\bibinfo
  {volume} {466}},\ \bibinfo {pages} {140} (\bibinfo {year}
  {2017})}\BibitemShut {NoStop}%
\bibitem [{\citenamefont {Thapliyal}\ \emph {et~al.}(2015)\citenamefont
  {Thapliyal}, \citenamefont {Banerjee}, \citenamefont {Pathak}, \citenamefont
  {Omkar},\ and\ \citenamefont {Ravishankar}}]{thapliyal2015quasiprobability}%
  \BibitemOpen
  \bibfield  {author} {\bibinfo {author} {\bibfnamefont {K.}~\bibnamefont
  {Thapliyal}}, \bibinfo {author} {\bibfnamefont {S.}~\bibnamefont {Banerjee}},
  \bibinfo {author} {\bibfnamefont {A.}~\bibnamefont {Pathak}}, \bibinfo
  {author} {\bibfnamefont {S.}~\bibnamefont {Omkar}}, \ and\ \bibinfo {author}
  {\bibfnamefont {V.}~\bibnamefont {Ravishankar}},\ }\href@noop {} {\bibfield
  {journal} {\bibinfo  {journal} {Ann. Phys.}\ }\textbf {\bibinfo {volume}
  {362}},\ \bibinfo {pages} {261} (\bibinfo {year} {2015})}\BibitemShut
  {NoStop}%
\end{thebibliography}%

\appendix

\section*{Appendix A: Various terms in the obtained solution \label{sec:Appendix-A:exp}}

\setcounter{equation}{0} \renewcommand{\theequation}{A.\arabic{equation}}
The functional form of the coefficients in the evolution of various
field modes given in Eq. (\ref{eq:solution}) is as follows.

\begin{equation}
\begin{array}{lcl}
f_{1} & = & \exp\left(-i\omega_{j}t\right),\\
f_{2} & = & -\frac{g^{*}f_{1}}{\Delta\omega_{1}}\left[\exp\left(-i\Delta\omega_{1}t\right)-1\right],\\
f_{3} & = & \frac{\chi f_{1}}{\Delta\omega_{2}}\left[\exp\left(i\Delta\omega_{2}t\right)-1\right],\\
f_{4} & = & -f_{5}=-f_{6}\\
 & = & -\frac{|g|^{2}f_{1}}{\Delta\omega_{1}^{2}}\left[\exp\left(-i\Delta\omega_{1}t\right)-1\right]-\frac{i|g|^{2}tf_{1}}{\Delta\omega_{1}},\\
f_{7} & = & \frac{-\chi^{*}g^{*}f_{1}}{\Delta\omega_{2}}\left[\frac{\exp\left[-i\left(\Delta\omega_{1}+\Delta\omega_{2}\right)t\right]-1}{\Delta\omega_{1}+\Delta\omega_{2}}-\frac{\exp\left(-i\Delta\omega_{1}t\right)-1}{\Delta\omega_{1}}\right],\\
f_{8} & = & \frac{-\chi g^{*}f_{1}}{\Delta\omega_{2}}\left[\frac{\exp\left[-i\left(\Delta\omega_{1}-\Delta\omega_{2}\right)t\right]-1}{\Delta\omega_{1}-\Delta\omega_{2}}-\frac{\exp\left(-i\Delta\omega_{1}t\right)}{\Delta\omega_{1}}\right]\\
 & - & \frac{\chi g^{*}f_{1}}{\Delta\omega_{1}}\left[\frac{\exp\left[-i\left(\Delta\omega_{1}-\Delta\omega_{2}\right)t\right]-1}{\Delta\omega_{1}-\Delta\omega_{2}}+\frac{\exp\left(i\Delta\omega_{2}t\right)}{\Delta\omega_{2}}\right],\\
f_{9} & = & \frac{-\chi gf_{1}}{\Delta\omega_{1}}\left[\frac{\exp\left[i\left(\Delta\omega_{1}+\Delta\omega_{2}\right)t\right]-1}{\Delta\omega_{1}+\Delta\omega_{2}}-\frac{\exp\left(i\Delta\omega_{2}t\right)-1}{\Delta\omega_{2}}\right],\\
f_{10} & = & f_{11}=-f_{12}\\
 & = & \frac{-|\chi|^{2}f_{1}}{\Delta\omega_{2}^{2}}\left[\exp\left(i\Delta\omega_{2}t\right)-1\right]+\frac{i|\chi|^{2}tf_{1}}{\Delta\omega_{2}},
\end{array}\label{eq:solutions of f}
\end{equation}
\begin{equation}
\begin{array}{lcl}
g_{1} & = & \exp\left(-i\omega_{b}t\right),\\
g_{2} & = & \frac{gg_{1}}{\Delta\omega_{1}}\left[\exp\left(i\Delta\omega_{1}t\right)-1\right],\\
g_{3} & = & \frac{\chi^{*}gg_{1}}{\Delta\omega_{2}}\left[\frac{\exp\left[i\left(\Delta\omega_{1}-\Delta\omega_{2}\right)t\right]-1}{\Delta\omega_{1}-\Delta\omega_{2}}-\frac{\exp\left(i\Delta\omega_{1}t\right)-1}{\Delta\omega_{1}}\right],\\
g_{4} & = & -g_{5}=\frac{-|g|^{2}g_{1}}{\Delta\omega_{1}^{2}}\left[\exp\left(i\Delta\omega_{1}t\right)-1\right]+\frac{i|g|^{2}tg_{1}}{\Delta\omega_{1}},\\
g_{6} & = & \frac{\chi gg_{1}}{\Delta\omega_{2}}\left[\frac{\exp\left[i\left(\Delta\omega_{1}+\Delta\omega_{2}\right)t\right]-1}{\Delta\omega_{1}+\Delta\omega_{2}}-\frac{\exp\left(i\Delta\omega_{1}t\right)-1}{\Delta\omega_{1}}\right],
\end{array}\label{eq:solutions of g}
\end{equation}
\begin{equation}
\begin{array}{lcl}
h_{1} & = & \exp\left(-i\omega_{c}t\right),\\
h_{2} & = & \frac{gh_{1}}{\Delta\omega_{1}}\left[\exp\left(i\Delta\omega_{1}t\right)-1\right],\\
h_{3} & = & \frac{\chi h_{1}}{\Delta\omega_{2}}\left[\exp\left(i\Delta\omega_{2}t\right)-1\right],\\
h_{4} & = & -h_{5}=-\frac{|g|^{2}h_{1}}{\Delta\omega_{1}^{2}}\left[\exp\left(i\Delta\omega_{1}t\right)-1\right]+\frac{i|g|^{2}th_{1}}{\Delta\omega_{1}},\\
h_{6} & = & \frac{\chi gh_{1}}{\Delta\omega_{2}}\left[\frac{\exp\left[i\left(\Delta\omega_{1}+\Delta\omega_{2}\right)t\right]-1}{\Delta\omega_{1}+\Delta\omega_{2}}-\frac{\exp\left(i\Delta\omega_{1}t\right)}{\Delta\omega_{1}}\right]\\
 & - & \frac{\chi gh_{1}}{\Delta\omega_{1}}\left[\frac{\exp\left[i\left(\Delta\omega_{1}+\Delta\omega_{2}\right)t\right]-1}{\Delta\omega_{1}+\Delta\omega_{2}}-\frac{\exp\left(i\Delta\omega_{2}t\right)}{\Delta\omega_{2}}\right],\\
h_{7} & = & -h_{8}=-\frac{|\chi|^{2}h_{1}}{\Delta\omega_{2}^{2}}\left[\exp\left(i\Delta\omega_{2}t\right)-1\right]+\frac{i|\chi|^{2}th_{1}}{\Delta\omega_{2}},
\end{array}\label{eq:eq:solutions of h}
\end{equation}
\begin{equation}
\begin{array}{lcl}
l_{1} & = & \exp\left(-i\omega_{d}t\right),\\
l_{2} & = & -\frac{\chi^{*}l_{1}}{\Delta\omega_{2}}\left[\exp\left(-i\Delta\omega_{2}t\right)-1\right],\\
l_{3} & = & \frac{\chi^{*}gl_{1}}{\Delta\omega_{1}}\left[\frac{\exp\left[i\left(\Delta\omega_{1}-\Delta\omega_{2}\right)t\right]-1}{\Delta\omega_{1}-\Delta\omega_{2}}+\frac{\exp\left(-i\Delta\omega_{2}t\right)-1}{\Delta\omega_{2}}\right],\\
l_{4} & = & \frac{\chi^{*}g^{*}l_{1}}{\Delta\omega_{1}}\left[\frac{\exp\left[-i\left(\Delta\omega_{1}+\Delta\omega_{2}\right)t\right]-1}{\Delta\omega_{1}+\Delta\omega_{2}}-\frac{\exp\left(-i\Delta\omega_{2}t\right)-1}{\Delta\omega_{2}}\right],\\
l_{5} & = & l_{6}=\frac{|\chi|^{2}l_{1}}{\Delta\omega_{2}^{2}}\left[\exp\left(-i\Delta\omega_{2}t\right)-1\right]+\frac{i|\chi|^{2}tl_{1}}{\Delta\omega_{2}},
\end{array}\label{eq:eq:solutions of l}
\end{equation}
where $\Delta\omega_{1}=(\omega_{b}+\omega_{c}-\sum_{i=1}^{k}\omega_{i})$
and $\Delta\omega_{2}=(\sum_{i=1}^{k}\omega_{i}+\omega_{c}-\omega_{d})$
are detuning in Stokes and anti-Stokes generation processes. As differential
equations of all the pump modes are similar, here, we have explicitly written 
the solution for $j$th pump mode only.

\section*{Appendix B: Sen Mandal Solution of the process \label{sec:Appendix-B:solution}}

\setcounter{equation}{0} \renewcommand{\theequation}{B.\arabic{equation}}

\setcounter{table}{0} \renewcommand{\thetable}{B.\Roman{table}}

We know that the evolution of an operator in Heisenberg picture can be
given by
\begin{equation}
a_{j}(t)=\exp(iHt)a_{j}(0)\exp(-iHt)\label{eq:heisenberg operator equ of motion-1}
\end{equation}
which on expansion gives

\begin{equation}
\begin{array}{lcl}
a_{j}(t) & = & a_{j}(0)+it[H,a_{j}(0)]+(it)^{2}[H,[H,a_{j}(0)]]\\
 & + & (it)^{3}[H,[H,[H,a_{j}(0)]]]+\ldots,
\end{array}\label{eq:EoM-2}
\end{equation}
where (in Sen-Mandal approach) we calculate different commutators until we obtain new terms
as functions of annihilation or creation operators of different modes
involved in the process. For instance, to obtain the evolution of
an arbitrary pump mode ($a_{j}$) we obtain

\begin{widetext}

\begin{equation}
\begin{array}{lcl}
[H,a_{j}(0)] & = & \prod_{i=1:i\neq j}^{k}\left(-\omega_{j}a_{j}(0)+g^{*}a_{i}^{\dagger}(0)b(0)c(0)+\chi a_{i}^{\dagger}(0)c^{\dagger}(0)d(0)\right)\\
{}[H,[H,a_{j}(0)]] & = & \prod_{i=1:i\neq j}^{k}\left(\omega_{j}^{2}a_{j}(0)+g^{*}\left(-\omega_{j}+\omega_{i}-\omega_{b}-\omega_{c}\right)a_{i}^{\dagger}(0)b(0)c(0)\right.\\
 & + & \chi\left(-\omega_{j}+\omega_{i}+\omega_{c}-\omega_{d}\right)a_{i}^{\dagger}(0)c^{\dagger}(0)d(0)-|g|^{2}\left(a_{j}(0)A_{l}b^{\dagger}(0)b(0)c^{\dagger}(0)c(0)\right.\\
 & - & \left.a_{i}^{\dagger}(0)a_{i}(0)a_{j}(0)b(0)b^{\dagger}(0)-a_{i}^{\dagger}(0)a_{i}(0)a_{j}(0)c^{\dagger}(0)c(0)\right)\\
 & - & |\chi|^{2}\left(a_{j}(0)a_{i}^{\dagger}(0)a_{i}(0)d^{\dagger}(0)d(0)+a_{j}(0)A_{l}c(0)c^{\dagger}(0)d^{\dagger}(0)d(0)-a_{i}^{\dagger}(0)a_{i}(0)a_{j}(0)c^{\dagger}(0)c(0)\right)\\
 & - & \left.\chi^{*}g^{*}a_{j}(0)A_{l}b(0)c^{2}(0)d^{\dagger}(0)-\chi ga_{j}(0)A_{l}b^{\dagger}(0)c^{\dagger2}(0)d(0)\right)\\
{}[H,[H,[H,a_{j}(0)]]] & = & \prod_{i=1:i\neq j}^{k}\sum_{i,m=1:i\neq j\neq m}^{k}\left(-\omega_{j}^{3}a_{j}(0)+g^{*}\left(\omega_{j}^{2}+\omega_{i}^{2}+\omega_{b}^{2}+\omega_{c}^{2}-\omega_{i}\omega_{j}\right.\right.\\
 & + & \left.\omega_{j}\omega_{b}+\omega_{j}\omega_{c}+2\omega_{i}\omega_{m}-2\omega_{i}\omega_{b}-2\omega_{i}\omega_{c}+2\omega_{b}\omega_{c}\right)a_{i}^{\dagger}(0)b(0)c(0)\\
 & + & \chi\left(\omega_{j}^{2}+\omega_{i}^{2}+\omega_{c}^{2}+\omega_{d}^{2}-\omega_{i}\omega_{j}-\omega_{j}\omega_{c}+\omega_{j}\omega_{d}+2\omega_{i}\omega_{c}\right.\\
 & + & \left.2\omega_{i}\omega_{m}-2\omega_{i}\omega_{d}-2\omega_{c}\omega_{d}\right)a_{i}^{\dagger}(0)c^{\dagger}(0)d(0)\\
 & - & |g|^{2}\left(2\omega_{j}+\omega_{i}-\omega_{b}-\omega_{c}\right)\left(a_{j}(0)A_{l}b^{\dagger}(0)b(0)c^{\dagger}(0)c(0)\right.\\
 & - & \left.a_{i}^{\dagger}(0)a_{i}(0)a_{j}(0)b(0)b^{\dagger}(0)-a_{i}^{\dagger}(0)a_{i}(0)a_{j}(0)c^{\dagger}(0)c(0)\right)\\
 & - & |\chi|^{2}\left(-2\omega_{j}+\omega_{i}+\omega_{c}-\omega_{d}\right)\left(a_{j}(0)a_{i}^{\dagger}(0)a_{i}(0)d^{\dagger}(0)d(0)\right.\\
 & + & \left.a_{j}(0)A_{l}c(0)c^{\dagger}(0)d^{\dagger}(0)d(0)-a_{i}^{\dagger}(0)a_{i}(0)a_{j}(0)c^{\dagger}(0)c(0)\right)\\
 & - & \chi^{*}g^{*}\left(-2\omega_{j}+\omega_{i}-2\omega_{b}-3\omega_{c}+\omega_{d}\right)a_{j}(0)A_{l}b(0)c^{2}(0)d^{\dagger}(0)\\
 & - & \chi g\left(-2\omega_{j}+\omega_{i}+\omega_{b}+3\omega_{c}-2\omega_{d}\right)a_{j}(0)A_{l}b^{\dagger}(0)c^{\dagger2}(0)d(0)\\
 & + & \left.\chi g^{*}\left(-\omega_{b}-2\omega_{c}+\omega_{d}\right)a_{j}^{\dagger}(0)a_{i}^{\dagger2}(0)b(0)d(0)\right).
\end{array}\label{eq:commutators}
\end{equation}

\end{widetext} 

\noindent It is important to note here that not a single new function
of creation and annihilation operators is obtained in all the commutators
after the third term in Eq. (\ref{eq:EoM-2}), after neglecting the
terms beyond quadratic powers of $g$, $\chi$ and their complex conjugates
to remain consistent with the perturbative method. 

All these different functions of creation and annihilation operators
occurring in these commutators may now be used to write the obtained solution
(\ref{eq:solution}) with unknown time dependent coefficients (such
as $f_{i}$). To obtain the functional form of these unknown coefficients
we substitute the obtained solution given in Eq. (\ref{eq:solution})
in Heisenberg's equations of motion (\ref{eq:field operators}) for
various field modes, one can easily obtain the coupled differential
equations for all $f_{i}$, $g_{i},$ $h_{i},$ and $l_{i}$ as follows
\begin{equation}
\begin{array}{lcl}
\dot{f}_{1} & = & -i\omega_{j}f_{1},\\
\dot{f}_{2} & = & -i\omega_{j}f_{2}+ig^{*}\prod_{i=2}^{k}\left(f_{1_{i}}^{*}\right)g_{1}h_{1},\\
\dot{f}_{3} & = & -i\omega_{j}f_{3}+i\chi\prod_{i=2}^{k}\left(f_{1_{i}}^{*}\right)h_{1}^{*}l_{1},\\
\dot{f}_{4} & = & -i\omega_{j}f_{4}+ig^{*}\prod_{i=2:i\neq l}^{k}\left(f_{1_{i}}^{*}f_{2_{l}}^{*}\right)g_{1}h_{1},\\
\dot{f}_{5} & = & -i\omega_{j}f_{5}+ig^{*}\prod_{i=2}^{k}\left(f_{1_{i}}^{*}\right)g_{1}h_{2},\\
\dot{f}_{6} & = & -i\omega_{j}f_{6}+ig^{*}\prod_{i=2}^{k}\left(f_{1_{i}}^{*}\right)g_{2}h_{1},\\
\dot{f}_{7} & = & -i\omega_{j}f_{7}+ig^{*}\prod_{i=2:i\neq l}^{k}\left(f_{1_{i}}^{*}f_{3_{l}}^{*}\right)g_{1}h_{1},\\
\dot{f}_{8} & = & -i\omega_{j}f_{8}+ig^{*}\prod_{i=2}^{k}\left(f_{1_{i}}^{*}\right)g_{1}h_{3}+i\chi\prod_{i=2}^{k}\left(f_{1_{i}}^{*}\right)h_{2}^{*}l_{1},\\
\dot{f}_{9} & = & -i\omega_{j}f_{9}+i\chi\prod_{i=2:i\neq l}^{k}\left(f_{1_{i}}^{*}f_{2_{l}}^{*}\right)h_{1}^{*}l_{1},\\
\dot{f}_{10} & = & -i\omega_{j}f_{10}+i\chi\prod_{i=2}^{k}\left(f_{1_{i}}^{*}\right)h_{3}^{*}l_{1},\\
\dot{f}_{11} & = & -i\omega_{j}f_{11}+i\chi\prod_{i=2:i\neq l}^{k}\left(f_{1_{i}}^{*}f_{3_{l}}^{*}\right)h_{1}^{*}l_{1},\\
\dot{f}_{12} & = & -i\omega_{j}f_{12}+i\chi\prod_{i=2}^{k}\left(f_{1_{i}}^{*}\right)h_{1}^{*}l_{2},
\end{array}\label{eq:differential equ for a1}
\end{equation}
 
\begin{equation}
\begin{array}{lcl}
\dot{g}_{1} & = & -i\omega_{b}g_{1},\\
\dot{g}_{2} & = & -i\omega_{b}g_{2}+ig\prod_{i=1}^{k}\left(f_{1_{i}}\right)h_{1}^{*},\\
\dot{g}_{3} & = & -i\omega_{b}g_{3}+ig\prod_{i=1}^{k}\left(f_{1_{i}}\right)h_{3}^{*},\\
\dot{g}_{4} & = & -i\omega_{b}g_{4}+ig\prod_{i=1}^{k}\left(f_{1_{i}}\right)h_{2}^{*},\\
\dot{g}_{5} & = & -i\omega_{b}g_{5}+ig\prod_{i=1:i\neq l}^{k}\left(f_{1_{i}}f_{2_{l}}\right)h_{1}^{*},\\
\dot{g}_{6} & = & -i\omega_{b}g_{6}+ig\prod_{i=1:i\neq l}^{k}\left(f_{1_{i}}f_{3_{l}}\right)h_{1}^{*},
\end{array}\label{eq:differential equ for b}
\end{equation}
\begin{equation}
\begin{array}{lcl}
\dot{h}_{1} & = & -i\omega_{c}h_{1},\\
\dot{h}_{2} & = & -i\omega_{c}h_{2}+ig\prod_{i=1}^{k}\left(f_{1_{i}}\right)g_{1}^{*},\\
\dot{h}_{3} & = & -i\omega_{c}h_{3}+i\chi\prod_{i=1}^{k}\left(f_{1_{i}}^{*}\right)l_{1},\\
\dot{h}_{4} & = & -i\omega_{c}h_{4}+ig\prod_{i=1}^{k}\left(f_{1_{i}}\right)g_{2}^{*},\\
\dot{h}_{5} & = & -i\omega_{c}h_{5}+ig\prod_{i=1:i\neq l}^{k}\left(f_{1_{i}}f_{2_{l}}\right)g_{1}^{*},\\
\dot{h}_{6} & = & -i\omega_{c}h_{6}+ig\prod_{i=1:i\neq l}^{k}\left(f_{1_{i}}f_{3_{l}}\right)g_{1}^{*}\\
 & + & i\chi\sum_{i=1:i\neq l}^{k}\left(f_{1_{i}}^{*}f_{2_{l}}^{*}\right)l_{1},\\
\dot{h}_{7} & = & -i\omega_{c}h_{7}+i\chi\prod_{i=1:i\neq l}^{k}\left(f_{1_{i}}^{*}f_{3_{l}}^{*}\right)l_{1},\\
\dot{h}_{8} & = & -i\omega_{c}h_{8}+i\chi\prod_{i=1}^{k}\left(f_{1_{i}}^{*}\right)l_{2},
\end{array}\label{eq:differential equ for c}
\end{equation}
\begin{equation}
\begin{array}{lcl}
\dot{l}_{1} & = & -i\omega_{d}l_{1}\\
\dot{l}_{2} & = & -i\omega_{d}l_{2}+i\chi^{*}\prod_{i=1}^{k}\left(f_{1_{i}}\right)h_{1}\\
\dot{l}_{3} & = & -i\omega_{d}l_{3}+i\chi^{*}\prod_{i=1}^{k}\left(f_{1_{i}}\right)h_{2}\\
\dot{l}_{4} & = & -i\omega_{d}l_{4}+i\chi^{*}\prod_{i=1:i\neq l}^{k}\left(f_{1_{i}}f_{2_{l}}\right)h_{1}\\
\dot{l}_{5} & = & -i\omega_{d}l_{5}+i\chi^{*}\prod_{i=1:i\neq l}^{k}\left(f_{1_{i}}f_{3_{l}}\right)h_{1}\\
\dot{l}_{6} & = & -i\omega_{d}l_{6}+i\chi^{*}\prod_{i=1}^{k}\left(f_{1_{i}}\right)h_{3}
\end{array}\label{eq:differential equ for d}
\end{equation}
The solutions of these differential equations are obtained using the
boundary conditions $F_{1}=1$ and $F_{i}=0\,\forall i\neq1$ with
$F=\left\{ f,g,h,l\right\} $ and are listed in Appendix A. The obtained
solution is expected to satisfy the equal time commutation relation
(ETCR) as 
\begin{equation}
\begin{array}{lcl}
\left[a_{j}(t),a_{j}^{\dagger}(t)\right] & = & \prod_{i=1}^{k}\left[1+\left\{ \left(f_{1}^{*}f_{4}+{\rm c.c.}\right)-|f_{2}|^{2}\right\} A_{l}\right.\\
 & \times & b^{\dagger}(0)b(0)c^{\dagger}(0)c(0)+a_{i}^{\dagger}(0)a_{i}(0)\left(b(0)b^{\dagger}(0)\right.\\
 & \times & \left\{ \left(f_{1}^{*}f_{5}+{\rm c.c.}\right)+|f_{2}|^{2}\right\} +c^{\dagger}(0)c(0)\\
 & \times & \left.\left\{ \left(f_{1}^{*}f_{6}+f_{1}^{*}f_{12}+{\rm c.c.}\right)+|f_{2}|^{2}+|f_{3}|^{2}\right\} \right)\\
 & + & \left\{ \left(f_{7}^{*}f_{1}+f_{1}^{*}f_{9}-f_{2}^{*}f_{3}\right)A_{l}b^{\dagger}(0)c^{\dagger2}(0)d(0)\right.\\
 & + & \left.{\rm c.c.}\right\} +\left(\left\{ \left(f_{1}^{*}f_{11}+{\rm c.c.}\right)-|f_{3}|^{2}\right\} A_{l}\right.\\
 & \times & c(0)c^{\dagger}(0)+a_{i}^{\dagger}(0)a_{i}(0)\\
 & \times & \left.\left.\left\{ \left(f_{1}^{*}f_{10}+{\rm c.c.}\right)+|f_{3}|^{2}\right\} \right)d^{\dagger}(0)d(0)\right]\\
 & = & 1.
\end{array}\label{eq:ETCR a1}
\end{equation}
Similarly, we can calculate ETCR for all other modes as follows 
\begin{equation}
\begin{array}{lcl}
\left[b(t),b^{\dagger}(t)\right] & = & \prod_{i=1}^{k}\left[1+\left\{ \left(g_{1}^{*}g_{4}+{\rm c.c.}\right)-|g_{2}|^{2}\right\} a_{i}^{\dagger}(0)a_{i}(0)\right.\\
 & + & \left.\left\{ \left(g_{1}^{*}g_{5}+{\rm c.c.}\right)+|g_{2}|^{2}\right\} A_{l}c^{\dagger}(0)c(0)\right]\\
 & = & 1,
\end{array}\label{eq:ETCR b}
\end{equation}
\begin{equation}
\begin{array}{lcl}
\left[c(t),c^{\dagger}(t)\right] & = & \prod_{i=1}^{k}\left[1+\left\{ \left(h_{1}^{*}h_{4}+h_{1}^{*}h_{8}+{\rm c.c.}\right)\right.\right.\\
 & - & \left.|h_{2}|^{2}+|h_{3}|^{2}\right\} a_{i}^{\dagger}(0)a_{i}(0)\\
 & + & \left\{ \left(h_{1}^{*}h_{5}+{\rm c.c.}\right)+|h_{2}|^{2}\right\} A_{l}b^{\dagger}(0)b(0)\\
 & + & \left.\left\{ \left(h_{1}^{*}h_{7}+{\rm c.c.}\right)-|h_{3}|^{2}\right\} A_{l}d^{\dagger}(0)d(0)\right]\\
 & = & 1,
\end{array}\label{eq:ETCR c}
\end{equation}
and
\begin{equation}
\begin{array}{lcl}
\left[d(t),d^{\dagger}(t)\right] & = & \prod_{i=1}^{k}\left[1+\left\{ \left(l_{1}^{*}l_{6}+{\rm c.c.}\right)+|l_{2}|^{2}\right\} a_{i}(0)a_{i}^{\dagger}(0)\right.\\
 & + & \left.\left\{ \left(l_{1}^{*}l_{5}+{\rm c.c.}\right)+|l_{2}|^{2}\right\} A_{l}c^{\dagger}(0)c(0)\right]\\
 & = & 1.
\end{array}\label{eq:ETCR d}
\end{equation}
We have also verified that various constants of motion for this system given in Ref. \cite{perina1991quantum} are satisfied by the present solution. Specifically, we have checked that
$C_{1}=\left(a_{j}^{\dagger}(t)a_{j}(t)+b^{\dagger}(t)b(t)+d^{\dagger}(t)d(t)\right)$
is a constant of motion \cite{perina1991quantum} for an arbitrary
pump mode, i.e., 
\begin{table}
\centering{}%
\begin{tabular}{|>{\centering}p{2.5cm}|>{\centering}p{2.5cm}|>{\centering}p{2.5cm}|}
\hline 
Multi-photon pump hyper-Raman case & Degenerate hyper-Raman case \cite{sen2007squeezing} & Raman case \cite{sen2005squeezed}\tabularnewline
\hline 
$g_{1}$ & $g_{1}^{'}$ & $g_{1}^{''}$\tabularnewline
\hline 
$g_{2}$ & $g_{2}^{'}$ & $g_{2}^{''}$\tabularnewline
\hline 
$g_{3}$ & $g_{5}^{'}$ & $g_{3}^{''}$\tabularnewline
\hline 
$g_{4}$ & $g_{8}^{'}$ & $g_{6}^{''}$\tabularnewline
\hline 
$g_{5}$ & $g_{6}^{'}$, $g_{7}^{'}$ & $g_{5}^{''}$\tabularnewline
\hline 
$g_{6}$ & $g_{3}^{'}$, $g_{4}^{'}$ & $g_{4}^{''}$\tabularnewline
\hline 
$h_{1}$ & $h_{1}^{'}$ & $h_{1}^{''}$\tabularnewline
\hline 
$h_{2}$ & $h_{2}^{'}$ & $h_{2}^{''}$\tabularnewline
\hline 
$h_{3}$ & $h_{3}^{'}$ & $h_{3}^{''}$\tabularnewline
\hline 
$h_{4}$ & $h_{8}^{'}$ & $h_{5}^{''}$\tabularnewline
\hline 
$h_{5}$ & $h_{6}^{'}$, $h_{7}^{'}$ & $h_{6}^{''}$\tabularnewline
\hline 
$h_{6}$ & $h_{4}^{'}$, $h_{5}^{'}$ & $h_{4}^{''}$\tabularnewline
\hline 
$h_{7}$ & $h_{9}^{'}$, $h_{10}^{'}$ & $h_{7}^{''}$\tabularnewline
\hline 
$h_{8}$ & $h_{11}^{'}$ & $h_{8}^{''}$\tabularnewline
\hline 
$l_{1}$ & $l_{1}^{'}$ & $l_{1}^{''}$\tabularnewline
\hline 
$l_{2}$ & $l_{2}^{'}$ & $l_{2}^{''}$\tabularnewline
\hline 
$l_{3}$ & $l_{3}^{'}$ & $l_{3}^{''}$\tabularnewline
\hline 
$l_{4}$ & $l_{4}^{'}$, $l_{5}^{'}$ & $l_{4}^{''}$\tabularnewline
\hline 
$l_{5}$ & $l_{6}^{'}$, $l_{7}^{'}$ & $l_{5}^{''}$\tabularnewline
\hline 
$l_{6}$ & $l_{8}^{'}$ & $l_{6}^{''}$\tabularnewline
\hline 
\end{tabular}\caption{\label{tab:non-pump} The relationship among the various functions
in the solution obtained here and the existing solutions for Raman
and degenerate hyper-Raman processes for the functions in the evolution
of modes except pump mode. Here, we have used a (two) prime(s) in
the superscript of the functions $F_{i}$s to distinguish the present
solution from the degenerate hyper-Raman (Raman) process. }
\end{table}
\begin{equation}
\begin{array}{lcl}
C_{1} & = & a_{j}^{\dagger}(t)a_{j}(t)+b^{\dagger}(t)b(t)+d^{\dagger}(t)d(t)\\
 & = & \prod_{i=1}^{k}\left[a_{j}^{\dagger}(0)a_{j}(0)+b^{\dagger}(0)b(0)+d^{\dagger}(0)d(0)\right.\\
 & + & \left\{ \left(f_{1}^{*}f_{2}+g_{2}^{*}g_{1}\right)a_{i}^{\dagger}(0)b(0)c(0)+{\rm c.c.}\right\} \\
 & + & \left\{ \left(f_{1}^{*}f_{3}+l_{2}^{*}l_{1}\right)a_{i}^{\dagger}(0)c^{\dagger}(0)d(0)+{\rm c.c.}\right\} \\
 & + & \left(f_{1}^{*}f_{4}+g_{1}^{*}g_{5}+{\rm c.c.}\right)A_{l}b^{\dagger}(0)b(0)c^{\dagger}(0)c(0)\\
 & + & \left\{ \left(f_{1}^{*}f_{6}+f_{1}^{*}f_{12}+{\rm c.c.}\right)+|g_{2}|^{2}+|l_{2}|^{2}\right\} \\
 & \times & a_{i}^{\dagger}(0)a_{i}(0)c^{\dagger}(0)c(0)\\
 & + & \left(f_{1}^{*}f_{5}+g_{1}^{*}g_{4}+{\rm c.c.}\right)a_{i}^{\dagger}(0)a_{i}(0)b^{\dagger}(0)b(0)\\
 & + & \left\{ \left(f_{1}^{*}f_{5}+{\rm c.c.}\right)+|g_{2}|^{2}\right\} a_{i}^{\dagger}(0)a_{i}(0)\\
 & + & \left\{ \left(f_{1}^{*}f_{7}+f_{9}^{*}f_{1}+g_{6}^{*}g_{1}+l_{4}^{*}l_{1}\right)\right.\\
 & \times & \left.A_{l}b(0)c^{2}(0)d^{\dagger}(0)+{\rm c.c.}\right\} \\
 & + & \left\{ \left(f_{1}^{*}f_{8}+g_{3}^{*}g_{1}+l_{3}^{*}l_{1}\right)a_{i}^{\dagger2}(0)b(0)d(0)+{\rm c.c.}\right\} \\
 & + & \left(f_{1}^{*}f_{10}+l_{1}^{*}l_{6}+{\rm c.c.}\right)a_{i}^{\dagger}(0)a_{i}(0)d^{\dagger}(0)d(0)\\
 & + & \left.\left(f_{1}^{*}f_{11}+l_{1}^{*}l_{5}+{\rm c.c.}\right)A_{l}c(0)c^{\dagger}(0)d^{\dagger}(0)d(0)\right]\\
 & = & a_{j}^{\dagger}(0)a_{j}(0)+b^{\dagger}(0)b(0)+d^{\dagger}(0)d(0)\\
 & = & {\rm constant}.
\end{array}\label{eq:Constant of motion 1}
\end{equation}
Similarly, the present solution satisfies another constant of motion
$C_{2}=a_{j}^{\dagger}(t)a_{j}(t)-a_{r}^{\dagger}(t)a_{r}(t)$ for
two arbitrary pump modes, as follows 
\begin{equation}
\begin{array}{lcl}
C_{2} & = & a_{j}^{\dagger}(t)a_{j}(t)-a_{r}^{\dagger}(t)a_{r}(t)\\
 & = & \prod_{i=1}^{r}\left[a_{j}^{\dagger}(0)a_{j}(0)-a_{r}^{\dagger}(0)a_{r}(0)\right.\\
 & + & \left\{ \left(f_{1}^{*}f_{2}-r_{1}^{*}r_{2}\right)a_{i}^{\dagger}(0)b(0)c(0)+{\rm c.c.}\right\} \\
 & + & \left\{ \left(f_{1}^{*}f_{3}-r_{1}^{*}r_{3}\right)a_{i}^{\dagger}(0)c^{\dagger}(0)d(0)+{\rm c.c.}\right\} \\
 & + & \left(|f_{2}|^{2}-|r_{2}|^{2}\right)A_{l}b^{\dagger}(0)b(0)c^{\dagger}(0)c(0)\\
 & + & \left(f_{1}^{*}f_{5}-r_{1}^{*}r_{5}+{\rm c.c.}\right)a_{i}^{\dagger}(0)a_{i}(0)b(0)b^{\dagger}(0)\\
 & + & \left(f_{1}^{*}f_{6}+f_{1}^{*}f_{12}-r_{1}^{*}r_{6}-r_{1}^{*}r_{12}+{\rm c.c.}\right)\\
 & \times & a_{i}^{\dagger}(0)a_{i}(0)c^{\dagger}(0)c(0)\\
 & + & \left\{ \left(f_{1}^{*}f_{7}+f_{9}^{*}f_{1}-r_{3}^{*}r_{2}\right)A_{l}b(0)c^{2}(0)d^{\dagger}(0)+{\rm c.c.}\right\} \\
 & + & \left\{ \left(f_{1}^{*}f_{8}-r_{1}^{*}r_{8}\right)a_{i}^{\dagger2}(0)b(0)d(0)+{\rm c.c.}\right\} \\
 & + & \left(f_{1}^{*}f_{10}-r_{1}^{*}r_{10}+{\rm c.c.}\right)a_{i}^{\dagger}(0)a_{i}(0)d^{\dagger}(0)d(0)\\
 & + & \left.\left(|f_{3}|^{2}-|r_{3}|^{2}\right)A_{l}c(0)c^{\dagger}(0)d^{\dagger}(0)d(0)\right]\\
 & = & a_{j}^{\dagger}(0)a_{j}(0)-a_{r}^{\dagger}(0)a_{r}(0)\\
 & = & {\rm constant}.
\end{array}\label{eq:constant of motion 3}
\end{equation}
The last constant of motion $C_{3}=c^{\dagger}(t)c(t)+d^{\dagger}(t)d(t)-b^{\dagger}(t)b(t)$
is also verified as follows 
\begin{equation}
\begin{array}{lcl}
C_{3} & = & c^{\dagger}(t)c(t)+d^{\dagger}(t)d(t)-b^{\dagger}(t)b(t)\\
 & = & \prod_{i=1}^{k}\left[c^{\dagger}(0)c(0)+d^{\dagger}(0)d(0)-b^{\dagger}(0)b(0)\right.\\
 & + & \left\{ \left(h_{1}^{*}h_{2}-g_{1}^{*}g_{2}\right)a_{i}(0)b^{\dagger}(0)c^{\dagger}(0)+{\rm c.c.}\right\} \\
 & + & \left\{ \left(h_{1}^{*}h_{3}+l_{2}^{*}l_{1}\right)a_{i}^{\dagger}(0)c^{\dagger}(0)d(0)+{\rm c.c.}\right\} \\
 & + & \left\{ \left(h_{1}^{*}h_{4}+h_{1}^{*}h_{8}+{\rm c.c.}\right)-|g_{2}|^{2}+|l_{2}|^{2}\right\} \\
 & \times & a_{i}^{\dagger}(0)a_{i}(0)c^{\dagger}(0)c(0)\\
 & + & \left(h_{1}^{*}h_{5}-g_{1}^{*}g_{5}+{\rm c.c.}\right)A_{l}b^{\dagger}(0)b(0)c^{\dagger}(0)c(0)\\
 & + & \left(h_{1}^{*}h_{6}-g_{1}^{*}g_{6}+l_{4}^{*}l_{1}+{\rm c.c.}\right)A_{l}b^{\dagger}(0)c^{\dagger2}(0)d(0)\\
 & + & \left(l_{1}^{*}l_{9}-h_{1}^{*}h_{9}+k_{5}^{*}k_{1}+{\rm c.c.}\right)a_{2}^{\dagger}(0)a_{2}(0)\\
 & \times & b^{\dagger}(0)c^{\dagger2}(0)d(0)+\left(h_{1}^{*}h_{7}+l_{1}^{*}l_{5}+{\rm c.c.}\right)A_{l}\\
 & \times & c^{\dagger}(0)c(0)d^{\dagger}(0)d(0)+\left(|h_{2}|^{2}-|g_{2}|^{2}\right)a_{i}^{\dagger}(0)a_{i}(0)\\
 & + & \left\{ \left(l_{1}^{*}l_{6}+{\rm c.c.}\right)+|h_{3}|^{2}\right\} a_{i}(0)a_{i}^{\dagger}(0)d^{\dagger}(0)d(0)\\
 & + & \left.\left(h_{2}^{*}h_{3}-g_{3}^{*}g_{1}+l_{3}^{*}l_{1}+{\rm c.c.}\right)a_{i}^{\dagger2}(0)b(0)d(0)\right]\\
 & = & c^{\dagger}(0)c(0)+d^{\dagger}(0)d(0)-b^{\dagger}(0)b(0)\\
 & = & {\rm constant}.
\end{array}\label{eq:constant of motion 4}
\end{equation}

It is previously mentioned that the present solution is general in
nature and the solutions reported earlier for degenarate hyper-Raman process \cite{sen2007squeezing} and Raman process \cite{sen2005squeezed,sen2008amplitude,sen2011sub,sen2007quantum}  can be obtained from the present solution as special cases. In Tables \ref{tab:pump} and \ref{tab:non-pump}, 
we have established
a one-to-one correspondence between the functions obtained in the present solution and the same reported in the 
solutions for Raman and degenerate hyper-Raman processes. \end{document}